\documentclass[fleqn,usenatbib]{mnras}

\usepackage{newtxtext,newtxmath}

\usepackage[T1]{fontenc}

\DeclareRobustCommand{\VAN}[3]{#2}
\let\VANthebibliography\thebibliography
\def\thebibliography{\DeclareRobustCommand{\VAN}[3]{##3}\VANthebibliography}

\usepackage{graphicx}	
\usepackage{amsmath}	
\usepackage{pdflscape}
\usepackage{soul}

\newcommand{\cubi}{$\Delta_{\rm C\, U,B,I}$}

\newcommand{\dbi}{$\Delta_{\rm B,I}$}

\newcommand{\teff}{\mbox{T$_{\rm eff}$}}
\newcommand{\logg}{\mbox{log~{\it g}}}
\newcommand{\vmicro}{\mbox{$\xi_{\rm t}$}}
\newcommand{\kmsec}{\mbox{km~s$^{\rm -1}$}}
\newcommand{\loggf}{\mbox{$\log gf$}}

\newcommand{\x}{\mbox{$\Delta_{\tiny{\mathrm{F275W,F814W}}}$}}
\newcommand{\y}{\mbox{$\Delta_{\tiny{C~\mathrm{ F275W,F336W,F438W}}}$}}

\title[ngc 2808]{The chemical compositions of multiple stellar populations in the globular cluster NGC\,2808}

\author[M. Carlos et al.]{
M. Carlos,$^{1}$\thanks{E-mail: mariliagabriela.carlos@unipd.it} 
 A. F. Marino,$^{2,3}$ A. P. Milone,$^{1,3}$ E. Dondoglio,$^{1}$ S. Jang,$^{1}$  M. V. Legnardi,$^{1}$ A. Mohandasan,$^{1}$ \newauthor 
 G. Cordoni,$^{1}$ E. P. Lagioia$^{1}$, A. M. Amarsi,$^4$ and H. Jerjen$^{5}$
\\
$^{1}$Dipartimento di Fisica e Astronomia “Galileo Galilei”, Università di Padova, Vicolo dell’Osservatorio 3, 35122, Padova, Italy\\
$^{2}$Istituto Nazionale di Astrofisica - Osservatorio Astrofisico di Arcetri, Largo Enrico Fermi, 5, I-50125 Firenze, Italy\\
$^{3}$Istituto Nazionale di Astrofisica - Osservatorio Astronomico di Padova, Vicolo dell’Osservatorio 5, I-35122 Padua, Italy\\
$^{4}$ Theoretical Astrophysics, Department of Physics and Astronomy, Uppsala University, Box 516, SE-751 20 Uppsala, Sweden\\
$^{5}$  Research School of Astronomy and Astrophysics, Australian National University, Canberra, ACT 2611, Australia\\
}

\date{Accepted 2022 November 30. Received 2022 November 30; in original form 2022 August 29}

\pubyear{2022}

\begin{document}
\label{firstpage}
\pagerange{\pageref{firstpage}--\pageref{lastpage}}
\maketitle

\definecolor{green}{RGB}{59,142,65}

\begin{abstract}

Pseudo two-colour diagrams or Chromosome maps (ChM) indicate that NGC\,2808 host five different stellar populations. The existing ChMs have been derived by the Hubble Space Telescope photometry, and comprise of stars in a small field of view around the cluster centre. To overcome these limitations, we built a ChM with $U,B,I$ photometry from ground-based facilities that disentangle the multiple stellar populations of NGC\,2808 over a wider field of view. We used spectra collected by GIRAFFE@VLT in a sample of 70 red giant branch (RGB) and seven asymptotic giant branch (AGB) stars to infer the abundances of C, N, O, Al, Fe, and Ni, which combined with literature data for other elements (Li, Na, Mg, Si, Ca, Sc, Ti, Cr and Mn), and together with both the classical and the new ground-based ChMs, provide the most complete chemical characterisation of the stellar populations in NGC\,2808 available to date. As typical of the multiple population phenomenon in globular clusters, the light elements vary from one stellar population to another; whereas the iron peak elements show negligible variation between the different populations (at a level of $\lesssim0.10$~dex). Our AGB stars are also characterised by the chemical variations associated with the presence of multiple populations, confirming that this phase of stellar evolution is affected by the phenomenon as well. Intriguingly, we detected one extreme O-poor AGB star (consistent with a high He abundance), challenging stellar evolution models which suggest that highly He-enriched stars should avoid the AGB phase and evolve as AGB-manqué star.

\end{abstract}




\section{Introduction} \label{sec:intro}
In the past fifteen years, the overwhelming observational evidence of multiple stellar populations in globular clusters (GCs) has  challenged the idea that a GC consists of stars born at the same time out of the same material \citep[see][for recent reviews]{bastian/2018,gratton/2019,milone/2022}. 



Thanks to the introduction of 
 new and powerful tools to investigate the different populations of stars in GCs, we now know that the phenomenon of  multiple stellar populations is complex, and we still lack a convincing satisfactory explanation.
Studies based on the {\it Hubble Space Telescope} ($HST$) high precision photometry have revealed that the interweave of multiple stellar populations is variegate.
 Nevertheless, some general patterns have been identified, possibly pointing to a general formation scenario. Indeed, the cluster-to-cluster differences may be ascribed to {\it external} conditions, such as the cluster mass or the birth-site \citep[e.g.][]{milone/17, lagioia/2019}.


The ``chromosome map'' (ChM) diagram, introduced by \cite{milone/15}, is a formidable tool to enhance our knowledge about the multiple stellar population phenomenon. This photometric diagram is able to maximise the separation of different stellar populations on a plane constructed by properly combining the 
 F275W, F336W, F438W, and F814W $HST$ filters. As discussed by Milone and coauthors, the {\it x} axis of this map (\x) is mostly sensitive to 
   effective temperature variations associated with helium differences while the {\it y} axis (\y) is highly sensitive to atmospheric chemical abundances. The major players in shaping the maps are light elements, primarily nitrogen, whose abundances impact on the atmospheres of the stars, together with He and the overall metallicity, as stars with distinct He and/or metals have  different internal structure, being indeed described by distinct isochrones \citep{milone/15,milone/18}.

A great variety of morphologies can be identified in the ChMs, with the GCs displaying different sub-structures and minor populations never observed before. The existence of a separate class of clusters with variations in the overall metallicity has been clearly assessed, and defined as the group of objects displaying a Type~II ChM morphology, to be distinguished from the Type~I clusters (see \citealt{milone/17} for details). 

In an effort of providing a chemical {\it key to read} ChMs, \citet{marino/19} exploited spectroscopic elemental abundances from the literature to assign the chemical composition to the distinct populations as identified on the ChMs of different GCs.
Among the investigated chemical species, those with a larger number of measurements in the literature, the element 
that was found to best correlate with the ChM pattern was Na, and a general empirical relation was even provided between \y\ and the abundances of this element.
However, although Na is probably the most studied element in the context of multiple stellar populations, the Na abundance itself does not directly affect any of the fluxes in the passbands used to construct ChMs. Instead these fluxes are affected by nitrogen coming from the destruction of carbon and oxygen. 

A first limit in the analysis by \citet{marino/19} was that not many abundances of N were available on the ChM, so we still lack a direct spectroscopic investigation on how this species, and possibly C, affects the distribution of stars in these photometric diagrams.
A second limit 
is that the assignment of stars to different stellar populations is basically confined to the small field observed with $HST$ cameras, from which ChMs have been constructed.
Hence, no information about the behaviour of stellar populations in the outer parts of the GCs was obtained.  

In this work we analyse chemical abundances of different stellar populations in the GC NGC\,2808, 
attempting a {\it population assignment} to  our entire sample of stars by exploiting new photometric diagrams constructed from ground-based photometry in a larger field of view.

Among GCs with a Type~I ChM morphology, 
NGC\,2808 displays the most spectacular map in terms of extension and number of stellar populations.
At least five main red giant branch (RGB) clumps of stars have been identified in the ChM of this GC. Its first population hosts two groups of stars designated as population A and B, whereas the second population is composed of three distinct sub-populations, namely C, D, and E \citep{milone/15}. 
NGC\,2808 is also one of the GCs with larger He internal variations with the E population being enhanced by $\Delta\rm{(Y)}=0.089\pm0.010$ with respect to the first population \citep{milone/18}. 
 Since NGC\,2808 has a relatively large mass ($\rm{M}=7.42\pm0.05 \times 10^5~\rm {M_{\odot}}$, \citealt{baumgardt/18}),
  its wide helium spread corroborates the evidence that massive GCs exhibit extreme chemical composition.
Stellar populations enhanced in He have also been detected among stars in different evolutionary stages, namely from its split main sequence \citep{dantona/05,piotto/07,milone/15}, and then, through direct analysis of He spectral lines, in its blue horizontal branch (HB) 
\citep{marino/14}. 

Being among the most fascinating clusters from the multiple stellar populations perspective, NGC\,2808 
has a large number of spectroscopic studies in literature, from high resolution in-depth analysis that might include the light, iron-peak and {\it s-}process elements in stars at different life stages (e.g., \citealt{marino/17} for asymptotic giant branch stars (AGB), and \citealt{carretta/15} and \citealt{meszaros/20} for RGB stars), passing through the detailed inspection of specific elements (such as the observations of Na-O anti-correlation in the RGB and HB in \citealt{carretta/06} and \citealt{gratton/11}, respectively, or the work on Li in RGB stars from \citealt{dorazi/15}, and the Mg-Al anti-correlation from \citealt{pancino/2017} and \citealt{carretta/18})
to low resolution spectra studies \citep[e.g.,][]{latour/19,hong/21}. 


Our work aims to provide a full chemical characterisation of the distinct stellar populations observed on the ChM of this intriguing cluster.
 To do this, we use high-resolution spectra to infer new abundances for C, N, along with O, Al, Ni and Fe, for 77 giants (70 RGB + 7 AGB) and take advantage of the literature abundances for other elements.
Our results, based on the largest number of stars with ChM information, will add valuable constraints to the scenarios of formation and enrichment of this GC. 

The paper is organised as follows: Section 2 presents the photometric and spectroscopic dataset, Section 3 contains the  analysis of the detailed spectroscopic study, results and discussion are in Section 4, and Section 5 provides the summary and conclusions.

\section{Data}\label{sec:sample}
We present a photometric and spectroscopic analysis of giant stars in NGC\,2808, to inspect the chemical abundances of the different stellar populations identified on the ChM.
Our analysis includes stars in the cluster central region for which we exploit {\it HST} photometry and in the external field by using ground based photometry. In the following subsections we describe the photometric and spectroscopic dataset employed here.

\subsection{Photometric dataset}
Our photometric dataset includes both ground-based and $HST$  photometry. 
  The ground-based photometry is provided by the  catalogue by\,\citet{stetson/19} and is obtained from images collected with different ground-based facilities. It consists in $U$, $B$, $V$, and $I$ photometry for stars in a $6.7$~arcmin radius centred on NGC\,2808.
  The photometry has been corrected for differential reddening as in \citet[][]{milone/2012}.
  In addition, we used stellar proper motions from the Gaia early data release 3 \citep[Gaia eDR3][]{gaia_col_edr3/21} to separate the bulk of cluster members from field stars \citep[see][for details on the procedure to select the probable cluster members]{cordoni/18}.
  
\begin{figure*}
    \centering
    \includegraphics[width=12cm]{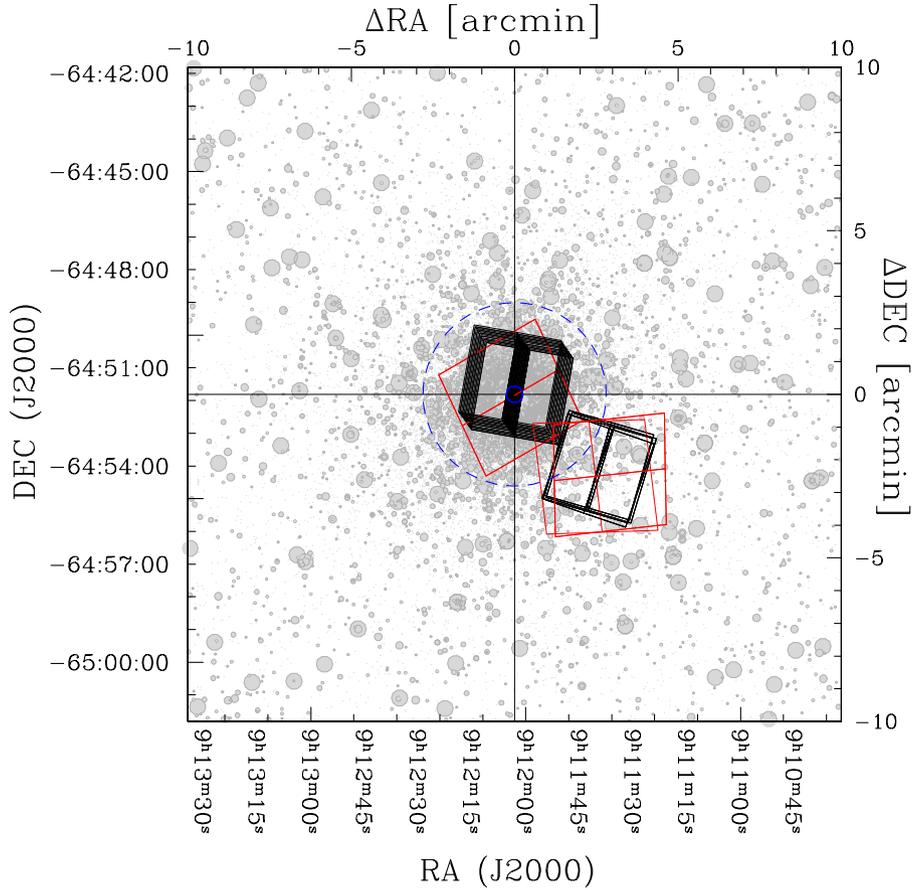} 
	\caption{Footprints of the UVIS/WFC3 (black) and WFC/ACS (red) images used in this paper. The circles represented with blue solid and dashed lines indicate the core and the half-light radii of NGC\,2808 \citep[from the 2010 version of the][catalogue]{harris/96}. }
	\label{fig:footprints}
\end{figure*}

 Multi-band {\it HST} photometry is available for stars in two distinct fields of view, namely a central field, and an outer field, which is located $\sim$5.5 arcmin south-west from the cluster centre.  Photometry in the F275W, F336W, and F438W bands of both fields is derived from images collected in the Utraviolet and Visual Channel of the Wide Field Camera 3 (UVIS/WFC3), while photometry in F814W is obtained from images taken with the Wide Field Channel of the Advanced Camera for Surveys (WFC/ACS). The footprints of the images used in this paper are shown in Figure\,\ref{fig:footprints}.
 
 The photometric catalogues of stars in the central field are provided by \citet[][]{milone/15, milone/18} and we refer to these papers for details on the data and the data reduction. The external field comprises 3$\times$360s$+$3$\times$350s$+$10s WFC/ACS images in F814W (GO\,10992, PI Piotto) and 12$\times$905s, 2$\times$592s$+ 4 \times 342$s, and 3$\times$217s$+$213s, in F275W, F336W, and F438W, respectively (GO\,15857, PI Bellini).
 We used the computer program {\it img2xym} to measure stellar positions and magnitudes \citep[][]{anderson/06}. Briefly, we build a catalogue of candidate stars that comprises all point-like sources where the central pixel has more than 50 counts within its $3\times3$ pixels and with no brighter pixels within a radius of 0.2 arcsec. We used the best available effective point spread function model to derive the magnitude and the position of all candidate stars.

\begin{figure*}
	\includegraphics[width=0.7\textwidth]{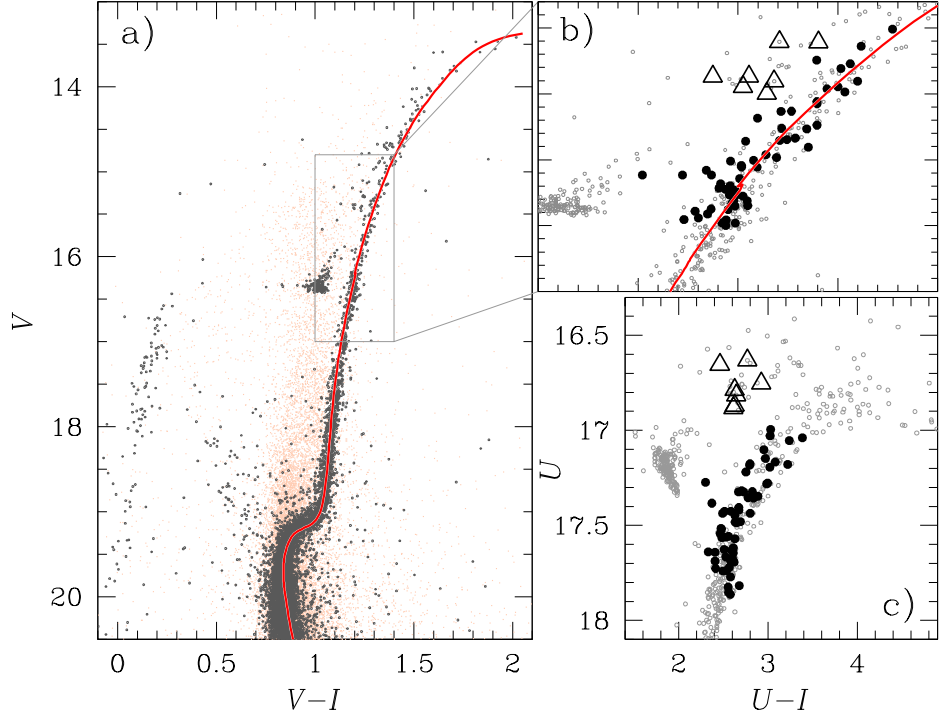} 
    \caption{CMDs of stars in the field of view of NGC\,2808 based on \protect\cite{stetson/19} photometry.
    Panel a shows the differential-reddening corrected $V$ vs.\,$V-I$ CMD of probable cluster members (grey points) and field stars (orange points). The red line represents the best-fit isochrone from the Dartmouth stellar evolution database \citep{dotter/08}.  We assumed $\rm{[Fe/H]}=-1.17$, $\rm{[}\alpha/\rm{Fe]}=0.40$, $\rm{Age}=12.5$ Gyr, $\rm{E(B-V)}=0.17$ mag and distance modulus, (m$-$M)$_{0}=14.98$ mag.
    Panel b and c show a zoom of the panel-a CMD and of the $U$ vs.\,$U-I$, respectively, around the RGB and the AGB region studied with spectroscopy.
   The RGB and AGB stars in our spectroscopic sample are indicated as black filled circles and open triangles, respectively. 
    }
    \label{cmd}
\end{figure*}

Ground-based photometry has been used to estimate the atmospheric parameters of the spectroscopic targets while both ground-based and {\it HST} photometry are used to identify the stellar populations within NGC\,2808 through the ChMs. The $V$ versus ($V-I$) colour-magnitude diagram (CMD) constructed with ground-based photometry is shown in Figure \ref{cmd}.






\subsubsection{Chromosome maps}

We associated the spectroscopic targets to the five stellar populations identified by \citet[][]{milone/15}, based on their positions in the $\Delta_{C \rm F275W, F336W, F438W}$ vs.\,$\Delta_{\rm F275W,F814W}$ ChMs derived from stars in the central and external {\it HST} fields illustrated in Figure\,\ref{fig:footprints}.

\begin{figure*}
    \centering
    \includegraphics[height=6.0cm,trim={0cm 13.5cm 0cm 0cm}]{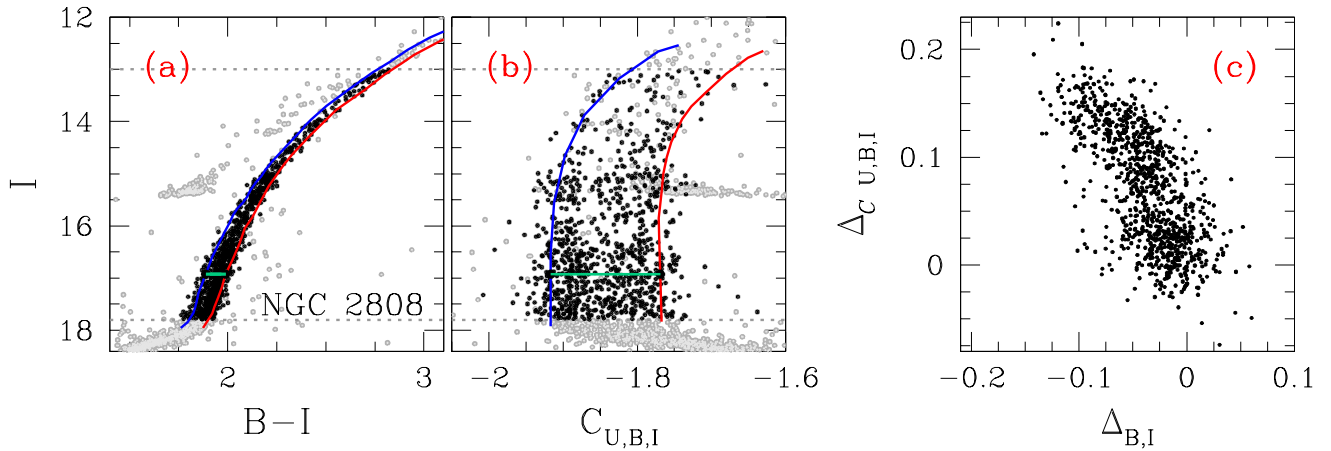}
    \caption{
This figure summarises the procedure to derive the ChM from ground based photometry. Panels a and b show the $I$ vs.\,$B-I$ and the $I$ vs.\,$C_{\rm U,B,I}$ pseudo CMD, respectively, of probable NGC\,2808 cluster members zoomed around the RGB. The red and blue fiducials mark the red and blue RGB boundaries, respectively, while the aqua segments indicates the RGB colour widths. Only the RGB stars between the two horizontal lines and marked with black dots are used to derive the ChM. The resulting \cubi\, vs.\,\dbi \, ChM is plotted in panel c.}
    \label{chm_built}
\end{figure*}

To associate the stars without {\it HST} photometry with the distinct stellar populations we introduced a new ChM that is based on ground-based photometry alone (see also \citealt{jang/22}). 
 We used the $I$ vs. $B-I$ CMD, which is sensitive to the helium content of the distinct stellar populations, and the $I$ vs.\,$C_{\rm U,B,I}$ pseudo-CMD, which separates stellar populations with different nitrogen content. These diagrams are plotted in the panels a and b of Figure \ref{chm_built}, 
 where we illustrate the main steps of the procedure to derive the ChM. 
 In a nutshell, we first build the red and blue boundaries of RGB stars in the diagrams of Figure \ref{chm_built}a and \ref{chm_built}b  \citep[see][for details]{milone/15, milone/17} and used them to derive the \dbi\, and \cubi\, pseudo colours of RGB stars. To do this, we adapted to the ground-based photometry the equations by \citet[][]{milone/17}:
 \begin{equation}
    \Delta_{B,I} = W_{\rm B,I}  \frac{X - X_{\rm fiducial R}}{X_{\rm fiducial R} - X_{\rm fiducial B}}
	\label{eq:quadratic}
\end{equation}

\begin{equation}
    \Delta_{C \rm U,B,I} = W_{C \rm U,B,I}  \frac{Y - Y_{\rm fiducial B}}{Y_{\rm fiducial R} - Y_{\rm fiducial B}}
	\label{eq:quadratic2}
\end{equation} 

where $X = B - I$ and $Y = C_{\rm U,B,I}$ and ‘fiducial R’ and ‘fiducial B’ correspond to the red and the blue fiducial lines, respectively. The adopted values for RGB widths $W_{\rm B,I}$ and $W_{C \rm U,B,I}$ correspond to the colour separation between the RGB boundaries 2.0 $I$ mag above the MS turn off.

The resulting $\Delta_{C \rm U,B,I}$ vs.\,$\Delta_{\rm B,I}$ is plotted in panel c of Figure\,\ref{chm_built}. In this diagram, the first population (or 1P) stars are clustered around the origin of the ChM, whereas second population (2P) stars define a sequence on the ChM that extends towards large values of \cubi\, and \dbi. We refer to the paper by \cite{jang/22} for extensive discussion of the \cubi\, vs.\, \dbi\, ChM of RGB stars in Galactic GCs.

\subsection{Spectroscopic dataset}

The initial sample consists of 81 giant stars observed with the spectrograph FLAMES/GIRAFFE \citep{pasquini/02} at the Very Large Telescope (observation program 094.D-0455).
The stars were observed in two different configurations, the HR4 setup has wavelength coverage  from 4188 to 4297~\AA \, with $\rm{R}\approx24000$ and the HR18 setup with wavelength coverage from 7468 to 7889~\AA\, and $\rm{R}\approx20150$. 

We identified 26 stars in the {\it HST} ChM, including 24 stars in the central field and 2 in the external field. The  ChM derived from ground-based photometry comprises 26 spectroscopic targets. 
 
The data reduction was done using the EsoReflex-based GIRAFFE pipeline \citep{ballester/00}, which includes bias subtraction, flat-field correction and wavelength calibration.
The individual exposures were then corrected to the rest-frame system and eventually combined by using {\sc iraf}\footnote{\url{http://iraf.noao.edu/}} routines. The final signal-to-noise ratio (S/N) of the co-added spectra is $\rm{S/N}\sim 200$. 

The list of the observed spectroscopic targets is reported in Table \ref{abundances_table}, together with coordinates, radial velocities, stellar parameters and chemical abundances obtained as explained in the next sections, and proper motions from Gaia.

\begin{table*}
	\centering
	\caption{Right ascension and declination ($J2000$, degrees), radial velocities (\kmsec), proper motions (mas~y$^{-1}$), effective temperature (K), surface gravity (dex), microturbulence velocities (\kmsec), chemical abundances inferred in this work (dex) for the spectroscopic sample, and a flag indicating the ChM information.
}
	\label{abundances_table}
	\begin{tabular}{llccccccccccccr} 
	\hline

Star & RA & DEC & V$_{\rm rad}$  & ${\mu}_{\alpha,\delta}$ & T$_{\rm{eff}}$  & $\log \rm{g}$ & $\xi_{\rm{t}}$  & [Fe/H] & [C/Fe] & [N/Fe] & [O/Fe] & [Al/Fe] & [Ni/Fe] & ChM$^\star$\\

    \hline \hline
\multicolumn{14}{l}{RGB spectroscopic sample:} \\  

N2808\_2\_36  &  137.97362  &  -64.84250  &  120.7  &  (1.39,0.05)  &  4909  &  2.56  &  1.28  &  -1.13  &  -0.13  &  0.30  &  0.56  &  -0.03  &  0.19 & 1 \\ 
...\\






\hline
\multicolumn{14}{l}{AGB spectroscopic sample:} \\

N2808\_2\_9\_wf  &  137.85825  &  -64.89367  &  100.1  &  (0.95,0.36)  &  4692  &  1.97  &  1.43  &  -1.20  &  -0.54  &  0.43  &  $<$-0.73  &  1.01  &  0.08 & 0  \\ 

...\\

    \hline
    \end{tabular}
    
    {\bf Note.} This table is available in its entirety in machine-readable form in the online journal. $^\star$0 for no ChM information, 1 for target in the $HST$ ChM, and 2 for objects in the ground based ChM.
\end{table*}

\section{Spectroscopic Data Analysis}\label{sec:analysis}

We derived chemical abundances for cluster stars only. The sample of probable cluster members has been selected based on stellar proper motions and radial velocities (RVs). 

We consider proper motions from  GAIA eDR3  \citep{gaia_col_edr3/21}.  After an inspection of the vector-point diagram of proper motions four stars were identified as outliers and removed from our sample.

The RVs were obtained with the aid of the {\sc fxcor} task from {\sc iraf} using a synthetic spectrum template obtained through the March, 2014 version of MOOG \citep{sneden/73}. This spectrum was computed with a stellar model atmosphere interpolated from the \cite{castelli/04} grid, adopting the parameters (\teff, \logg, \vmicro, [Fe/H]) = (4800~K, 2.5, 1.5~\kmsec, $-$1.2), and cross-correlated with the observed spectra. Each spectrum was corrected to the rest-frame system, and the observed RVs were then corrected to the heliocentric system. 
The average RV found in our work for NGC\,2808 is $102.9\pm1.1$\,\kmsec, in good agreement with the value of  $101.6\pm1.2$\,\kmsec\ from the 2010 version of \cite{harris/96}. 


\subsection{Stellar parameters}

The stellar parameters effective temperature (\teff), surface gravity (\logg), microturbulent velocity (\vmicro) and [Fe/H] were, thus, determined for 77 giant stars, 
 including 70 RGB stars and seven AGB stars.

For consistency, the photometry adopted here to determine stellar parameters for all the stars is from ground-based facilities \citep{stetson/19}. 
 The 
  corresponding $I$ vs.\,$V-I$ and $U$ vs.\,$U-I$ CMDs are shown in Figure \ref{cmd}.  
The 
 effective temperature, $\rm{T}_{\rm{eff}}$,
 of each star was determined using the ($V-I$), ($V-R$), ($B-V$), ($U-V$), and ($R-I$) colours and the calibration from \cite{alonso/99}, adopting the \cite{bessell/79} corrections from the Cousins to the Johnson systems.
We adopted a reddening of $\rm{E(B-V)}=0.17$ 
mag 
 and a distance modulus of 14.98
 mag,
 which are the values that best fit the CMD of Figure \ref{cmd}. 
 Our best determination of the effective temperature of each star corresponds to the average of the $\rm{T}_{\rm{eff}}$ values derived from the various colours.
  The typical 
  standard deviation 
is 65~K. 

We applied  bolometric corrections from \cite{alonso/99} with $\rm{M}=0.85\,\rm{M}_\odot$ and the $\rm{T}_{\rm{eff}}$ 
value
 calculated from each colour to infer the average $\log g$ value for all the stars (the standard deviation considering the distinct $\log g$ values is 0.03~dex). 
  To infer microturbulence, $\xi_{\rm{t}}$, we used the empirical relation from \cite{marino/08}, where the value  $\xi_{\rm{t}}$ depends on stellar gravity. 

Finally, we derive the Fe abundance of each star by the equivalent width (EW) method with the aid of the 1D LTE 2014 version of the code {\sc moog} \citep{sneden/73} and the Kurucz grid of {\sc ATLAS9} model atmospheres \citep{castelli/04} with $\rm{[}\alpha/\rm{Fe]}=0.40$. We adopted the line list presented in Table~\ref{line_list}, which includes 15 Fe {\sc i} lines and one Fe {\sc ii} line from \cite{heiter/21}. The number of lines considered to evaluate [Fe/H] varied from 6 to 16 depending on each star. The internal error associated with the [Fe/H] values was estimated considering the stellar parameters uncertainties as detailed below, and the statistical error from the several lines analysed. The obtained typical [Fe/H] error from our spectra is 0.08~dex (see Table~\ref{error_table}). 

\begin{table}
	\centering
	\caption{Line list used in the determination of Fe and Ni abundances.}
	\label{line_list}
	\begin{tabular}{llrr} 
	\hline

Wavelength (\AA) & Species &   $\chi_{\rm{exc}}$ (eV) & $\log \rm{gf}$ \\  

\hline \hline

7491.647   &  Fe {\sc i}   &  4.300 & -1.060 \\  
7495.066   &  Fe {\sc i}   &  4.220 & -0.100 \\  
7507.266   &  Fe {\sc i}   &  4.410 & -1.470 \\  
7511.019   &  Fe {\sc i}   &  4.180 &  0.119 \\  
7531.144   &  Fe {\sc i}   &  4.370 & -0.940 \\                                
7568.899   &  Fe {\sc i}   &  4.280 & -0.880 \\  
7583.796   &  Fe {\sc i}   &  3.020 & -1.880 \\  
7586.018   &  Fe {\sc i}   &  4.310 & -0.270 \\  
7710.363   &  Fe {\sc i}   &  4.220 & -1.070 \\  
7723.210   &  Fe {\sc i}   &  2.280 & -3.580 \\  
7748.284   &  Fe {\sc i}   &  2.950 & -1.750 \\  
7751.109   &  Fe {\sc i}   &  4.990 & -0.830 \\  
7780.568   &  Fe {\sc i}   &  4.470 & -0.090 \\  
7807.909   &  Fe {\sc i}   &  4.990 & -0.640 \\   
7832.196   &  Fe {\sc i}   &  4.430 & -0.019 \\  
7711.723   &  Fe {\sc ii}   &  3.900 & -2.540 \\
7522.758 & Ni {\sc i} & 3.655 & -0.570 \\
7525.111 & Ni {\sc i} & 3.633 & -0.550 \\
7555.598 & Ni {\sc i} & 3.844 & -0.050 \\
7574.043 & Ni {\sc i} & 3.830 & -0.530 \\
7714.314 & Ni {\sc i} & 1.934 & -2.200 \\
7727.613 & Ni {\sc i} & 3.676 & -0.170 \\
7788.936 & Ni {\sc i} & 1.949 & -2.420 \\
7797.586 & Ni {\sc i} & 3.895 & -0.260 \\

	\hline
	\end{tabular}
\end{table}

To identify the presence of possible systematics associated with our analysis,
we compared our stellar parameters with the ones obtained by \cite{carretta/15}. Eight out of 77 stars in our sample have been studied in \cite{carretta/15} as well. We find  an average offset of $\Delta\rm{T}_{\rm{eff}} = 24 \, \rm{K}$ between our values and the $\rm{T}_{\rm{eff}}$ from \cite{carretta/15}, with associated dispersion of 36 K. This difference is probably partially caused by the difference in the method to acquire $\rm{T}_{\rm{eff}}$ as well as the distinct reddening adopted. The largest discrepancy has been obtained for surface gravities, for which our values are $\sim 0.26$ dex higher than the \logg\ by \cite{carretta/15} and an associated dispersion of 0.03 dex. However, we can explain this difference as mostly due to the distinct distance modulus adopted; \cite{carretta/15} used a distance modulus of 15.59 (from \citealt{harris/96}, 2010 version), and if we consider the same distance modulus the difference between the \logg\ values would decrease to 0.02~dex. Our $\xi_{\rm{t}}$ values agree with those of Carretta and collaborators, being only marginally lower by $\sim 0.12$\,\kmsec\, with an associated dispersion of 0.10 dex.
The difference between our [Fe/H] and the ones of \cite{carretta/15} shows an average discrepancy of $\sim 0.08$~dex and an associated dispersion of 0.04 dex. The mean discrepancy in the [Fe/H] values is even lower if we consider the difference in the adopted solar constants (in this work we adopted A(Fe)$_{\odot}=7.50$ from \citealt{asplund/09}; while \citealt{carretta/15} adopted A(Fe)$_{\odot}=7.54$ from \citealt{gratton/03a}).  

This comparison between different datasets suggests that estimates of internal uncertainties associated with our stellar parameters are $\sigma\rm{T}_{\rm{eff}}\sim50\,\rm{K}$, $\sigma\log \rm{g}\sim0.10\,\rm{dex}$, $\sigma\xi_{\rm{t}} \sim 0.10\,\rm{km.s}^{-1}$, and $\sigma[\rm{Fe/H}]\sim0.1\,\rm{dex}$.
These uncertainties do not include possible systematics that could be present between our values and the literature ones, such as the systematic difference that we have found in the surface gravities between our values and  \cite{carretta/15}.


\subsection{Chemical abundances}

\begin{figure*}
	\includegraphics[width=1.1\textwidth]{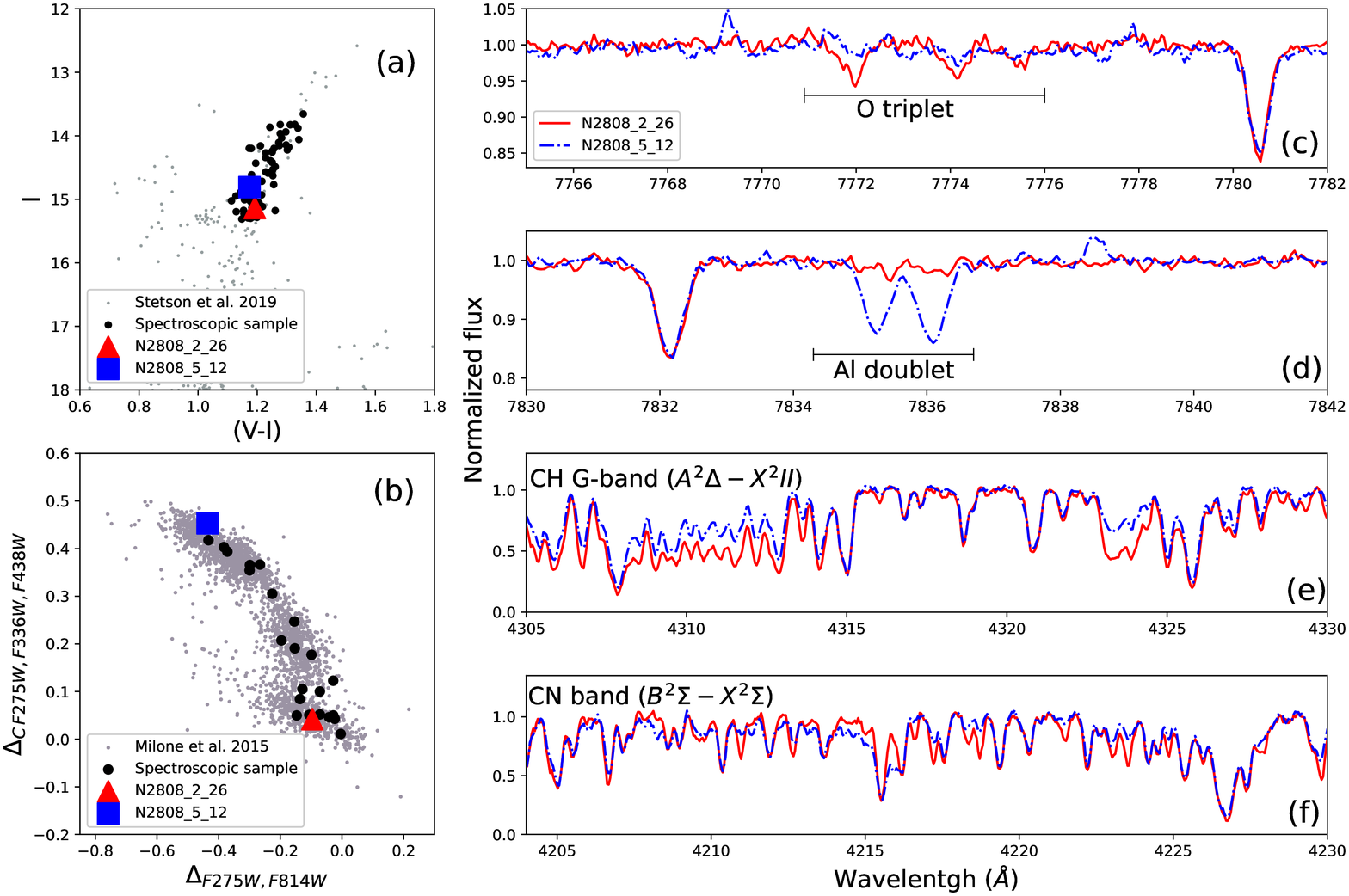} 
    \caption{
    Comparison between two stars with similar photometry (panel {\it a}) but from different populations according to their ChM position (panel {\it b}). Red solid lines show the spectra of N2808\_2\_26 ($\teff=4909$ K, $\logg=2.56$, $\vmicro=1.28$ \kmsec, $\rm{[Fe/H]}=-1.13$) and the dashed blue lines represent the spectra of N2808\_5\_12 ($\teff=4893$ K, $\logg=2.44$, $\vmicro=1.31$ \kmsec, $\rm{[Fe/H]}=-1.12$) for four distinct regions of their spectra: O triplet at 7770 \AA\, (panel {\it c}), Al doublet at 7835 \AA\, (panel {\it d}), CH G-band $\rm{A}^2\Delta-\rm{X}^2II$ (panel {\it e}) and CN band $\rm{B}^2\Sigma-\rm{X}^2\Sigma$ (panel {\it f}).}
      \label{cmd_chm_spectro}
\end{figure*}

Beside Fe, we inferred chemical abundances for the light elements C, N, O, Al, and for the iron-peak element Ni. 
For the light elements we adopted a spectral synthesis analysis using the 1D LTE 2014 version of the code {\sc moog} \citep{sneden/73} and the Kurucz grid of {\sc ATLAS9} model atmospheres \citep{castelli/04} with $\rm{[}\alpha/\rm{Fe]}=0.40$. Figure \ref{cmd_chm_spectro} (panels {\it c, d, e} and {\it f}) shows the spectral regions adopted to analyse each element. 

The heads of the CH G-band ($\rm{A}^2\Delta-\rm{X}^2II$) at $\sim4312$~\AA\, and $\sim4323$~\AA\,
were used in the C abundances determination by assuming the O abundances determined in this work (as explained below). Nitrogen abundances were inferred from the CN band ($\rm{B}^2\Sigma-\rm{X}^2\Sigma$) at $\sim4215$ \AA\, 
by assuming both the previously measured O and C abundances.

The triplet in the 7770~\AA\, region was used to determine the O abundances. 
Our line list in this spectral region is based on the Kurucz line compendium\footnote{Available at: {\sc http://kurucz.harvard.edu/}}, with \loggf\ values of the three O lines from \cite{wiese/96}.  
The 7771.94 \AA\, O line was mildly affected by the MgH molecules, which could introduce biases in the O abundances given the large star-to-star Mg variations observed in NGC\,2808 \citep[e.g.,][]{pancino/2017,carretta/18}.
Thus, for the O estimates we adopted values from $-$0.2 to 0.4 in [Mg/Fe] for each star according to their Al abundances and the [Al/Fe] vs. [Mg/Fe] anti-correlation from \citet[][see their Figure~3]{carretta/18}.
We also assessed the impact of CN abundances into the O determinations and found that in our stars a change in C and N by $\sim0.3$~dex 
affects the abundances from the 7774.17~\AA\, and 7775.39~\AA\, lines in the most O-poor stars by as much as 0.2~dex. 
To avoid the introduction of such systematics due to our assumptions in CN for the most O-poor population, for stars with  $\rm{[O/Fe]}\lesssim0.3$~dex, we infer O abundances 
from the first triplet line at 7771.94~\AA\ only. The O {\sc i} 7770 \AA\, triplet is known to suffer from large departures from LTE \citep{amarsi/18}, especially towards higher \teff. To account for this, the abundances inferred from the different components were individually corrected by interpolating the 1D non-LTE grid presented in \cite{amarsi/19}.

Al abundances were obtained by spectral synthesis of the doublet at 7835~\AA\, region.
The line list used in this region was generated by {\sc linemake}\footnote{\url{https://github.com/vmplacco/linemake}.} \citep{placco/21}. Since we want to investigate the [Al/Fe] versus [O/Fe] anti-correlation, the Al abundances were also corrected for departures from LTE, similarly to O abundances discussed above. This was done by interpolating the 1D non-LTE grid of abundances corrections presented in \cite{nordlander/17}. 


Ni was inferred at the same manner as the Fe abundances, e.g. through the assessment of equivalent widths by using the Ni lines listed in Table \ref{line_list}.  

An estimate of the final errors associated with our chemical abundances has been derived by determining the variation of each element after the stellar parameters have been changed, once at a time, by their assumed uncertainties (see discussion in the previous section). For the elements obtained through EWs, we also include 
the statistical error associated with the abundances inferred from the spectral lines that were measured.
For the other elements, derived by spectral synthesis, 
we consider
the rms deviation of the observed line profile relative to the synthetic spectra, and the continuum setting. 
Table~\ref{error_table} lists the contribution of each error source to the final estimated uncertainties due to the internal errors for our inferred abundances. We also list how our chemical abundances change by varying \teff, \logg, \vmicro\ by larger amounts, as possibly due to systematics. We emphasise that our interest is on the internal star-to-star variations in NGC\,2808 giants, so that the systematic variations may be considered as a reference when comparing with different analysis.

\begin{table*}
	\centering
	\caption{Sensitivity of derived abundances to the uncertainties in the stellar parameters, the signal to noise ($\sigma_{\rm fit}$), abundances of C and O (when applicable), and the subsequent total error ($\sigma_{\rm total}$). 
	}
	\label{error_table}
	\begin{tabular}{lcccccccr} 
	\hline
	 \multicolumn{7}{c}{Internal errors} \\ 
	 \hline 
	  & $\Delta{\rm T}_{\rm eff.}\pm50\,{\rm K}$ & $\Delta\logg\pm0.10$ & $\Delta\xi_{\rm t}\pm0.10\,\kmsec$ & $\Delta{\rm [Fe/H]}\pm0.10$ & $\Delta{\rm [O/Fe]}\pm0.25$ & $\Delta{\rm [C/Fe]}\pm0.20$ & $\sigma_{\rm fit}$ & $\sigma_{\rm total}$ \\
	  
	  \hline \hline
	 $[{\rm Fe/H}]$ & 0.03 & 0.01 & 0.03 & -- & -- & -- & 0.06 & 0.08 \\
	 $[{\rm C/Fe}]$ & 0.03 & 0.00 & 0.00 & 0.07 & 0.06 & -- & 0.06 & 0.11 \\
	 $[{\rm N/Fe}]$ & 0.08 & 0.01 & 0.00 & 0.05 & 0.08 & 0.23 & 0.11 & 0.28\\
	 $[{\rm O/Fe}]$ & 0.06 & 0.04 & 0.00 & 0.02 & -- & -- & 0.18 & 0.23\\
	 $[{\rm Al/Fe}]$ & 0.04 & 0.01 & 0.01 & 0.02 & -- & -- & 0.15 & 0.16\\
	 $[{\rm Ni/Fe}]$ & 0.03 & 0.01 & 0.04 & 0.01 & -- & -- & 0.10 & 0.11\\
	 
	 \hline
	 \multicolumn{7}{c}{Systematic errors} \\ 
	 \hline 
	 
       & $\Delta{\rm T}_{\rm eff.}\pm100\,{\rm K}$ & $\Delta\logg\pm0.30$ & $\Delta\xi_{\rm t}\pm0.30\,\kmsec$ &$\Delta{\rm [Fe/H]}\pm0.10$ &$\Delta{\rm [O/Fe]}\pm0.25$ & $\Delta{\rm [C/Fe]}\pm0.20$ & $\sigma_{\rm fit}$ & $\sigma_{\rm total}$ \\
	\hline \hline
     $[{\rm Fe/H}]$ & 0.07 & 0.01 & 0.09 & -- & -- & -- & 0.06 & 0.13 \\
	 $[{\rm C/Fe}]$ & 0.07 & 0.02 & 0.00 & 0.07 & 0.06 & -- & 0.06 & 0.13\\
	 $[{\rm N/Fe}]$ & 0.17 & 0.05 & 0.00 & 0.05 & 0.08 & 0.23 & 0.11 & 0.32\\
	 $[{\rm O/Fe}]$ & 0.12 & 0.10 & 0.01 & 0.02 & -- & -- & 0.18 & 0.25\\
	 $[{\rm Al/Fe}]$ & 0.04 & 0.01 & 0.01 & 0.02 & -- & -- & 0.15 & 0.17\\
	 $[{\rm Ni/Fe}]$ & 0.07 & 0.04 & 0.10 & 0.01 & -- & -- & 0.10 & 0.16\\

	\hline
	
	\end{tabular}
\end{table*}


The final abundances of the element ``X'' obtained in our study are shown as $\rm{[X/Fe]}=\rm{A}(\epsilon_{\rm{X}\star})-\rm{A}(\epsilon_{\rm{X}\odot})-\rm{[Fe/H]}$, where $\rm{A}(\epsilon_{\rm{X}\star})=\log \rm{N(X)}/\rm{N(H)}+12$ is the absolute abundance of a star and $\rm{A}(\epsilon_{\rm{X}\odot})$ is the solar photospheric abundances values from \cite{asplund/09}.

\section{Results and discussion} \label{sec:disc}

The  astrometric positions, stellar parameters and chemical abundances inferred in this work 
are listed in Table~\ref{abundances_table}.  The table is divided into  RGB and AGB stars.

In the following subsections we discuss the general chemical abundances of RGB and AGB stars in NGC\,2808, and finally we present the analysis of the chemical patterns associated to the distinct stellar populations as observed on the ChM.

\subsection{The chemical composition of RGB stars}

\begin{figure}
    \centering
    \includegraphics[width=\columnwidth]{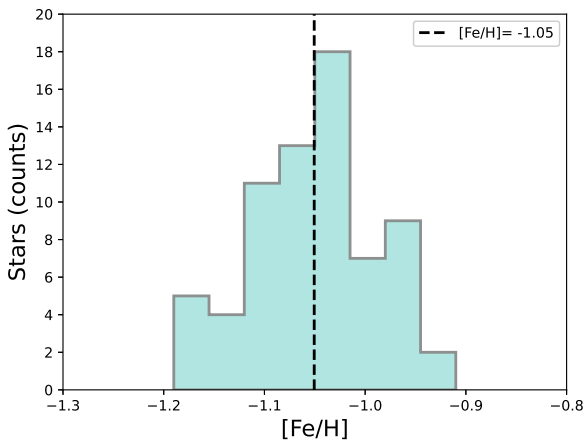} 
	\caption{[Fe/H] distribution for the 70 RGB stars in our sample. The black dashed line indicates the average [Fe/H] value.}
	\label{feh_hist}
\end{figure}

%
In Figure~\ref{feh_hist} we show the histogram distribution of [Fe/H]. Our analysed stars have 
a mean iron content of $\rm{[Fe/H]}=-1.05\pm0.01$. The associated dispersion of 0.07~dex, very similar to the estimated error of 0.08~dex, suggests that our sample of stars is consistent with an homogeneous Fe distribution. 
 

As expected for Milky Way GCs, the chemical content of the light elements C, N, O and Al, involved in the high-temperature $p$-capture reactions, differs from star to star \citep[e.g.,][]{gratton/04}.  The variation of this set of elements is visualised in Figure~\ref{cmd_chm_spectro}, which presents the comparison of two stars with 
  similar atmospheric parameters (panel {\it a}) but from different populations, according to the location on the ChM (panel {\it b}). 
  According to the classification of \cite{milone/15}, the two represented stars, namely N2808\_2\_26 (red triangle) and N2808\_5\_12 (blue square) belong to populations B and D, respectively. They show different abundances of C (panel {\it e}), N (panel {\it f}), O (panel {\it c}) and Al (panel {\it d}), that can be easily identified by eye from a quick inspection of their spectra. 

Specifically, for our spectroscopic sample of RGB stars, the [O/Fe] varies from  $\lesssim -0.73$ 
to $+0.73$~dex with an average value of $\rm{[O/Fe]}=0.13\pm0.05$ 
and an associated dispersion of 0.42 dex. [C/Fe] abundances range from $-0.93$ to $0.02$ with an average value of $\rm{[C/Fe]}=-0.42\pm0.03$ and an associated dispersion of 0.26~dex. [N/Fe] abundances vary from $-0.50$ to $1.35$ with an average of  $\rm{[N/Fe]}=0.74\pm0.06$ and an associated dispersion of 0.49 dex. The values of [Al/Fe] go from $-0.24$ to $1.14$ with average abundance of $\rm{[Al/Fe]}=0.38\pm0.06$ and an associated dispersion of 0.46 dex.  Differently, [Ni/Fe] has a smaller variation, ranging from $-0.05$ to $0.25$~dex, with average abundance of $\rm{[Ni/Fe]}=0.08\pm0.01$ and associated dispersion of 0.06~dex. 

\begin{figure*}
\begin{tabular}{cc}
    \includegraphics[width=0.5\textwidth]{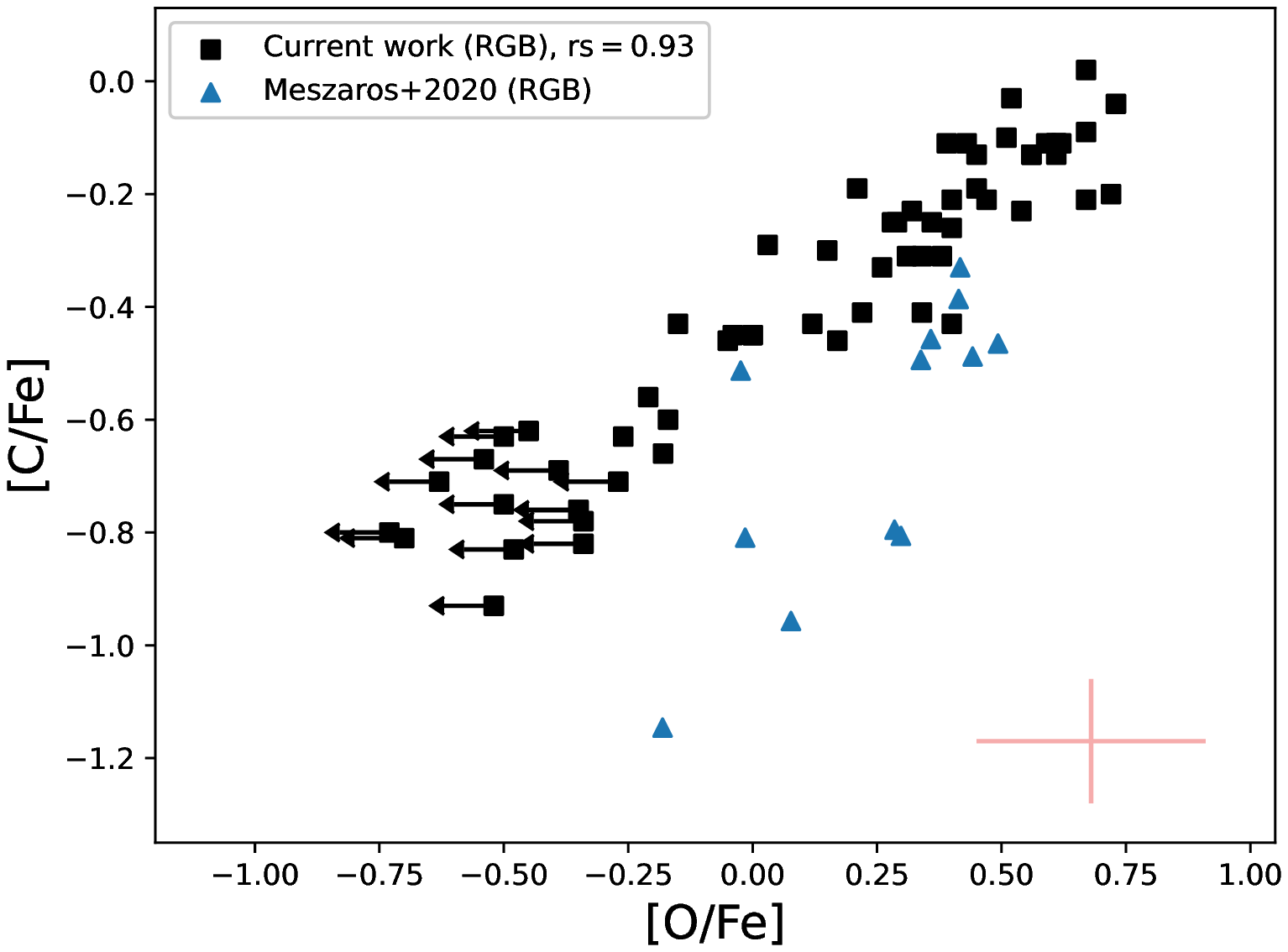}  & \includegraphics[width=0.5\textwidth]{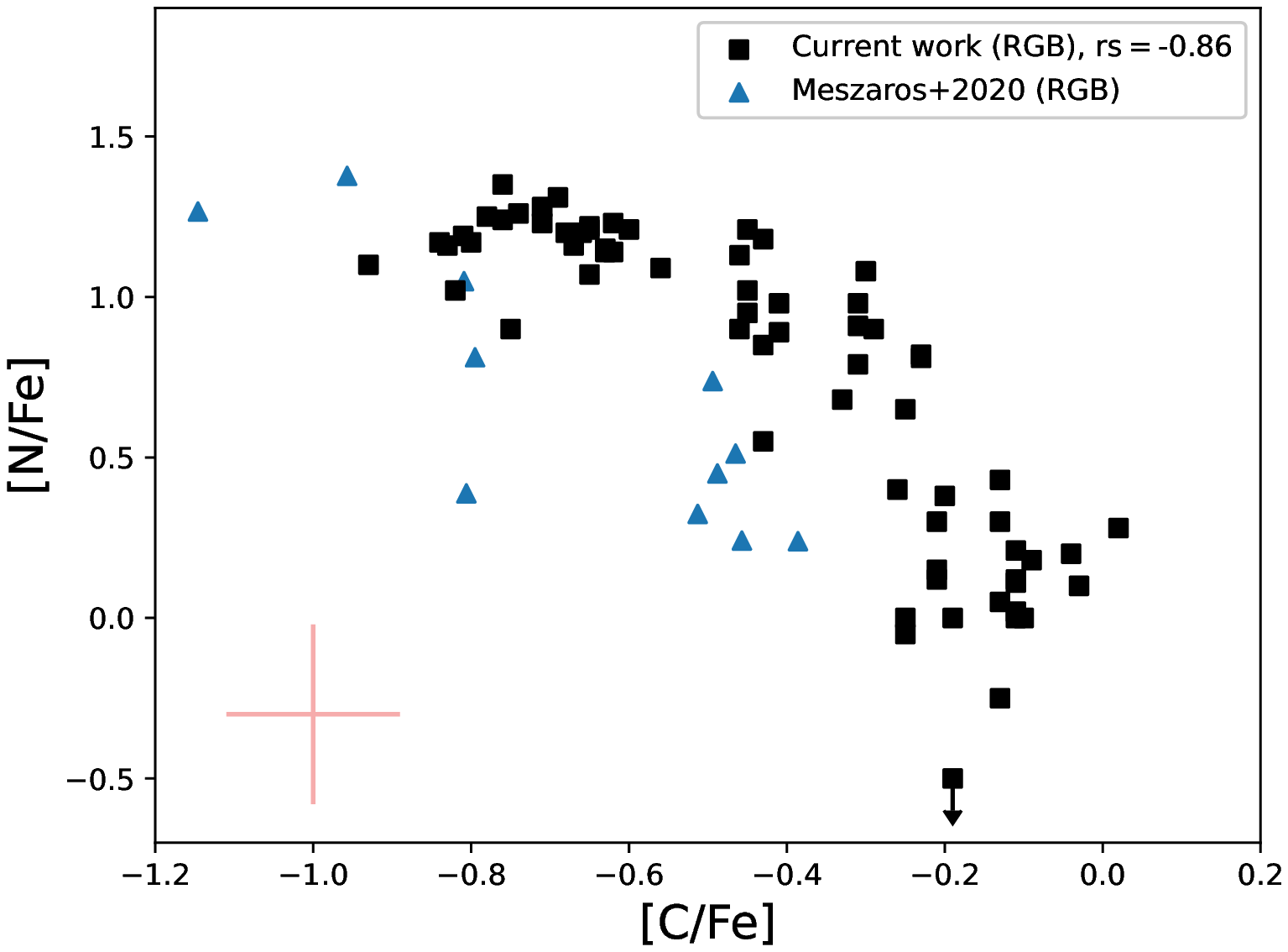}  \\
    \includegraphics[width=0.5\textwidth]{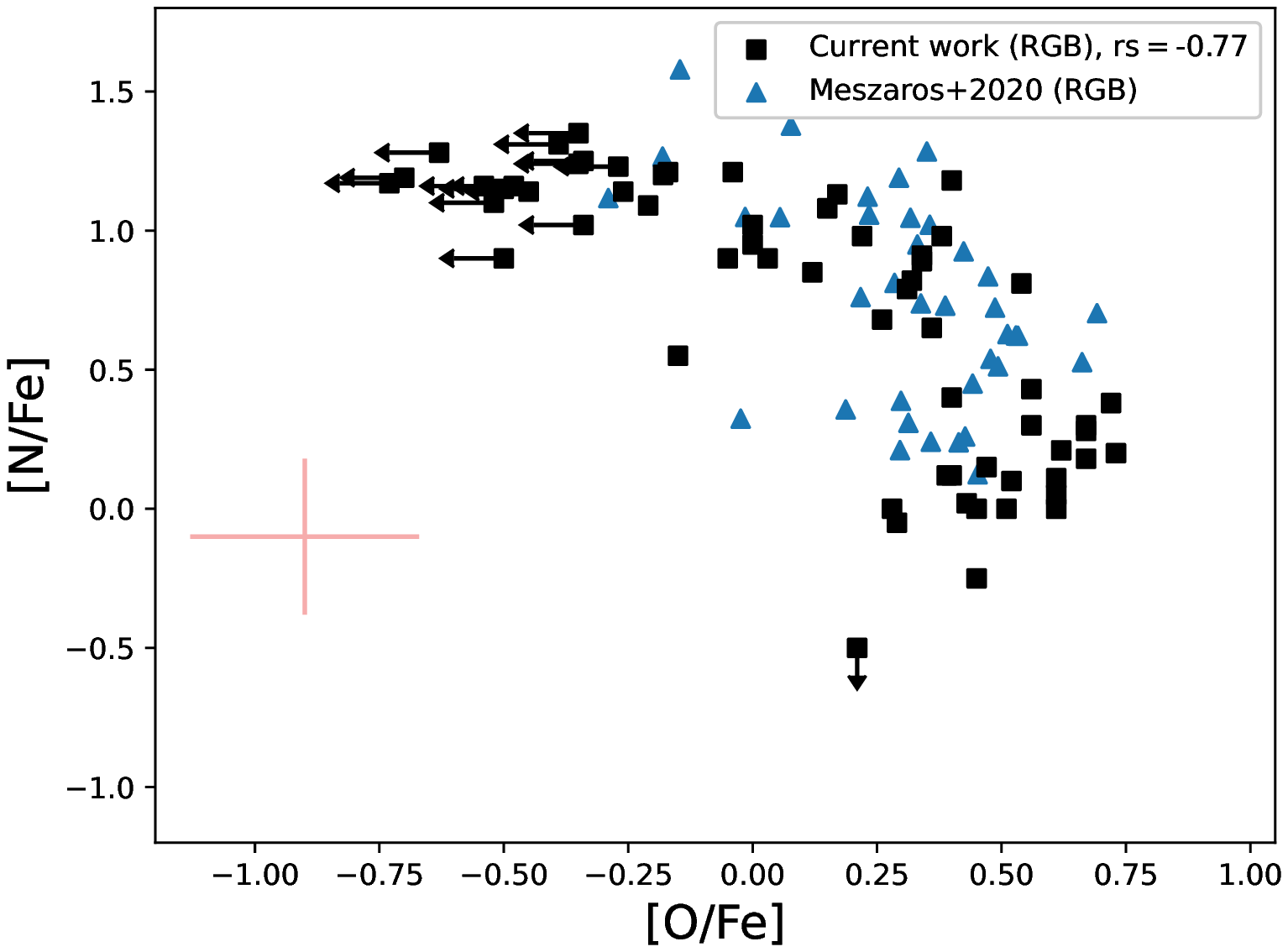} &
    \includegraphics[width=0.5\textwidth]{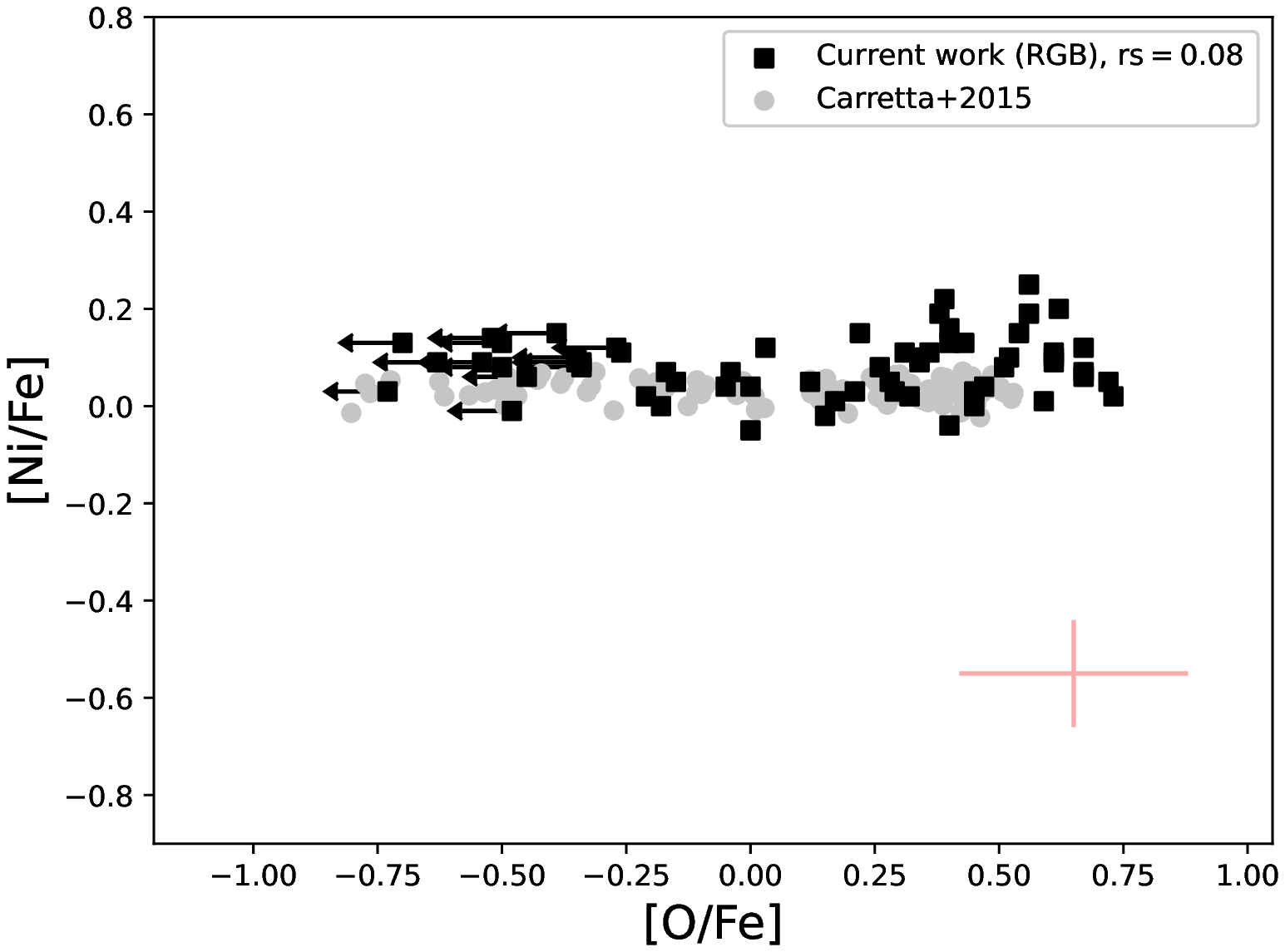} \\
    \includegraphics[width=0.5\textwidth]{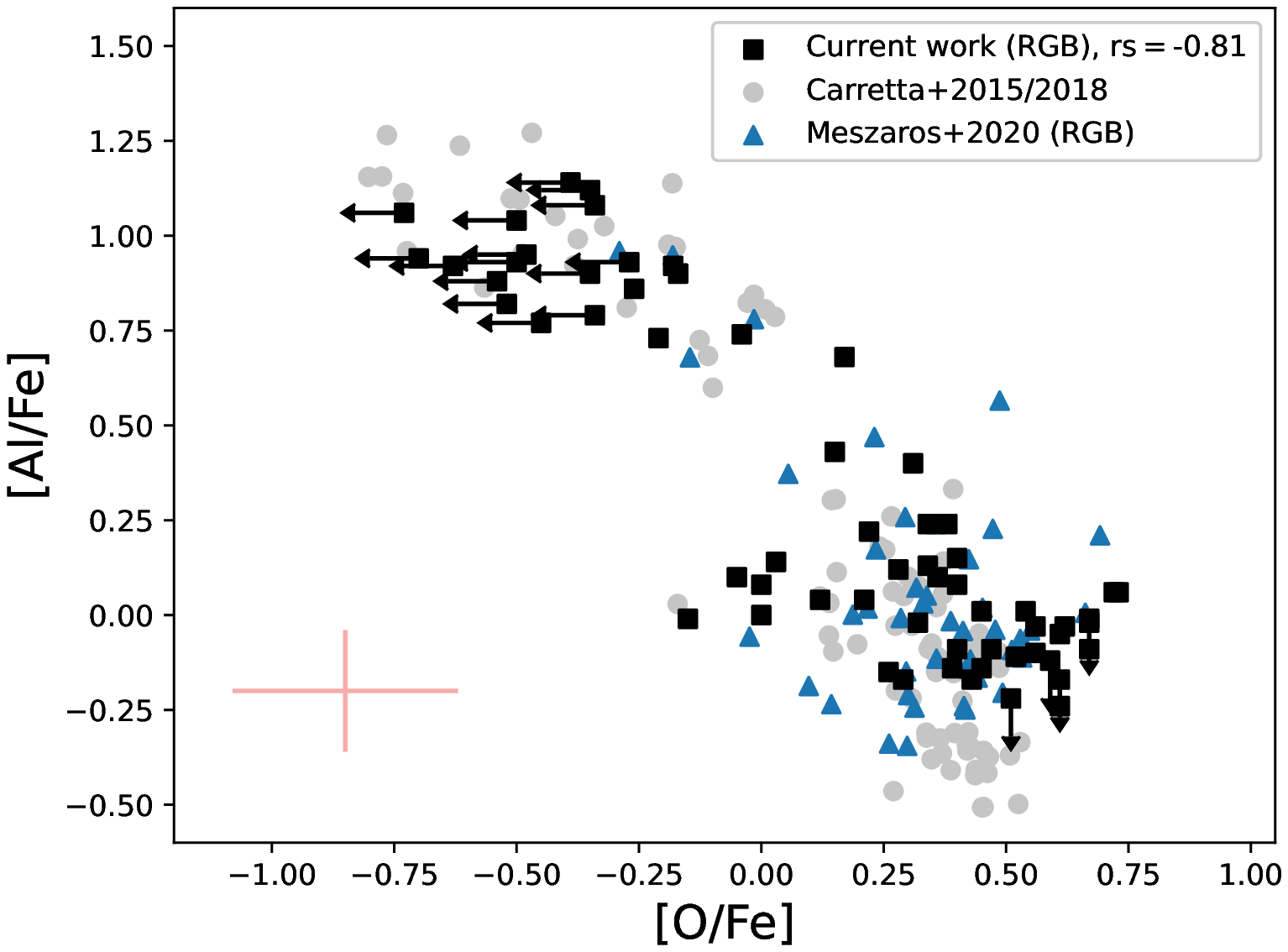} &
    \includegraphics[width=0.5\textwidth]{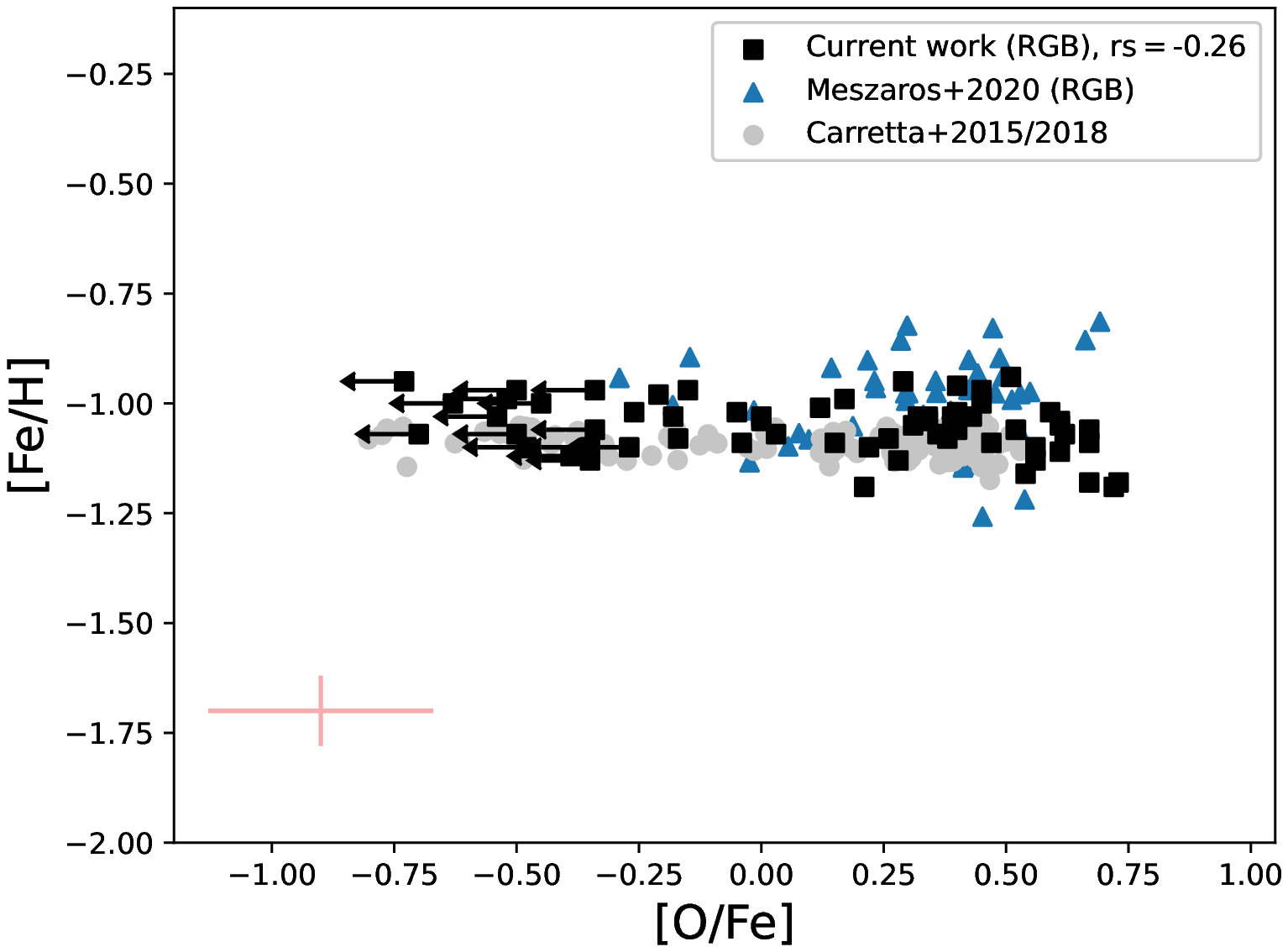} \\
\end{tabular}
	\caption{Abundances obtained in the current work for RGB stars (black squares) compared to data from \protect\cite{meszaros/20} (blue triangles) and \protect\cite{carretta/15,carretta/18} (grey circles). Top left panel shows [C/Fe] vs. [O/Fe], top right panel presents [N/Fe] vs. [C/Fe], [N/Fe] vs. [O/Fe] is in the left middle panel, [Ni/Fe] vs. [O/Fe] is presented in the right middle panel, [Al/Fe] vs. [O/Fe] is shown in the bottom left panel and [Fe/H] vs. [O/Fe] is in the bottom right panel. The typical error (mean value) is displayed in red. The Spearman coefficients ($rs$) for each  possible (anti-)correlation are reported in all panels. }
    \label{ofes}
\end{figure*}

The expected (anti-)correlations of [C/Fe] with 
[N/Fe], 
and [Al/Fe] 
with [O/Fe] are shown in Figure \ref{ofes}. For comparison purposes, we also show [Ni/Fe] 
and [Fe/H] 
abundances versus [O/Fe].
The classical anti-correlation between [N/Fe] and [C/Fe] is presented in the top right panel.

The variations in the abundances from C, N, O and Al show the intense interplay in the production and destruction of these elements in the different objects in NGC\,2808.  As a product of the CNO cycle, during hot hydrogen burning, C and O are reduced while N is highly enhanced. At the same time, Al is produced through the Mg-Al chain \citep{arnould/99,karakas_latanzio/03}. Meanwhile, [Ni/Fe] is constant with [O/Fe], and also constant in the different populations (see discussion in \S\, \ref{ss_chm}).

For our sample of stars with the full set of C, N, and O abundances available (excluding the upper limits), we
 find an average C$+$N$+$O overall abundance of $\rm{A(C+N+O)}=8.16\pm0.02$ with associated dispersion of 0.10~dex.
%
We also notice that the overall C$+$N$+$O content slightly decreases when moving from first-population stars 
 to populations C and D.

However, we notice that values of A(C$+$N$+$O) may be significantly affected by systematic uncertainties of the absolute abundances of C, N, and O. 
To estimate the robustness of our C$+$N$+$O abundances, we estimated the variations in  A(C$+$N$+$O) that we obtain by changing C, N, O, individually. To do this, we varied each element at time by an amount corresponding to its internal uncertainty   (Table~\ref{error_table}).
  We find that the C$+$N$+$O content is not significantly affected by systematic errors in C. 
   On the contrary, N and O variations correspond to significant C$+$N$+$O changes, which may have different effectes on the different stellar populations. 
   Specifically, a systematic variation of N by 0.28~dex mostly affect Populations C and D stars, with the latter varying up to $\sim$0.2~dex. 
   %
   Systematic errors in O mostly affect the most O-rich populations, (i.e. the 1P). 
   An oxygen variation by $\pm0.23$ dex,
   corresponds to a C$+$N$+$O increase by $\sim$0.2~dex in Population 1 stars, to a decrease by $\sim$0.1~dex and  $\sim$0.05 in Population C and D, respectively.
We conclude that the relative C$+$N$+$O content of the distinct stellar populations is significantly affected by systematic errors in the C, N, and O determinations.    This fact prevents us from firm conclusion 
 on C$+$N$+$O variations among the distinct stellar populations of NGC\,2808.

\subsection{The chemical composition of AGB stars}

Our spectroscopic sample comprises seven probable AGB stars, as clearly suggested by the location of these stars in the $U$ vs. ($U-I$) CMD (see Figure~\ref{cmd}). 
 In this sample, all the analysed light elements show a significant range, specifically: 
 [O/Fe] vary from  $\lesssim -0.73$ to $0.83$~dex with an average value of $\rm{[O/Fe]}=0.37\pm0.21$~dex (associated dispersion of 0.52~dex); [C/Fe] abundances range from $-0.98$ to $-0.25$~dex and average value of $\rm{[C/Fe]}=-0.47\pm0.10$~dex (associated dispersion of 0.25~dex); [N/Fe] varies from $+0.40$ to $+1.17$~dex with an average abundance of $\rm{[N/Fe]}=+0.64\pm0.10$~dex (associated dispersion of 0.25~dex); and [Al/Fe] values range from $-0.25$ to $1.01$~dex with average abundance of $\rm{[Al/Fe]}=0.16\pm0.19$~dex (associated dispersion of 0.46~dex). Additionally, similar to the RGB sample, [Ni/Fe] abundances span a smaller range from $-0.04$ to $0.13$~dex, with average abundance of $\rm{[Ni/Fe]}=0.05\pm0.02$~dex and dispersion of 0.06~dex. 

\begin{figure*}
\begin{tabular}{cc}
    \includegraphics[width=0.5\textwidth]{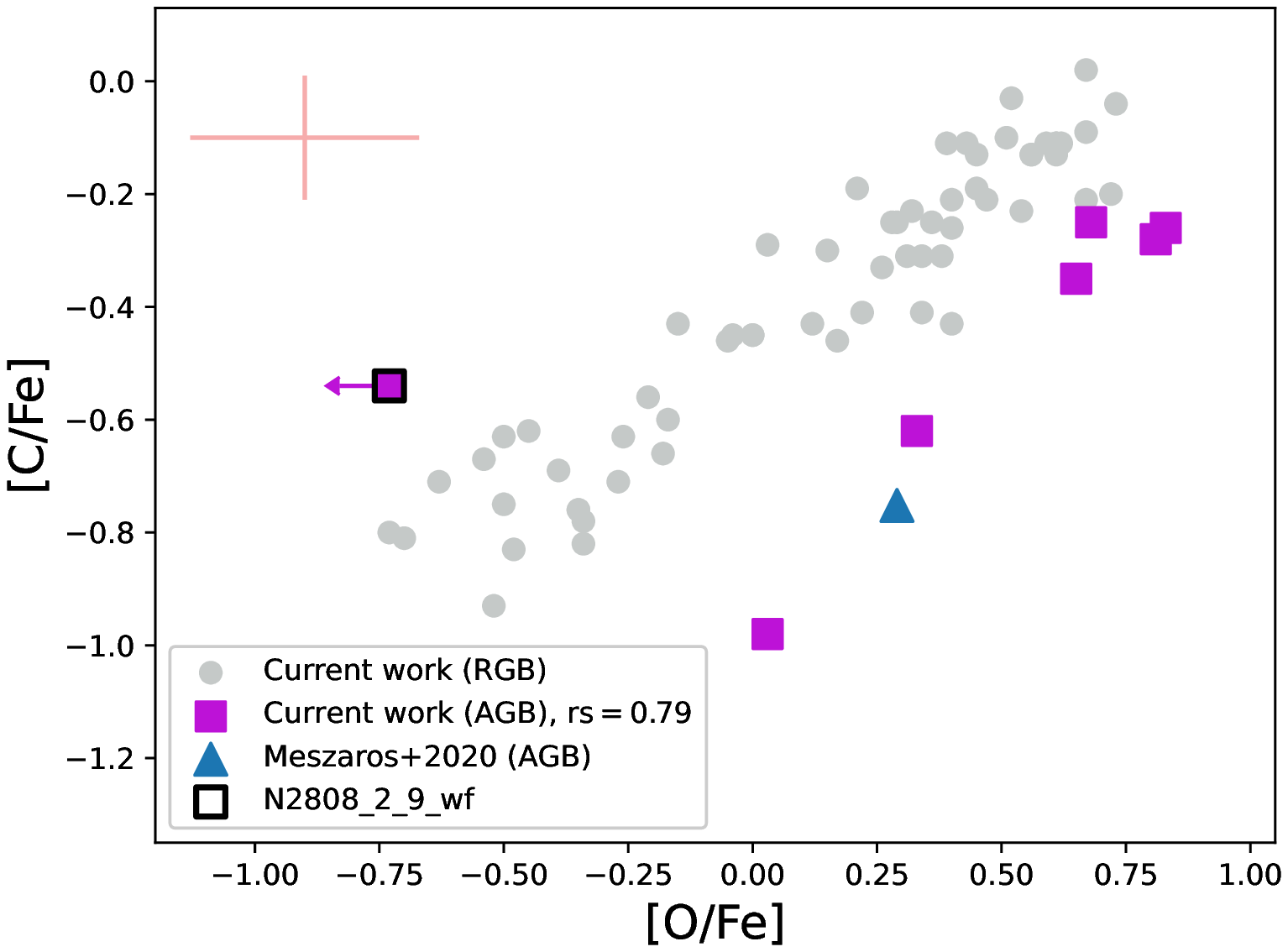}  & \includegraphics[width=0.5\textwidth]{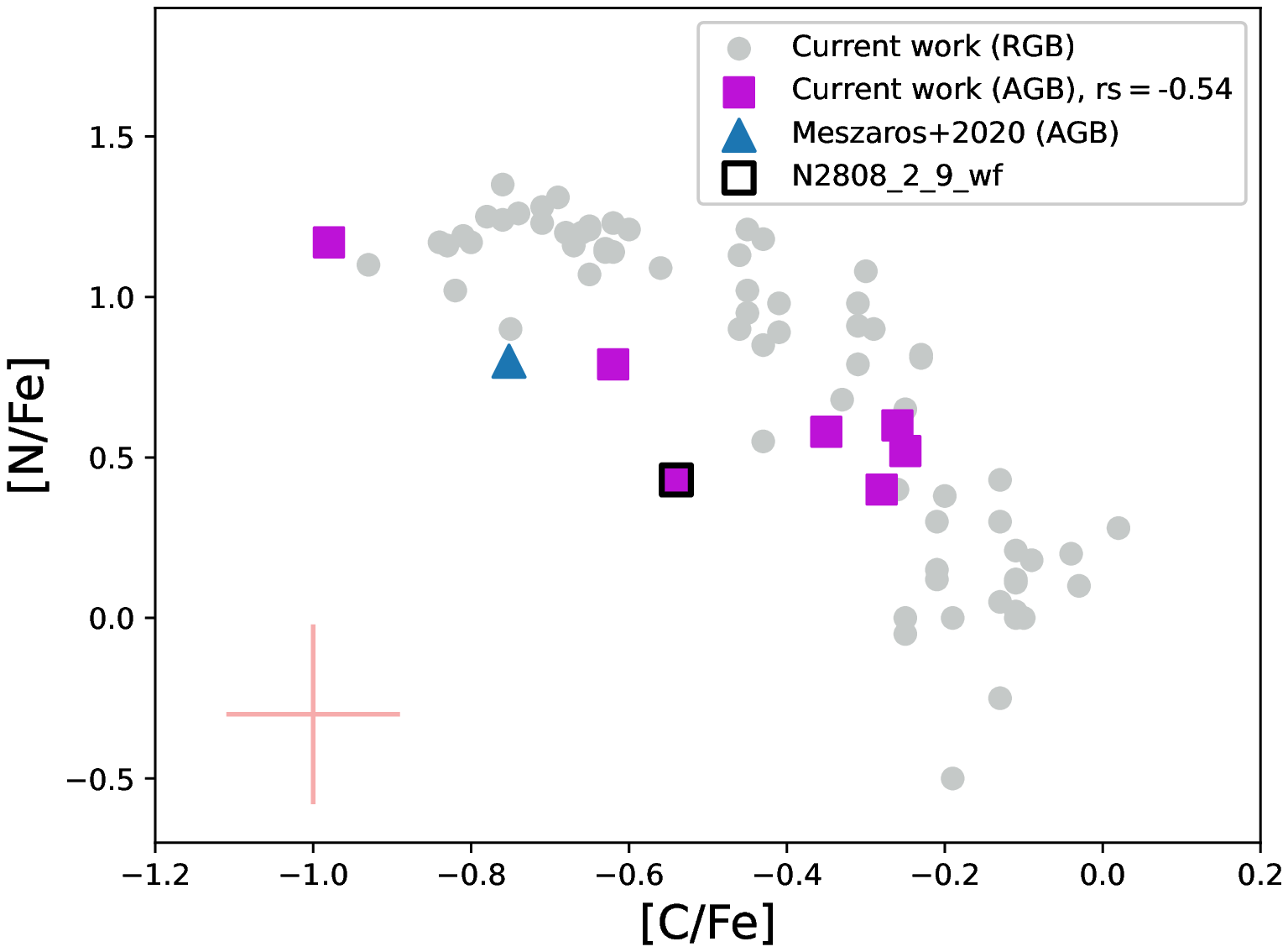}  \\
    \includegraphics[width=0.5\textwidth]{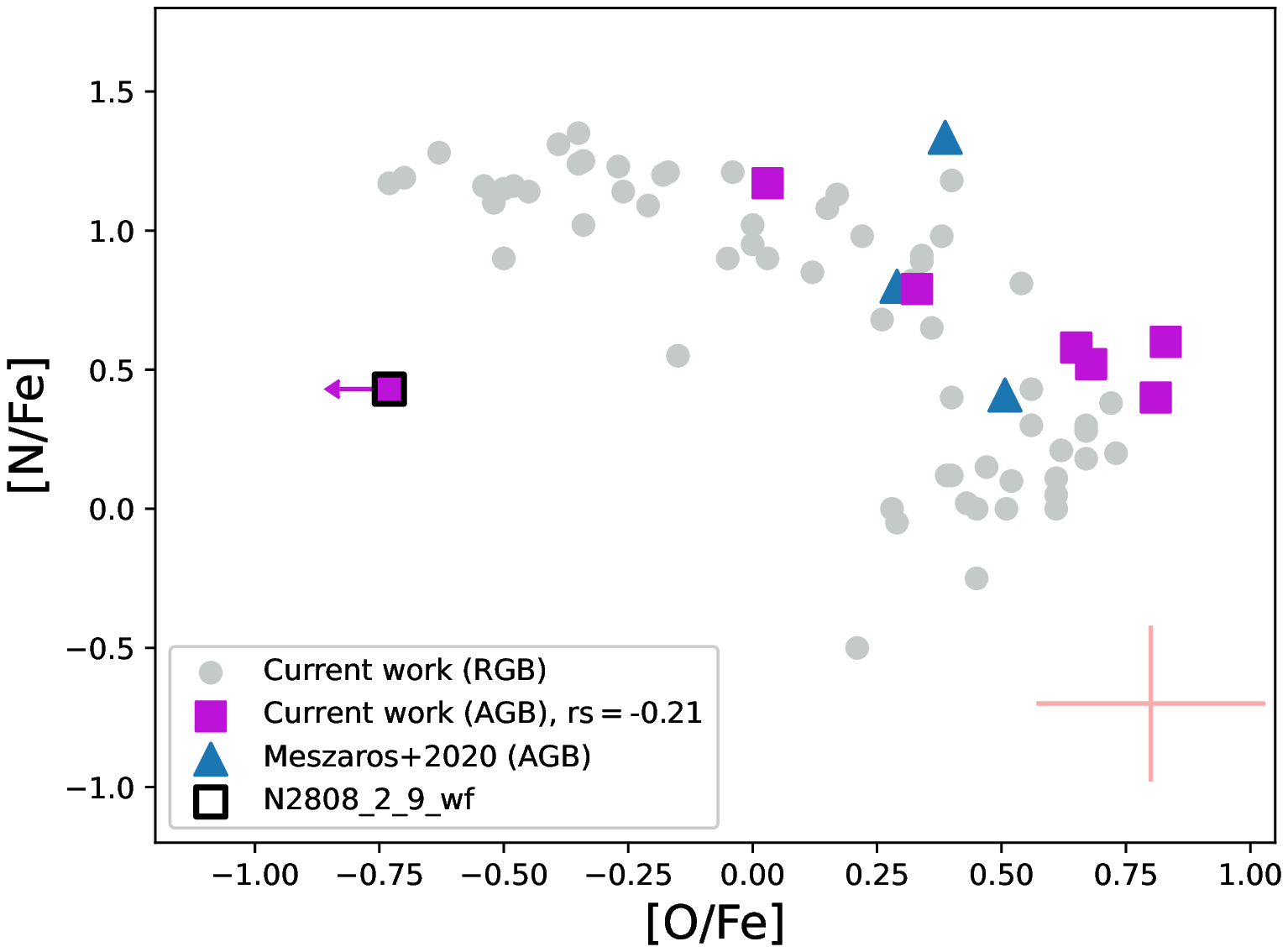} &
    \includegraphics[width=0.5\textwidth]{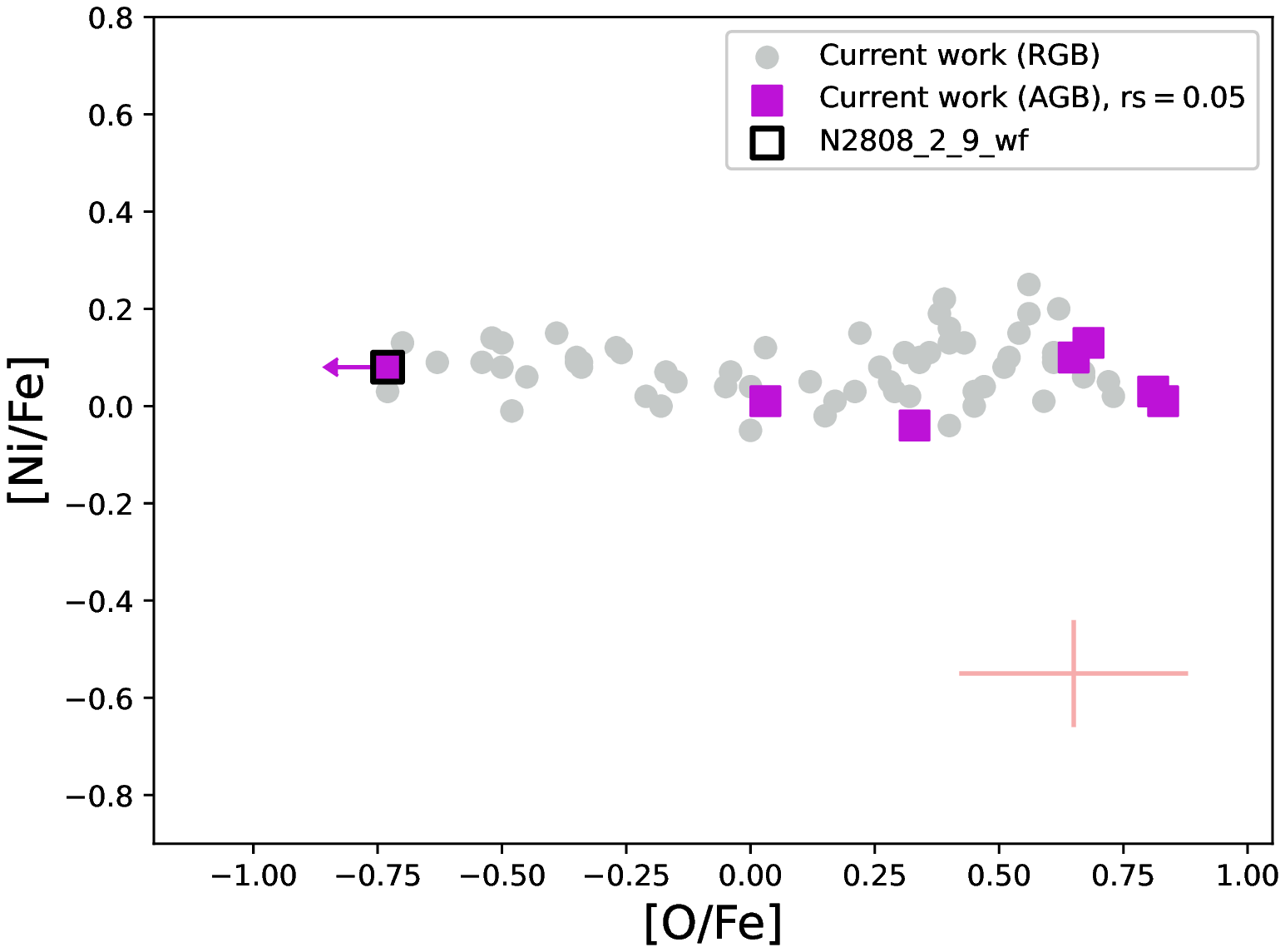} \\
    \includegraphics[width=0.5\textwidth]{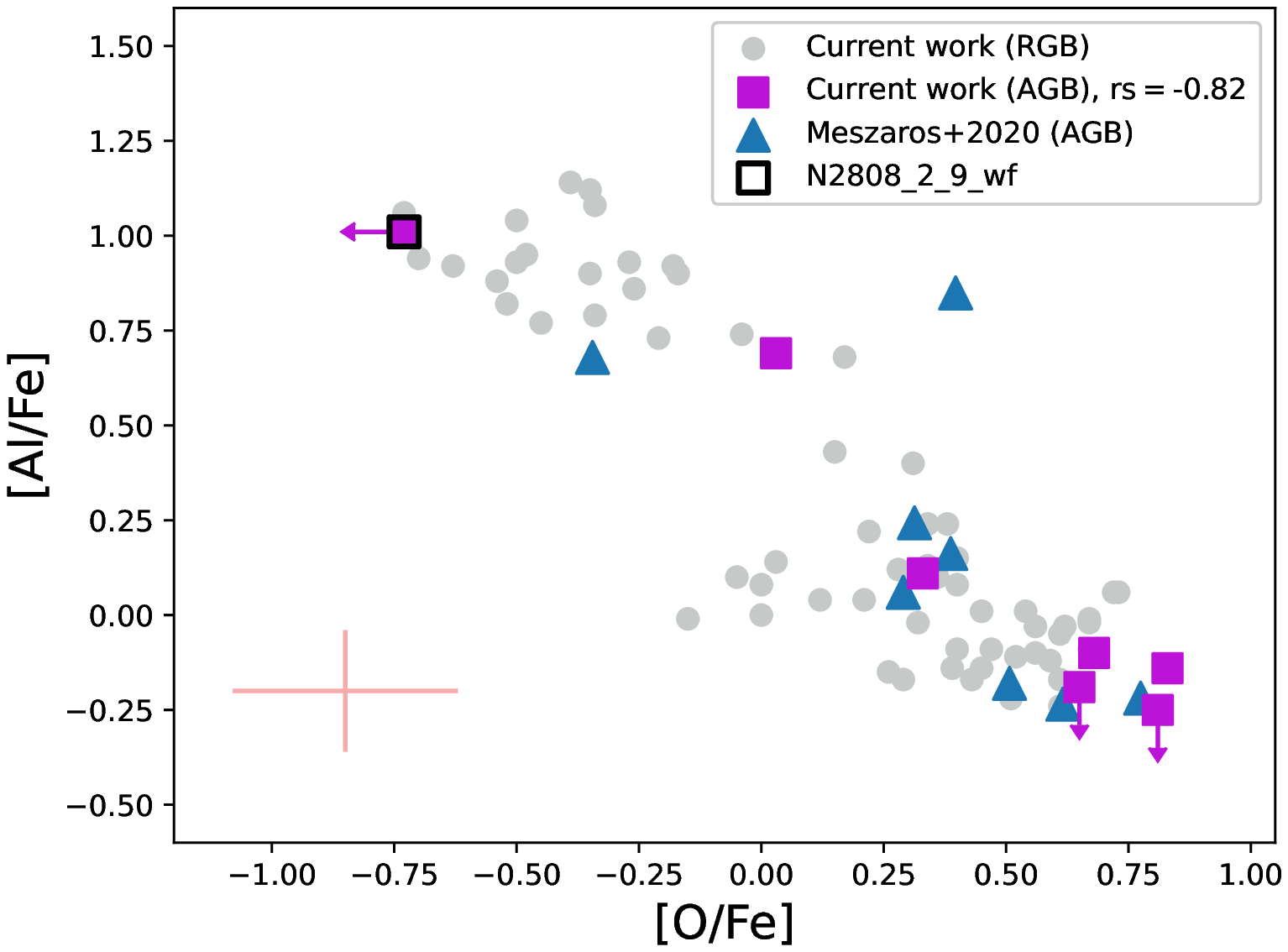} &
    \includegraphics[width=0.5\textwidth]{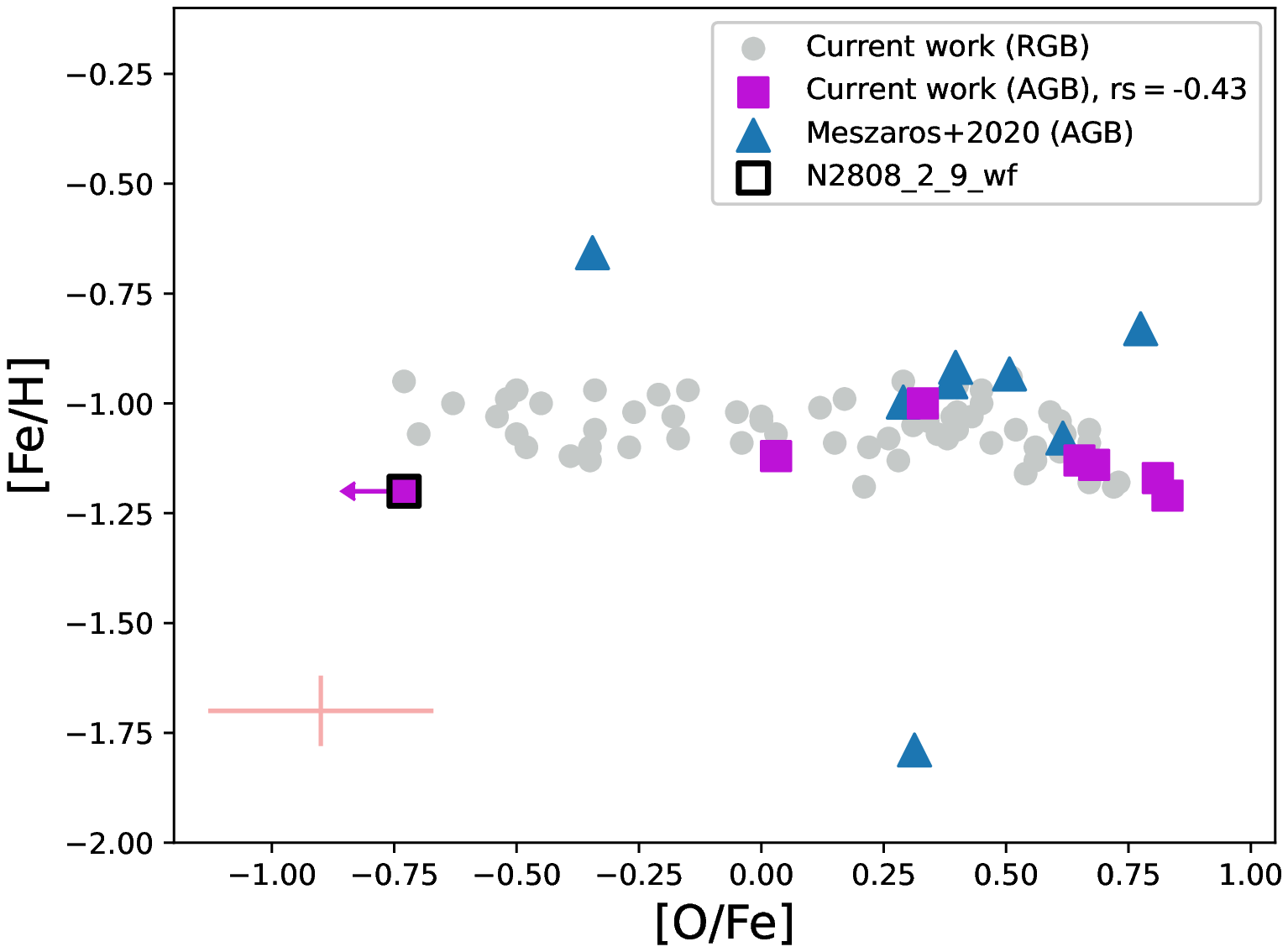} \\
\end{tabular}
	\caption{Abundances for AGB stars obtained in the current work (purple squares) compared to AGB stars from \protect\cite{meszaros/20} (blue triangles) and our study of RGB stars (light grey circles). Top left panel shows [C/Fe] vs. [O/Fe], top right panel presents [N/Fe] vs. [C/Fe], [N/Fe] vs. [O/Fe] is in the left middle panel, [Ni/Fe] vs. [O/Fe] is presented in the right middle panel, [Al/Fe] vs. [O/Fe] is shown in the bottom left panel and [Fe/H] vs. [O/Fe] is in the bottom right panel. The typical error (mean value) is displayed in red. The Spearman coefficients ($rs$) for each  possible (anti-)correlation are reported in all panels.}
    \label{ofes_agb}
\end{figure*}

 The abundances of our AGB stars in comparison with the results for RGB are shown in Figure \ref{ofes_agb}. 
  For completeness, we show results from AGB stars by \cite{meszaros/20}.
  An inspection of the [N/Fe] (middle left), [Ni/Fe] (middle right), [Al/Fe] (bottom left) and [Fe/H] (bottom right) versus [O/Fe] plots indicates that those stars have similar abundances when compared to the RGB stars in our sample. However, the upper panels of Figure \ref{ofes_agb} ([C/Fe] versus [O/Fe] and [N/Fe] versus [C/Fe]) might indicate that our AGB stars 
 have lower content of C, in general, when compared to the RGB sample (by a factor of $\sim 1.1$).
 This behaviour could be related to the different evolutionary stage of the AGB stars. In particular the C $\to \,{\rm{N}}$ processing occurring in the RGB envelopes could modify the CN surface abundances on the AGB \citep[e.g.,][]{smith_norris/93}.
 
 From the Al-O anti-correlation presented in Figure \ref{ofes_agb}, we observe AGB stars belonging to both first and second  populations, in agreement with \cite{marino/17} in the same cluster.
 Interestingly enough, we note that the O abundance inferred for one star is very low, with none of such extremely-O depleted stars, possibly associated to the population E on the ChM, included in the sample of \cite{marino/17}. 
 This star  (N2808\_2\_9\_wf) also displays high [Al/Fe], which is compatible with the star belonging to the extremely-helium rich population. On the other hand, we note that this object has more [C/Fe] and less [N/Fe] content than what is expected for the extremely-He enriched stellar population. 
 
 We regard this star as an interesting target for future observations with higher-resolution spectroscopy. 
 Indeed, the stars with the highest content of He \citep{dantona/05,milone/15} would in principle avoid the AGB phase becoming AGB-manqué stars \citep{greggio/90,gratton/10,chantereau/16}. Nonetheless, \cite{marino/17} 
  detected an AGB star 
   with the same chemical composition as population-D stars.
    This finding demonstrates that NGC\,2808 stars with Y$\sim$0.3 can evolve to the AGB phase; in addition, \cite{lagioia/21} found that several AGB stars can be associated with population D and possibly population E, based on their positions on the ChM. These findings might challenge the current stellar evolution models \citep[][]{chantereau/16}.

The evidence that He-rich stars in GCs skip the AGB phase and evolve as AGB manqu\'e, together with the presence of extremely blue He-rich HB stars, makes it tempting to link the phenomenon of multiple stellar populations with the X-UV Upturn phenomenon of elliptical galaxies. The latter consists of a ‘UV rising branch’ at wavelengths shorter than $\lambda \lesssim 2500$ \AA\, observed in the spectra of elliptical galaxies (e.g., \citealt{code/69}), whose origin is still controversial. By far, AGB manqu\'e stars and extreme HB stars are considered among sources that are likely to produce the largest contribution to the UV output from a galaxy (e.g., \citealt{renzini/90}). Hence, if elliptical galaxies host multiple stellar populations, similar to GCs, it is possible that helium-rich stars are responsible for the X-UV upturn phenomenon.

Recent works have provided robust evidence of a significant amount of AGB manqu\'e stars, among He-rich GC stars (e.g., \citealt{campbell/13}, \citealt{marino/17} and \citealt{lagioia/21}). The discovery of a He-rich AGB star would certainly indicate that a small fraction of AGB He-rich stars skip the AGB phase. However, a large statistic sample of He-rich AGB stars is needed to estimate their fraction and their contribution to the total X-UV flux of a GC.

 

\subsection{Literature comparison}
 
 Several studies of the chemical abundances
 in NGC\,2808 can be found in the literature, so it is worth discussing here a comparison with previous results. The most recent studies by \cite{carretta/15} and \cite{carretta/18} include the [O/Fe], [Al/Fe] and [Ni/Fe] abundances. In addition, \cite{meszaros/20} presented [C/Fe], [N/Fe], [O/Fe] and [Al/Fe] abundances, based on APOGEE spectra.
 
 Our average [Fe/H] for NGC\,2808 is $-1.05$~dex
 with a dispersion of 0.07~dex (considering only RGB stars). 
 This value is  0.09~dex higher than the one reported in \cite{harris/96} (2010 version, [Fe/H]$=-1.14$~dex that is an average result compiled with values found in the literature), 0.08~dex greater than the [Fe/H] average from \cite{carretta/15} ([Fe/H]$=-1.13$~dex) but if we consider the same solar constant for Fe the difference decreases to 0.04~dex, and 0.12~dex smaller than the average metallicity from \cite{meszaros/20} ([Fe/H]$=-0.93$~dex) and again, if we consider the same Fe solar constant the difference diminishes to 0.07~dex. 
 
 
 Figures \ref{ofes} and \ref{ofes_agb} show the literature comparison for the (anti-) correlations between [C/Fe], [N/Fe], [Al/Fe] and [O/Fe], as well as [Ni/Fe] versus [O/Fe] and [Fe/H] versus [O/Fe] for RGB and AGB stars, respectively. The classical anti-correlation between [C/Fe] and [N/Fe] and their comparison with literature studies is also shown. 
 For a better comparison, the abundances from \cite{carretta/15,carretta/18} and \cite{meszaros/20} presented in these  figures are also corrected to the same solar constants used in our work.
 
 Although the methods and spectral regions used to analyse the chemical abundances of different elements from \cite{carretta/15}, \cite{carretta/18} and \cite{meszaros/20} are distinct from our work, we found a reasonable concordance of our findings when compared to these studies. \cite{carretta/15} used the [O {\sc i}] forbidden lines at $\sim6300$ \AA\, and several Ni {\sc i} lines from 4900 to 6800 \AA. For Al determination, \cite{carretta/18} used the method presented in \cite{carretta/12} that analysed the Al {\sc i} doublet at 8772-73 \AA. Lastly, \cite{meszaros/20} used infrared features to determine the abundances of Al, C, N and O. 
 
 Our RGB C abundances are higher than the ones from \cite{meszaros/20} by a factor of $\sim 1.3$, while the N from \cite{meszaros/20} are higher than our results by a factor of $\sim 1.7$. We analysed  RGB stars that are below or above the luminosity bump (with magnitudes in the interval 
  $15.0\lesssim V \lesssim 16.6$) 
  while \cite{meszaros/20} only have stars above the RGB bump (with stars as bright as $V\sim13.5$), so a mild difference is expected. Although we have stars both below and above the RGB bump, our abundances of [C/Fe] and [N/Fe] do not show any trend with luminosity; probably because our sample does not include very bright RGB stars.
 
 

\subsection{Chemical abundances along the ChMs}
\label{ss_chm}

\begin{figure*}
    \includegraphics[width=1.0\textwidth]{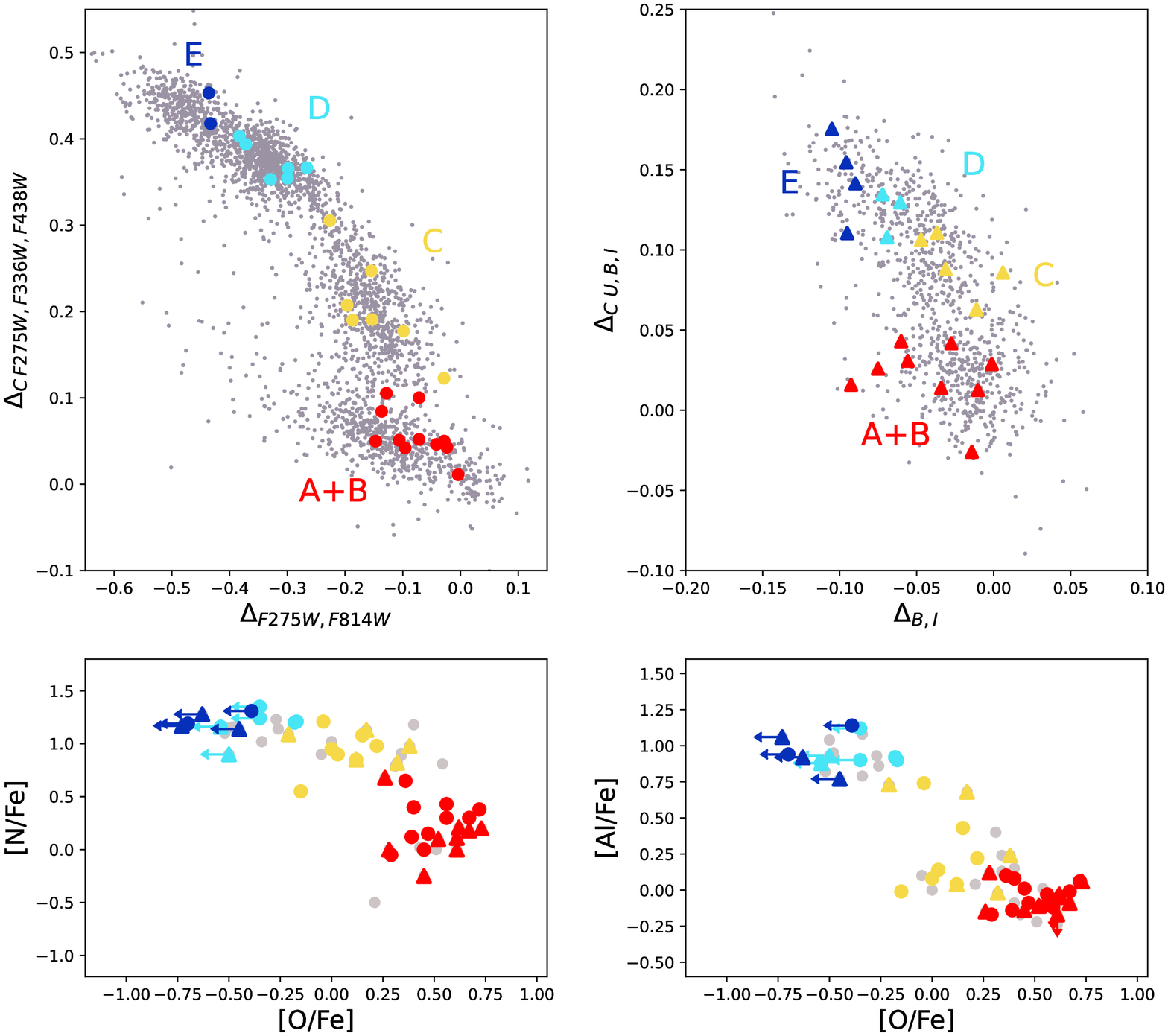} 
	\caption{The upper panels shows the {\it HST} ChM from the two field of view shown in Figure \ref{fig:footprints} that include data from \protect\cite{milone/15} (left) and our ChM from ground based photometry (right), with the spectroscopic data divided in populations  A+B (red), C (yellow), D (cyan) and E (dark blue). [N/Fe] (bottom left panel) and [Al/Fe] (bottom right panel) versus [O/Fe] for our RGB sample (grey circles) with the populations colour coded in agreement with the ChM panels. Spectroscopic data with {\it HST} ChM information are represented with filled circles, while filled triangles show spectroscopic data with ground based ChM information.}
    \label{chm_xfe_hst_gb}
\end{figure*}

The upper panels of Figure \ref{chm_xfe_hst_gb}  show the ChMs of NGC\,2808 from {\it HST} and ground based photometry. The spectroscopic sample is colour-coded according to their populations based on the classification in \cite{milone/15}. In the following, we investigate the chemical composition of four stellar populations: the first population 1P, and the three main groups of second-population stars, namely C, D, and E\footnote{1P stars of NGC\,2808 host two main stellar sub-populations, named A and B by \cite{milone/15}. 
 However, one spectroscopic target only belongs to the Population A based on it position in the {\it HST} ChM. Moreover, the ground-based ChM does not provide a clear separation between Population A and B stars. Based on ChM analysis, Population A and B stars share nearly the same chemical composition, but have slightly different metallicity  \citep[corresponding to a variation in iron abundance by less than 0.1 dex][]{marino/2019b, legnardi/22}.
 For these reasons, we will consider all 1P stars in the same sample, without distinguishing between Population A and Population B stars.
 }.
 Figure \ref{chm_xfe_hst_gb} shows [N/Fe] (bottom left panel) and [Al/Fe] (bottom right panel) versus [O/Fe] for the sample of RGB stars and the highlighted stars for which we have ChM information\footnote{
 We note that abundances of N and O are not available for all spectroscopic targets. Hence, the number of colored stars in the upper and lower panels of Figure \ref{chm_xfe_hst_gb} do not match with each other.}.

N is the most sensitive element (presented in our work) for the location of a star in the ChM, as shown by 
 \cite{marino/19}. Due to the filters used in the construction of a ChM, variations in N would translate in a different position for the stars in this diagram. 
 Oxygen variation also affect, to a minor extent, the position of a star along the ChM.
  Thus, in our study, the [N/Fe] versus [O/Fe] plot is an  efficient tool to analyse multiple populations in a GC with spectroscopic data. 
   The left bottom panel of Figure \ref{chm_xfe_hst_gb} shows this anti-correlation and the clear separation between 1P stars and Population C stars. Populations D and E share similar N contents, whereas the most extreme star from population E present the lowest O abundances. 
We note a clear separation among the distinct populations also in the right bottom panel of Figure \ref{chm_xfe_hst_gb}, which shows the anti-correlation of [Al/Fe] versus [O/Fe].


Table \ref{mean_ab_pops} summarises the average abundances of each element studied here for each ChM population, both for {\it HST} and ground based photometry. [Fe/H] is 
constant in the distinct populations, at a level of $\lesssim$0.10~dex.
The abundances of [C/Fe] and [O/Fe] decrease from 1P (i.e. Populations A$+$B) stars to E, while [N/Fe] and [Al/Fe] abundances increase.
Except for O
 and Li,
 which are substantially lower in the E population with respect to population D stars, we do not observe large differences in the other light elements between these two stellar populations. 

\begin{table*}
\centering
\caption{Average abundances for each population of stars, with their respective mean errors, dispersion ($\sigma$), and number of stars (\#).}
\label{mean_ab_pops}
\begin{tabular}{rrrrrrrrrrrrr}
\hline
\multicolumn{13}{c}{Current work} \\
\hline
 & Pop. A+B & $\sigma$ & \# & Pop. C & $\sigma$ & \# & Pop. D & $\sigma$  & \# & Pop. E  & $\sigma$ & \#  \\

\hline
\hline                            

$\left \langle \rm{[Fe/H]} \right \rangle$ & $-1.07\pm0.02$ & 0.07 & 20 & $-1.04\pm0.01$ & 0.04 & 12 & $-1.05\pm0.02$ & 0.06 & 10 &  $-1.02\pm0.02$ & 0.06 & 6 \\
$\left \langle \rm{[C/Fe]} \right \rangle$ & $-0.16\pm0.02$ & 0.08 & 20 & $-0.40\pm0.03$ & 0.09 & 12 & $-0.68\pm0.02$ & 0.05 & 10 &  $-0.72\pm0.03$ & 0.07 & 6 \\
$\left \langle \rm{[N/Fe]} \right \rangle$ & $0.21\pm0.05$ & 0.23 & 19 & $0.95\pm0.05$ & 0.17 & 12 & $1.17\pm0.04$ & 0.11 & 10 & $1.22\pm0.02$ & 0.06 & 6 \\
$\left \langle \rm{[O/Fe]} \right \rangle$ & $0.51\pm0.03$ & 0.14 & 20 & $0.09\pm0.05$ & 0.17 & 12 & $\lesssim-0.39$ & 0.14 & 8 & $\lesssim-0.58$ & 0.14 & 5 \\
$\left \langle \rm{[Al/Fe]} \right \rangle$ & $-0.05\pm0.02$ & 0.09 & 20 & $0.28\pm0.08$ & 0.28 & 12 & $0.90\pm0.03$ & 0.08 & 10 & $0.95\pm0.05$ & 0.12 & 6 \\
$\left \langle \rm{[Ni/Fe]} \right \rangle$ & $0.08\pm0.02$ & 0.08 & 20 & $0.06\pm0.02$ & 0.07 & 12 & $0.06\pm0.02$ & 0.05 & 10 & $0.09\pm0.02$ & 0.04 & 6 \\
\hline

\multicolumn{13}{c}{Literature data$\dagger$}\\
\hline
 & Pop. A+B & $\sigma$ & \# & Pop. C & $\sigma$ & \# & Pop. D & $\sigma$  & \# & Pop. E  & $\sigma$ & \#  \\
\hline
\hline 
 $\left \langle \rm{[Fe/H]} \right \rangle$ & $-1.12\pm0.01$ & 0.02 & 27 & $-1.12\pm0.01$ & 0.02 & 14 & $-1.10\pm0.01$ & 0.01 & 4 & $-1.11\pm0.01$ & 0.02 & 6 \\
$\left \langle \rm{A(Li)} \right \rangle$ & $1.16\pm0.02$ & 0.05 & 5 & $1.10\pm0.03$ & 0.08 & 6 & $1.05\pm0.02$ & 0.05 & 6 & $0.82\pm0.11$ & 0.16 & 3 \\
$\left \langle \rm{[O/Fe]} \right \rangle$ & $0.27\pm0.02$ & 0.09 & 25 & $-0.16\pm0.08$ & 0.30 & 11 & $-0.35\pm0.11$ & 0.19 & 3 & $-0.46\pm0.07$ & 0.16 & 2 \\
$\left \langle \rm{[Na/Fe]} \right \rangle$ & $0.02\pm0.02$ & 0.08 & 27 & $0.32\pm0.03$ & 0.12 & 14 & $0.39\pm0.04$ & 0.06 & 4 & $0.44\pm0.04$ & 0.10 & 6 \\
$\left \langle \rm{[Mg/Fe]} \right \rangle$ & $0.37\pm0.01$ & 0.05 & 27 & $0.27\pm0.03$ & 0.11 & 14 & $0.12\pm0.07$ & 0.12 & 4 & $0.09\pm0.06$ & 0.14 & 6 \\
$\left \langle \rm{[Al/Fe]} \right \rangle$ & $-0.08\pm0.04$ & 0.18 & 21 & $0.72\pm0.13$ & 0.47 & 9 & $1.09\pm0.04$ & 0.08 & 4 & $1.13\pm0.09$ & 0.21 & 6 \\
$\left \langle \rm{[Si/Fe]} \right \rangle$ & $0.27\pm0.01$ & 0.03 & 27 & $0.31\pm0.01$ & 0.04 & 14 & $0.34\pm0.02$ & 0.03 & 4 & $0.36\pm0.02$ & 0.05 & 6 \\
$\left \langle \rm{[Ca/Fe]} \right \rangle$ & $0.32\pm0.01$ & 0.03 & 27 & $0.33\pm0.01$ & 0.02 & 14 & $0.32\pm0.01$ & 0.02 & 4 & $0.32\pm0.01$ & 0.03 & 6 \\
$\left \langle \rm{[Sc/Fe]} \right \rangle$ & $-0.03\pm0.01$ & 0.04 & 27 & $0.01\pm0.01$ & 0.04 & 14 & $0.01\pm0.01$ & 0.03 & 4 & $0.05\pm0.01$ & 0.02 & 6 \\
$\left \langle \rm{[Ti/Fe]} \right \rangle$ & $0.22\pm0.01$ & 0.04 & 27 & $0.21\pm0.01$ & 0.02 & 14 & $0.211\pm0.01$ & 0.02 & 4 & $0.18\pm0.01$ & 0.03 & 6 \\
$\left \langle \rm{[Cr/Fe]} \right \rangle$ & $-0.04\pm0.01$ & 0.03 & 27 & $-0.03\pm0.01$ & 0.03 & 14 & $-0.03\pm0.01$ & 0.01 & 4 & $-0.01\pm0.01$ & 0.01 & 6 \\
$\left \langle \rm{[Mn/Fe]} \right \rangle$ & $-0.38\pm0.01$ & 0.01 & 5 & $-0.36$ & -- & 1 & $-0.38$ & -- & 1 & $-0.40$ & -- & 1 \\
$\left \langle \rm{[Ni/Fe]} \right \rangle$ & $-0.07\pm0.01$ & 0.02 & 27 & $-0.07\pm0.01$ & 0.02 & 14 & $-0.07\pm0.02$ & 0.03 & 4 & $-0.07\pm0.01$ & 0.02 & 6 \\

\hline
\end{tabular}

{\bf Note.} $\dagger$Data from \protect\cite{carretta/15}, \protect\cite{carretta/18} and \protect\cite{dorazi/15}.

\end{table*}

\begin{figure*}
\begin{tabular}{cc}
    \includegraphics[width=0.5\textwidth]{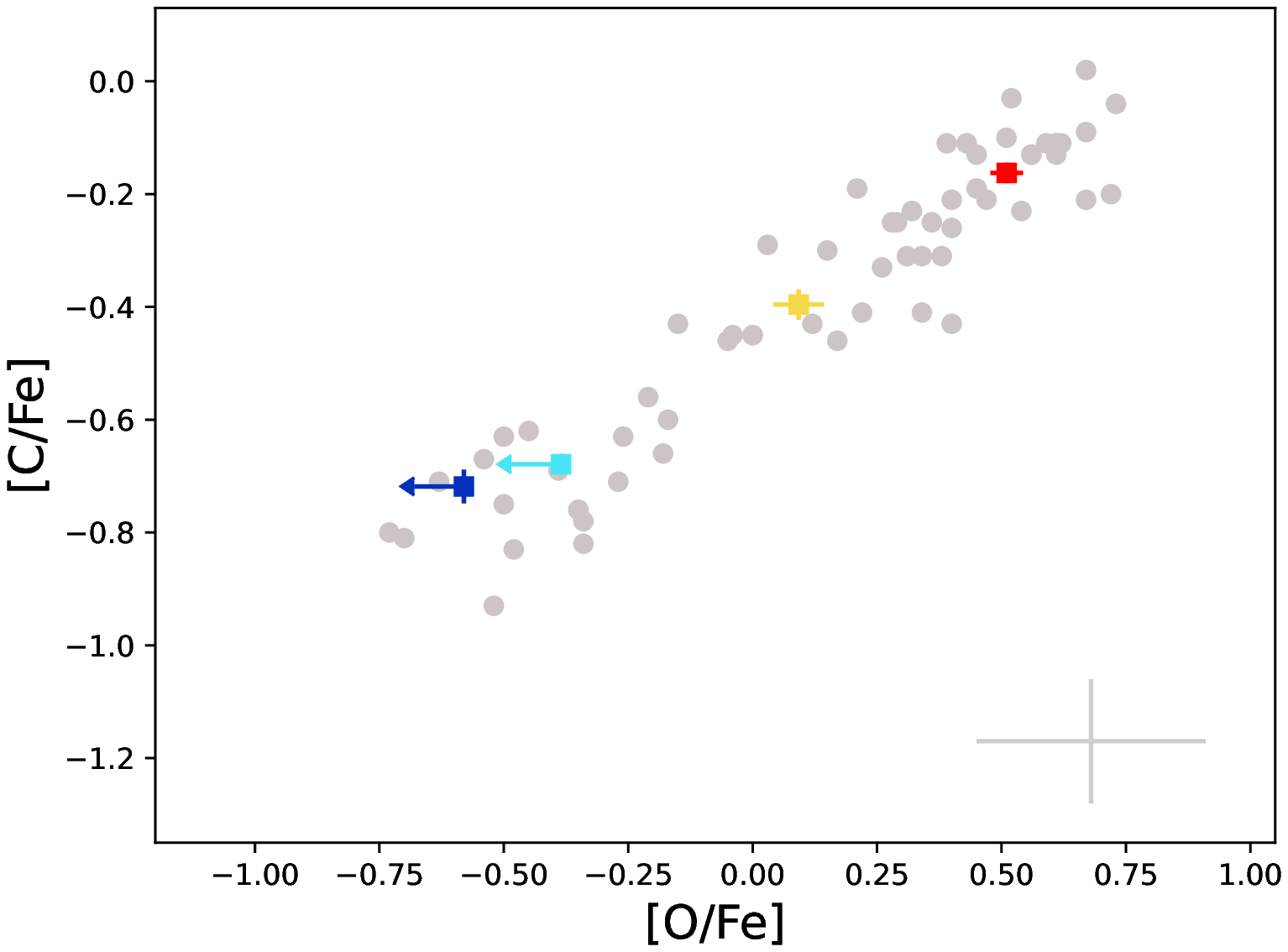}  & \includegraphics[width=0.5\textwidth]{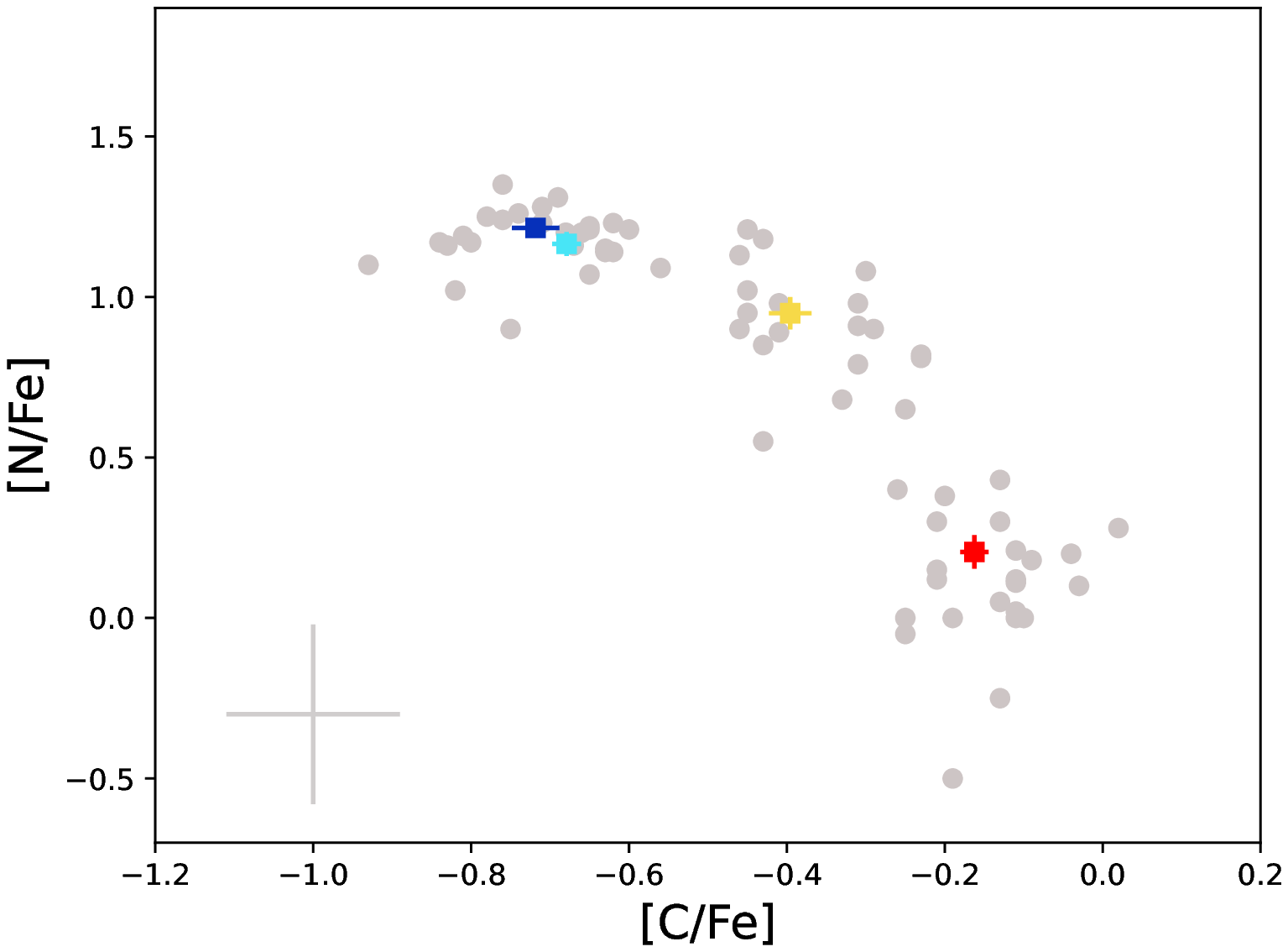}  \\
    \includegraphics[width=0.5\textwidth]{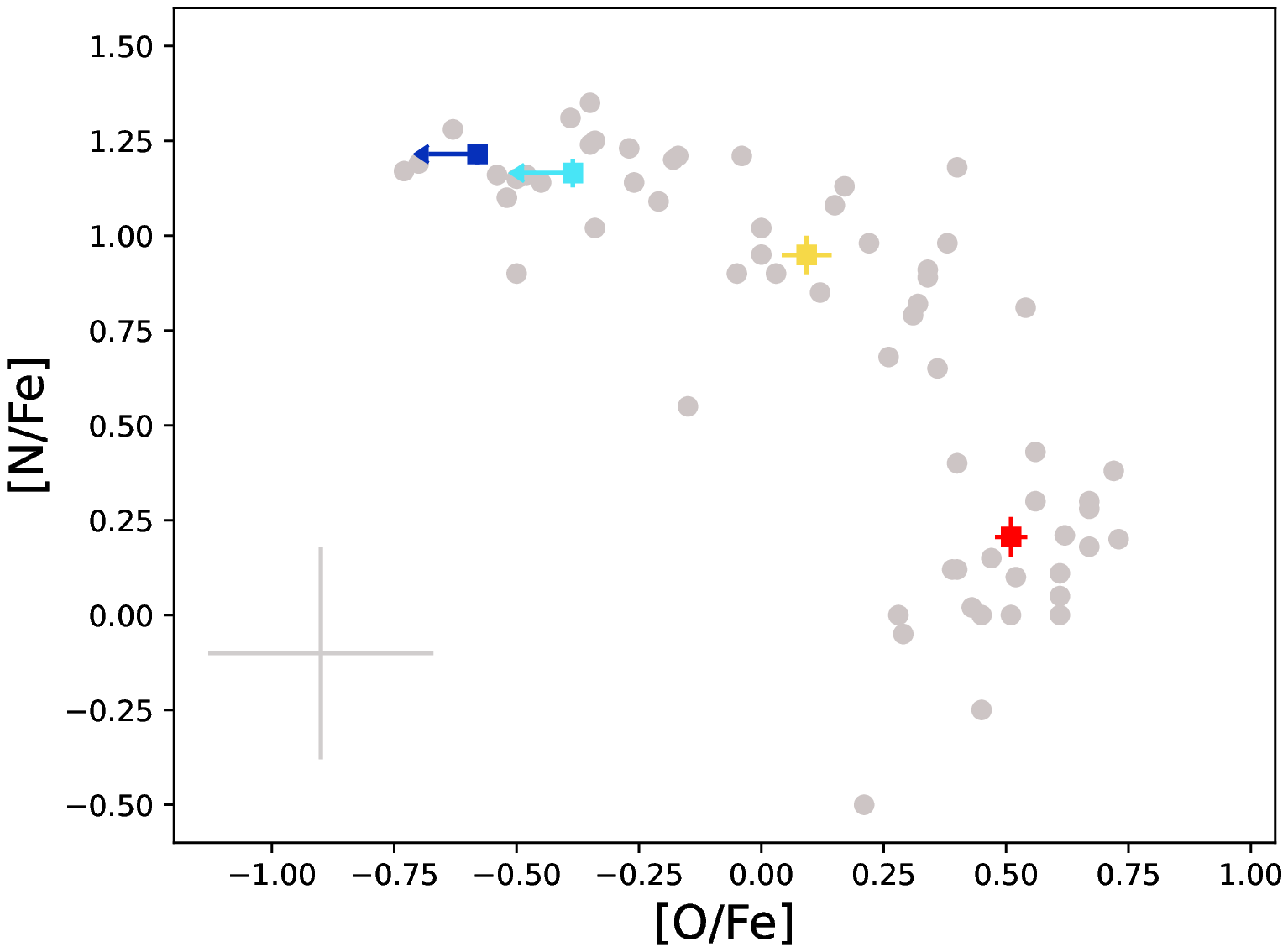} &
    \includegraphics[width=0.5\textwidth]{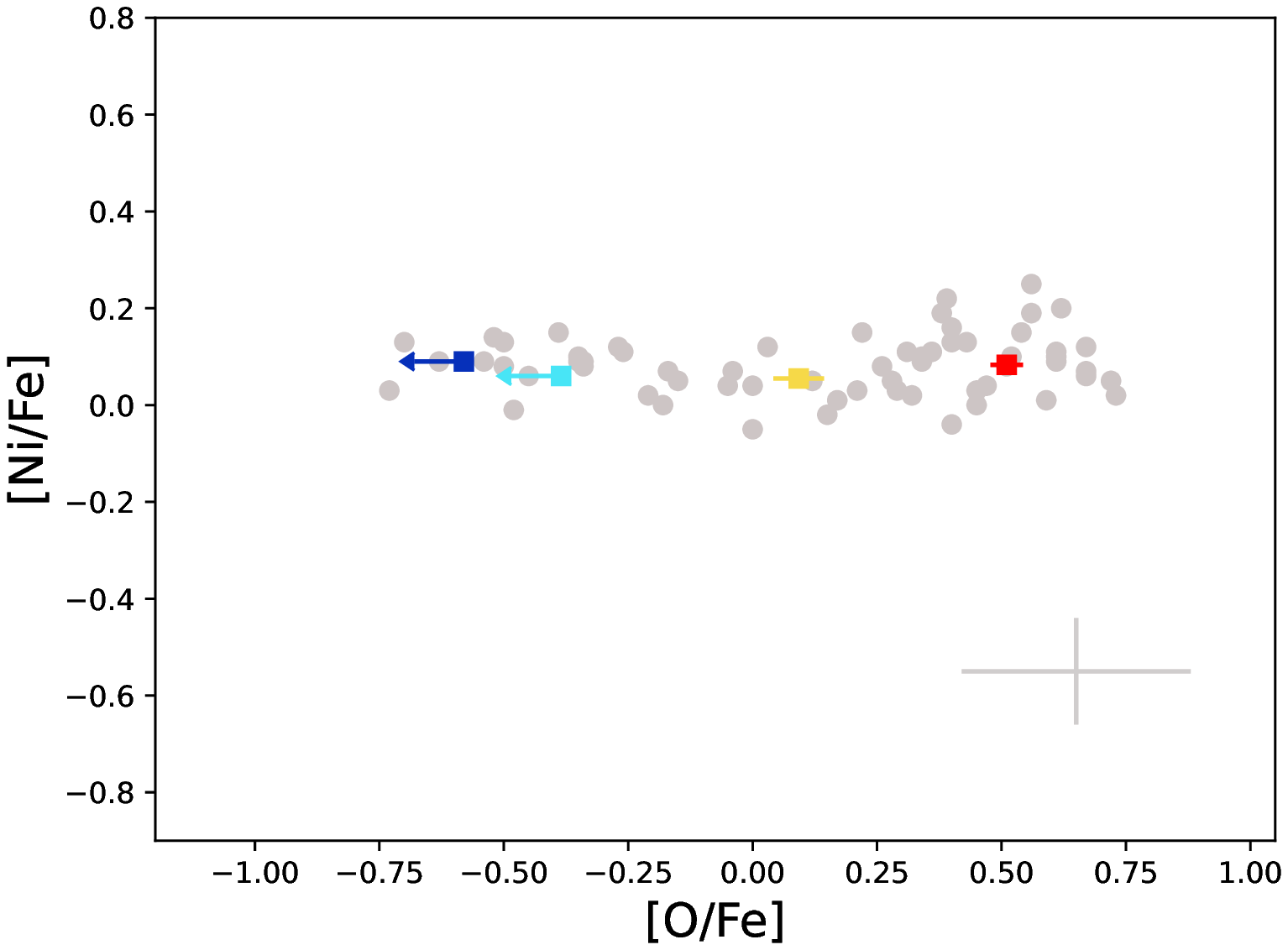} \\
    \includegraphics[width=0.5\textwidth]{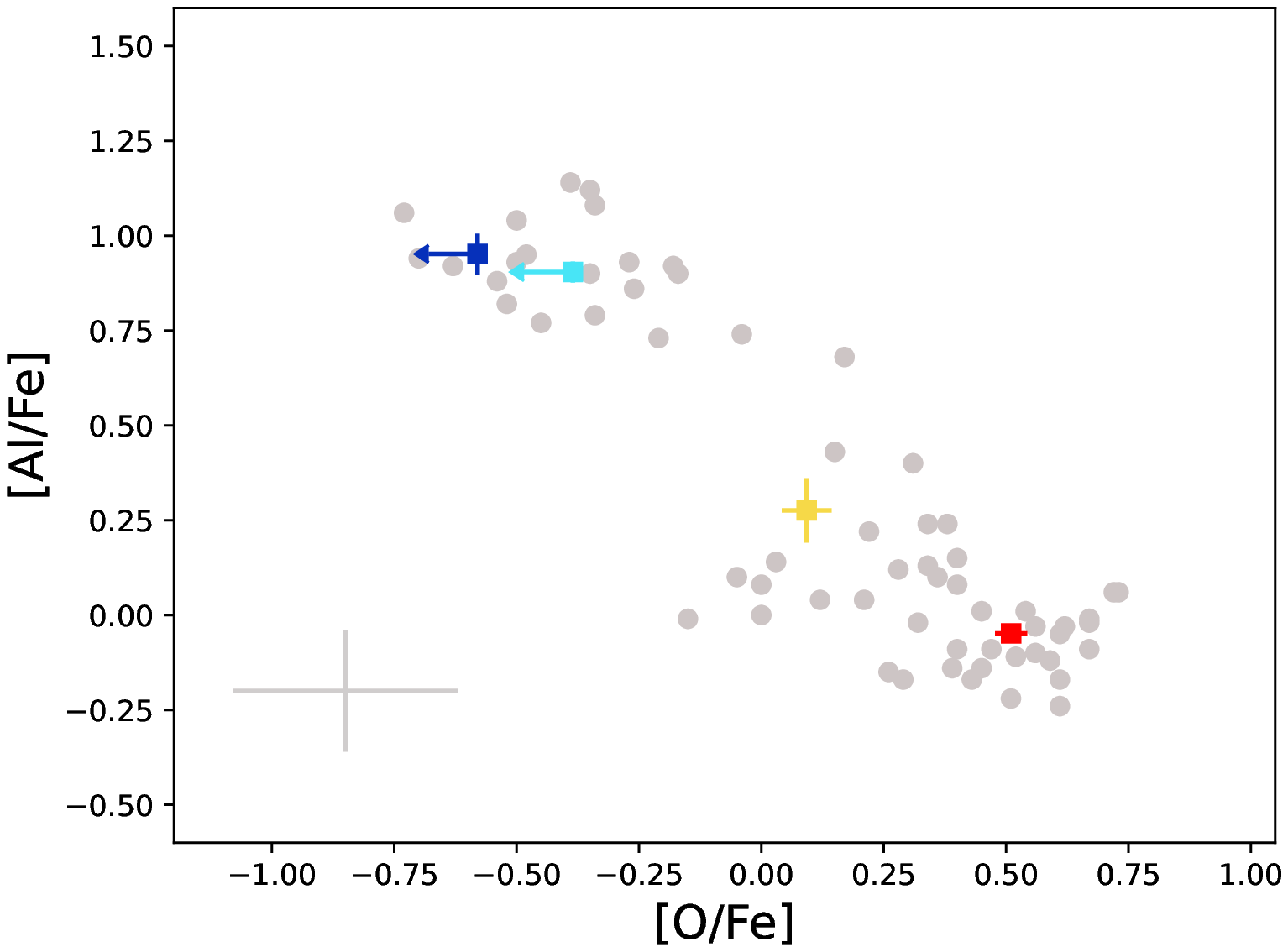} &
    \includegraphics[width=0.5\textwidth]{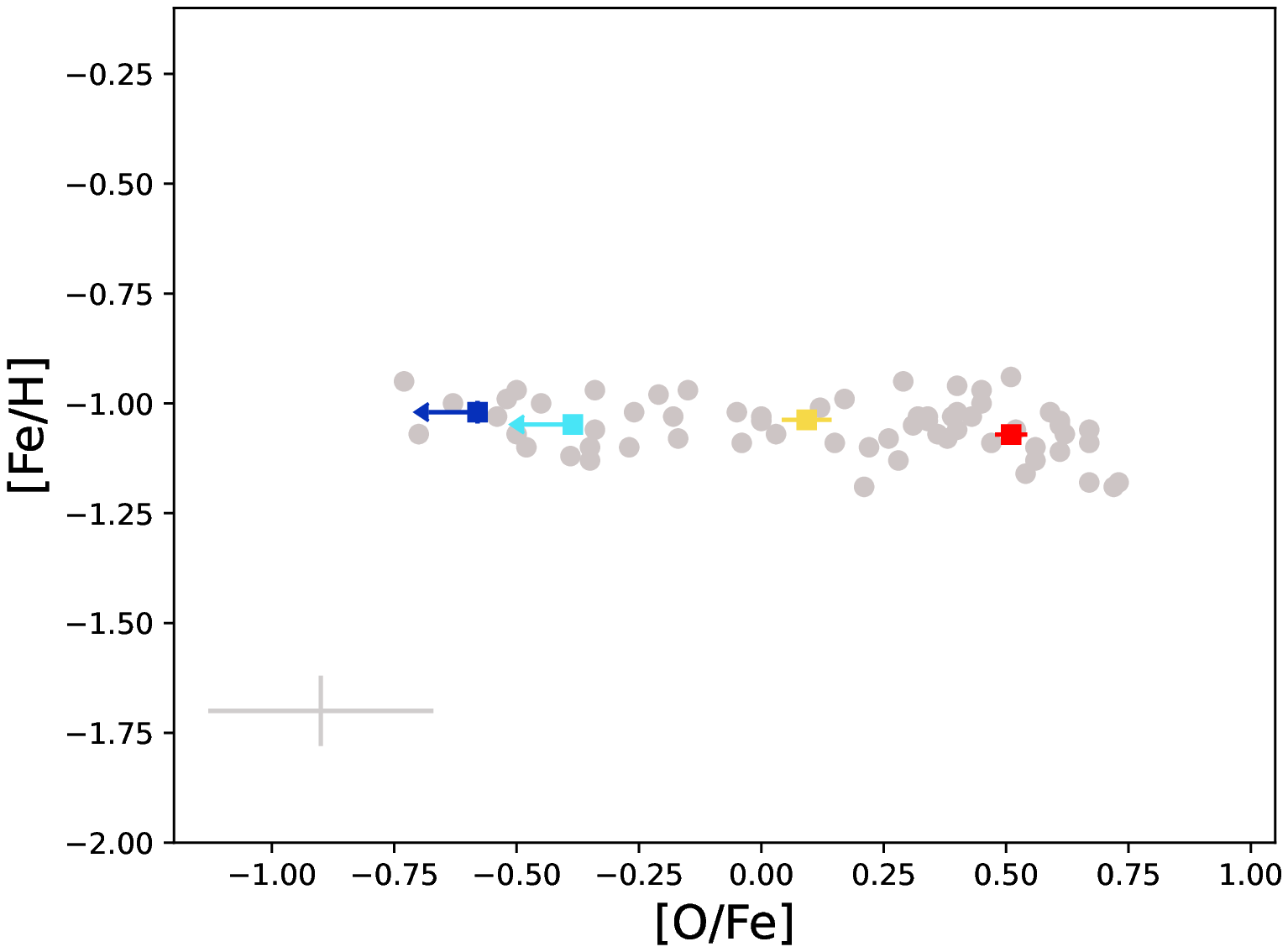} \\
\end{tabular}
	\caption{Abundance panels for RGB stars (grey circles) with their average abundances for each population highlighted. Population A+B is represented by red squares, the yellow squares illustrate population C, averages for population D are shown as cyan squares and dark blue squares display values for population E.  Top left panel shows [C/Fe] vs. [O/Fe], top right panel presents [N/Fe] vs. [C/Fe], [N/Fe] vs. [O/Fe] is in the left middle panel, [Ni/Fe] vs. [O/Fe] is presented in the right middle panel, [Al/Fe] vs. [O/Fe] is shown in the bottom left panel and [Fe/H] vs. [O/Fe] is in the bottom right panel.}
    \label{chm_xfe_mean}
\end{figure*}

Figure~\ref{chm_xfe_mean} shows the same (anti-)correlations presented in Figure~\ref{ofes}, but now with the average abundances for each population of the ChM illustrated by coloured circles (red for population A+B, C in yellow, D in cyan and dark blue for population E).
We find distinct average abundances for the different stellar populations, 
with population E (blue circles) showing the lowest O abundances (with some stars having only upper limits). 



\begin{figure*}
    \centering
    \begin{tabular}{c c}
        \includegraphics[width=0.5\textwidth]{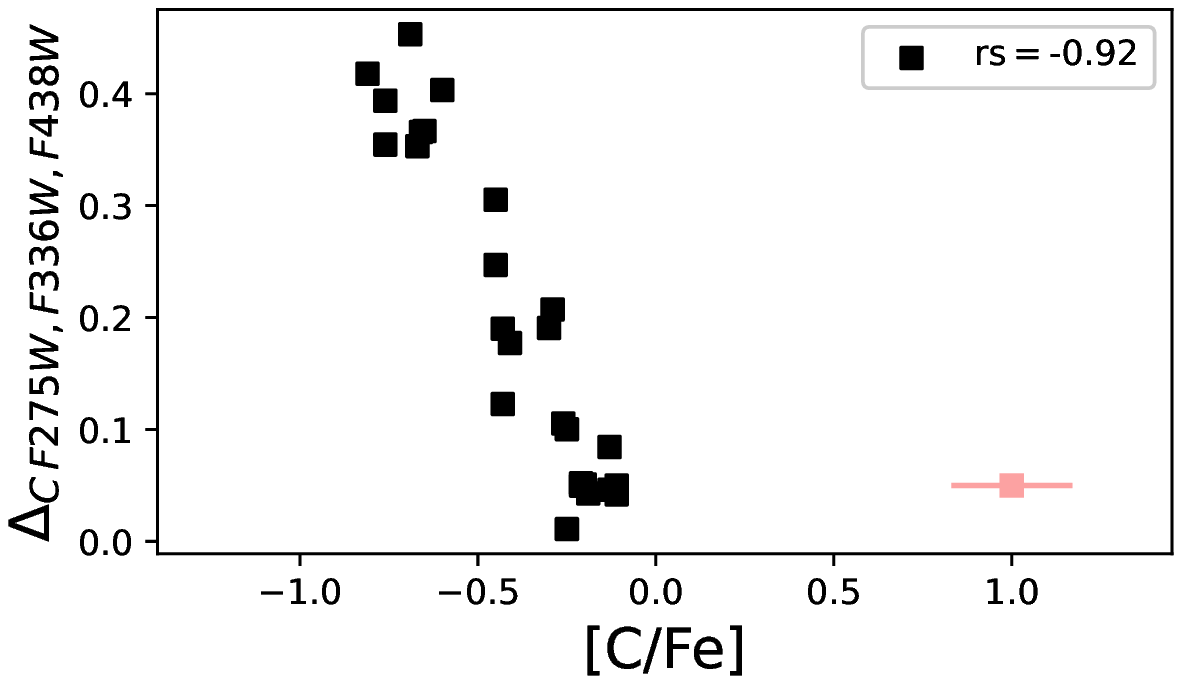}  &
        \includegraphics[width=0.5\textwidth]{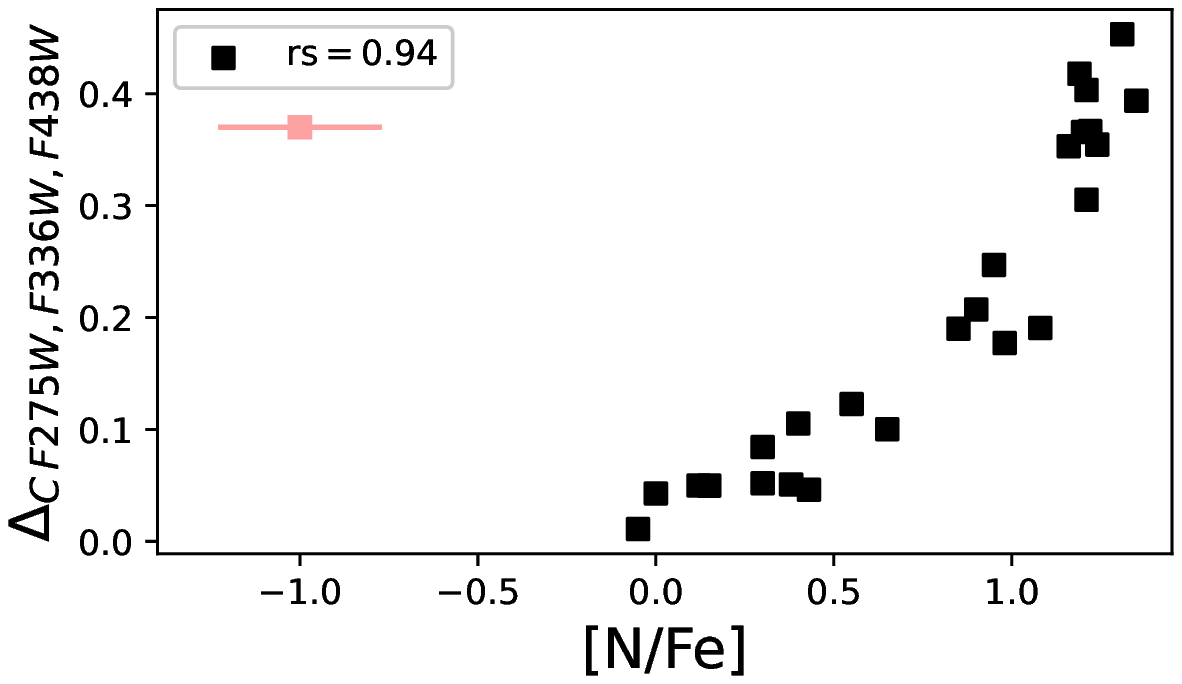} \\
        \includegraphics[width=0.5\textwidth]{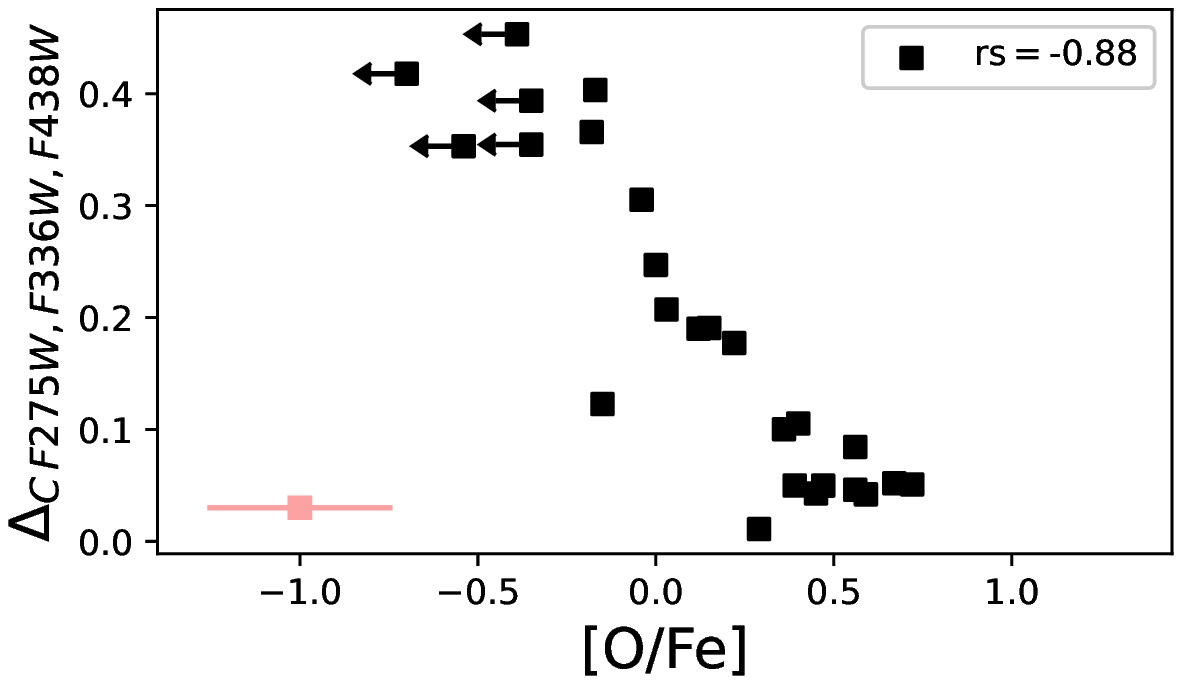} &
        \includegraphics[width=0.5\textwidth]{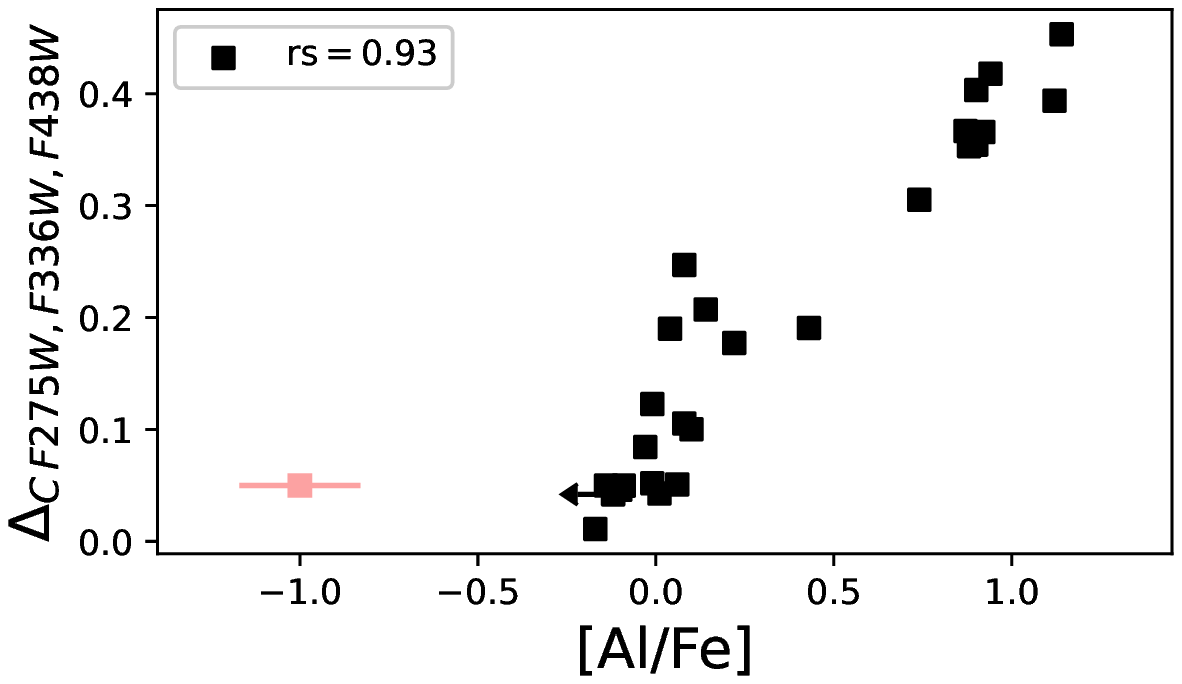} \\
        \includegraphics[width=0.5\textwidth]{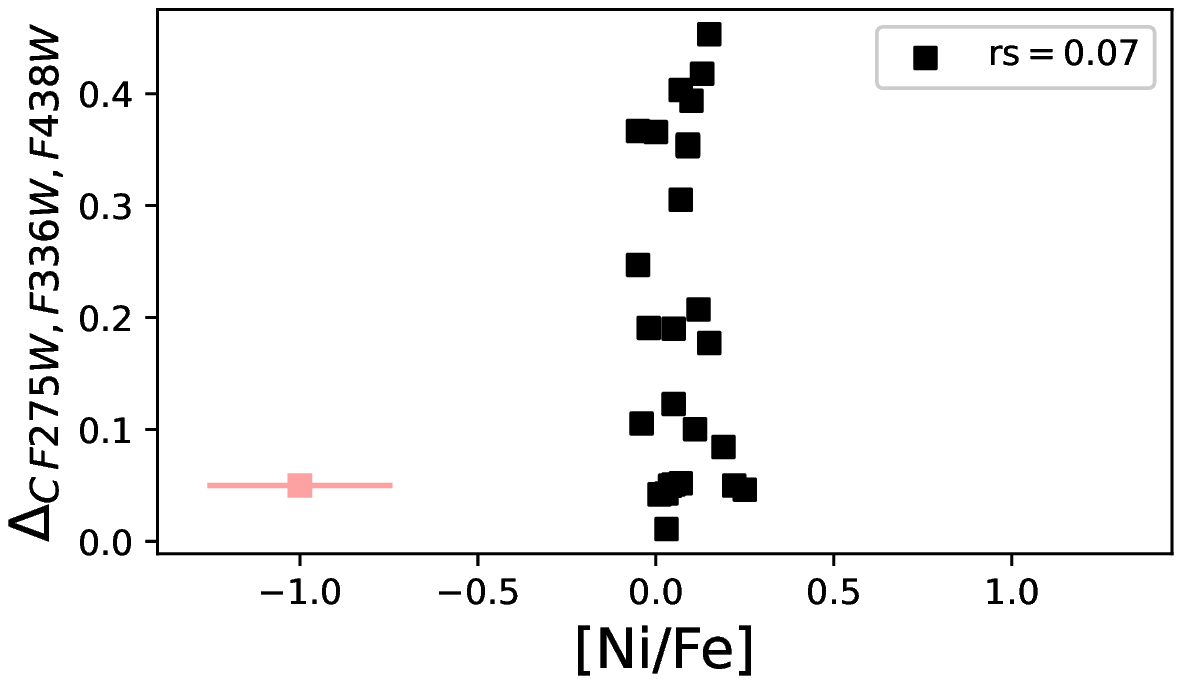} &
        \includegraphics[width=0.5\textwidth]{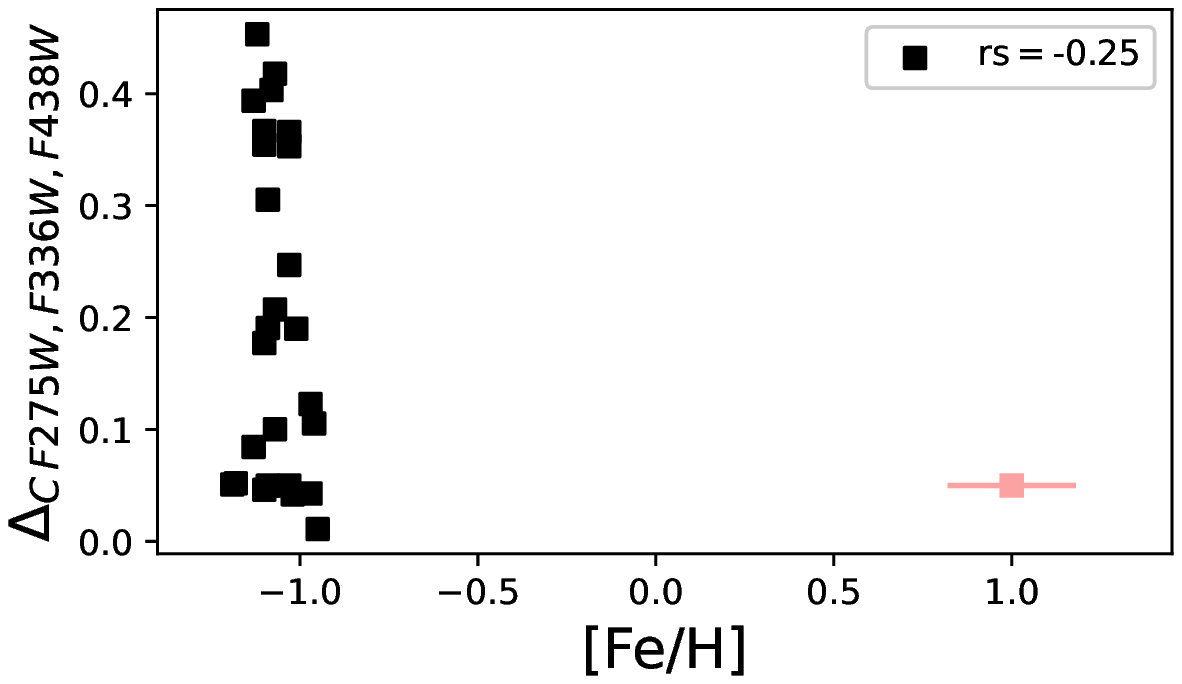} \\
    \end{tabular}
    \caption{\y \,versus [C/Fe] (top left panel), [N/Fe] (top right panel), [O/Fe] (middle left panel), [Al/Fe] (middle righgt panel), [Ni/Fe] (bottom left panel) and [Fe/H] (bottom right panel). The typical errors are marked in red.}
    \label{elements_dy_hst}
\end{figure*}

\begin{figure*}
    \centering
    \begin{tabular}{c c}
        \includegraphics[width=0.5\textwidth]{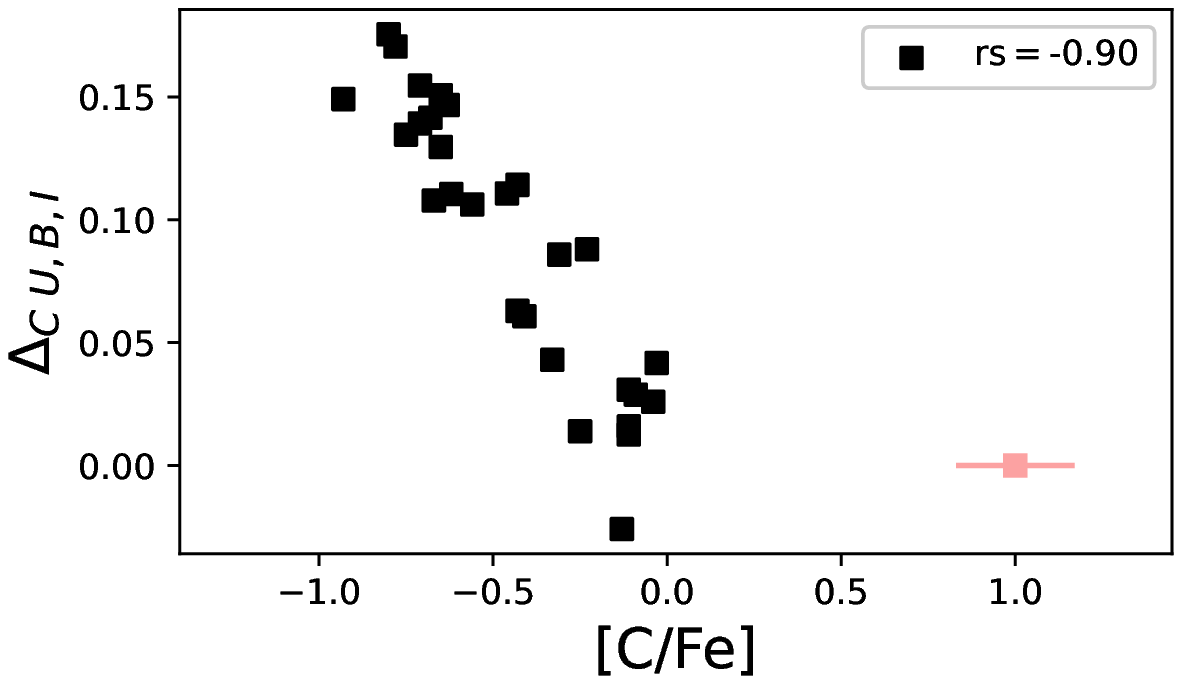}  &
        \includegraphics[width=0.5\textwidth]{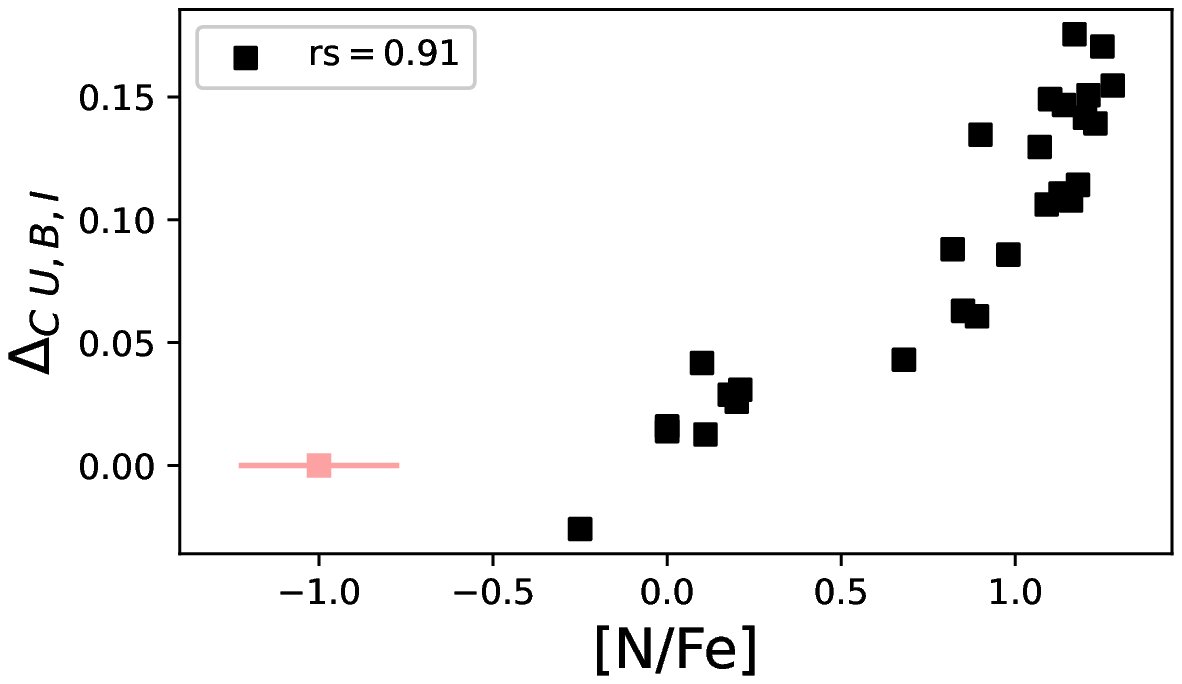} \\
        \includegraphics[width=0.5\textwidth]{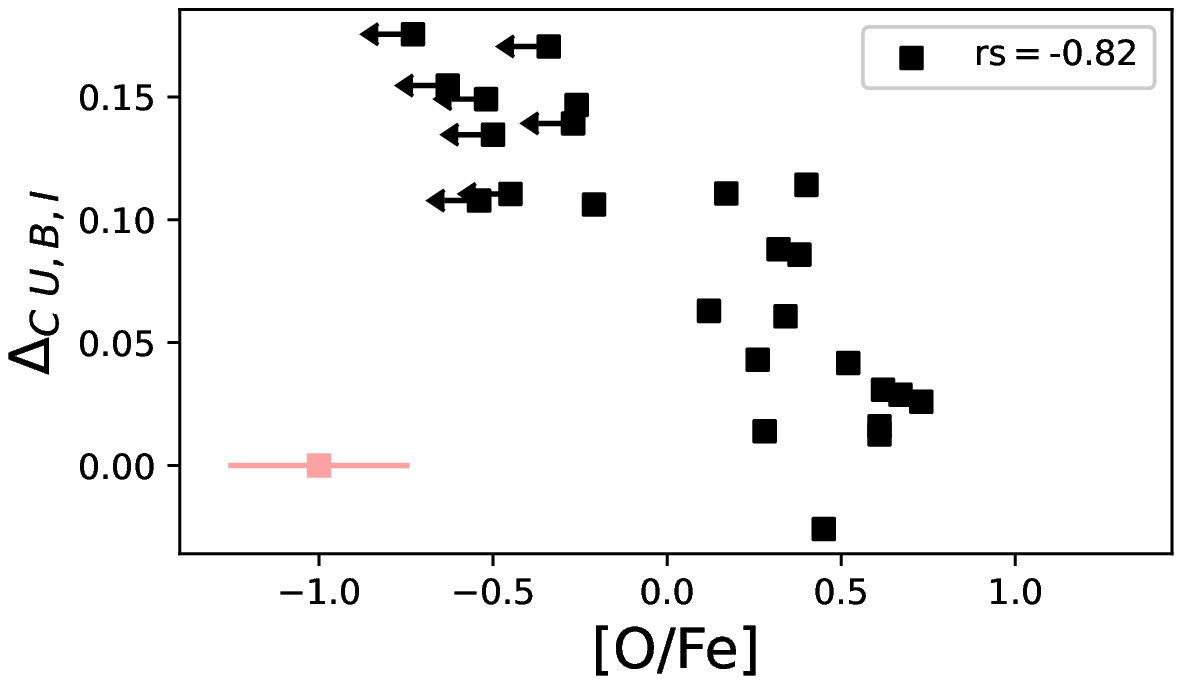} &
        \includegraphics[width=0.5\textwidth]{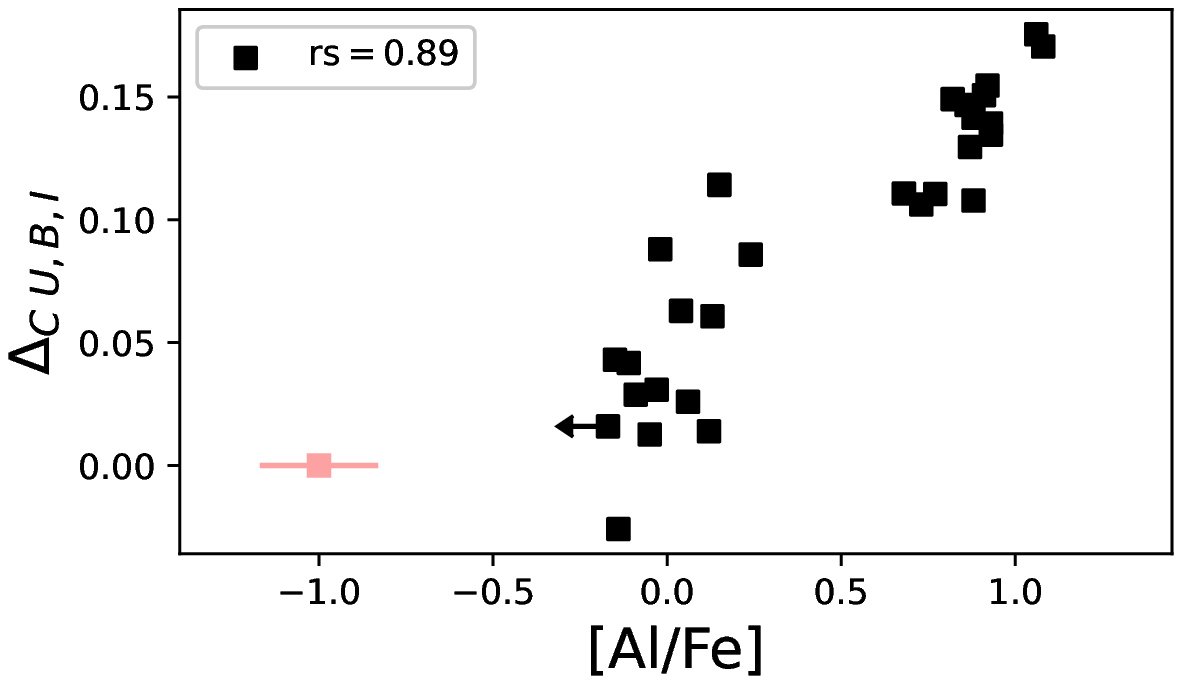} \\
        \includegraphics[width=0.5\textwidth]{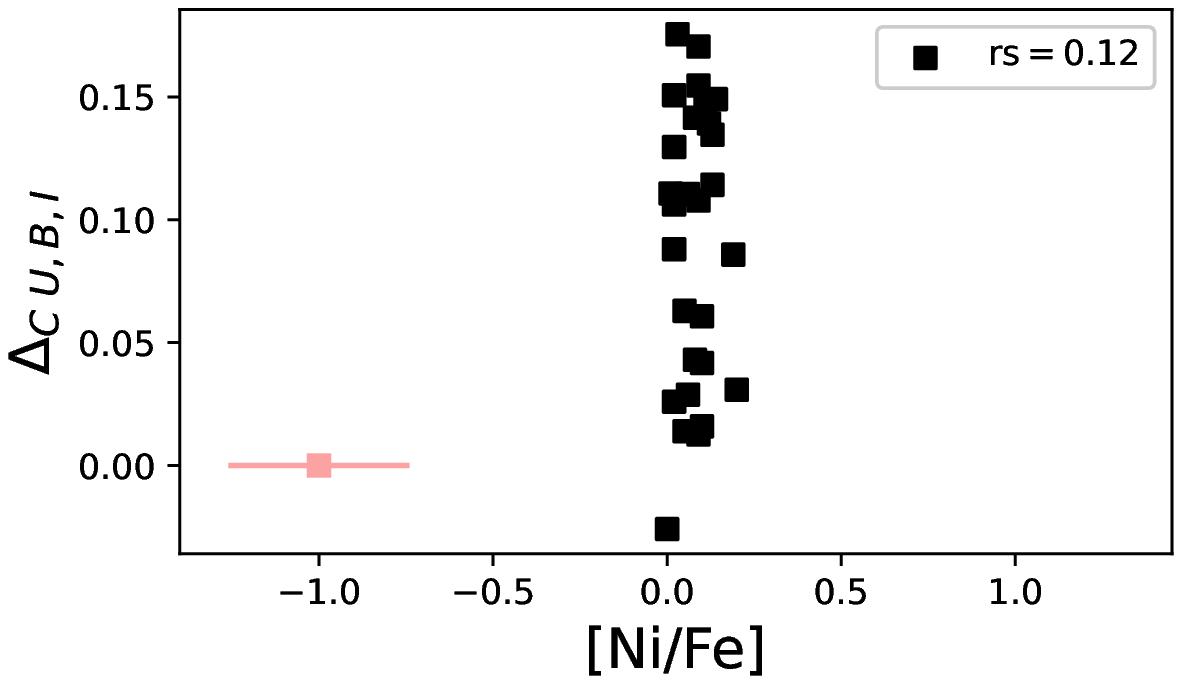} &
        \includegraphics[width=0.5\textwidth]{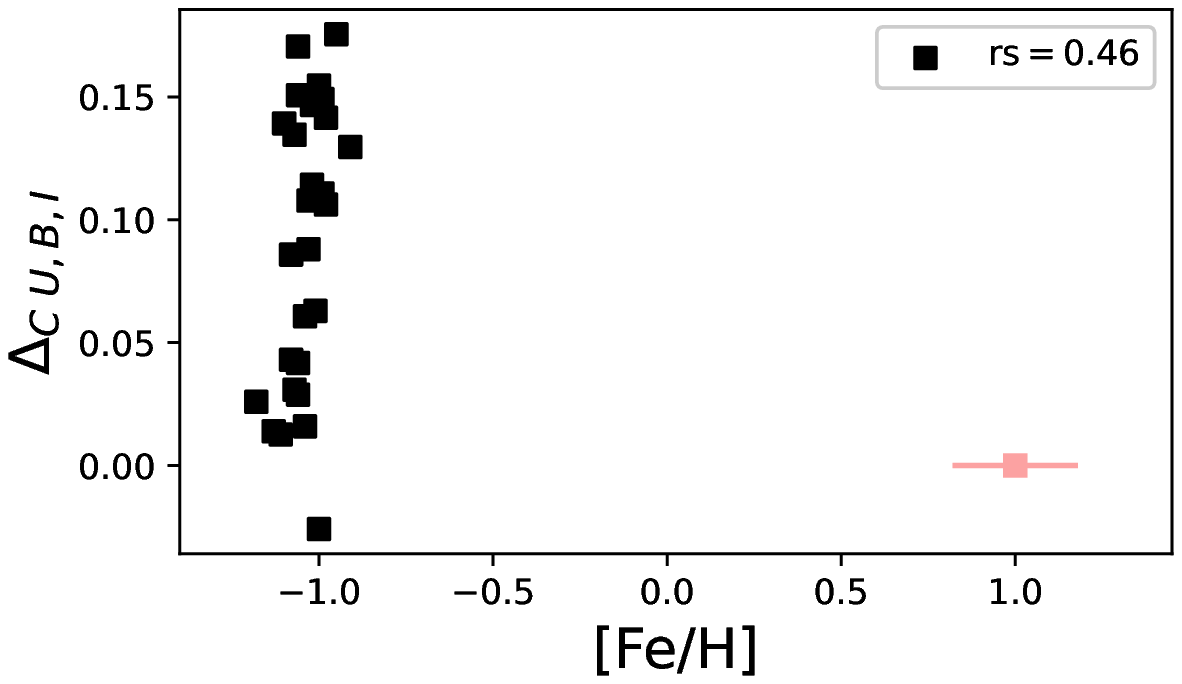} \\
    \end{tabular}
    \caption{\cubi\,versus [C/Fe] (top left panel), [N/Fe] (top right panel), [O/Fe] (middle left panel), [Al/Fe] (middle right panel), [Ni/Fe] (bottom left panel) and [Fe/H] (bottom right panel). The typical errors are marked in red.}
    \label{elements_dy_gb}
\end{figure*}

To further analyse the chemical abundances along the ChM of NGC\,2808, in Figures~\ref{elements_dy_hst} and \ref{elements_dy_gb} we plot the chemical abundances as a function of the \y\ (from $HST$ photometry) and \cubi\ (from ground-based photometry) values, respectively.

Carbon and oxygen both show a strong anti-correlation with \y, with Spearman coefficients $r_S = -0.92$ and $r_S = -0.88$, respectively; which indicates that 1P stars, located at lower values of \y, have higher content of these elements. 
On the other hand, [N/Fe] and [Al/Fe] are positively correlated with \y, with Spearman coefficients equal to $+0.94$ and $+0.93$, meaning that A+B stars have less [N/Fe] and [Al/Fe] in comparison with populations C, D and E stars. Meanwhile, [Ni/Fe] and [Fe/H] do not present any (anti-)correlation with \y\ (Spearman coefficients $r_S = 0.07$ and $r_S = -0.25$, respectively) which indicates that there are no variations of those abundances, within our uncertainties, through the different populations hosted in NGC\,2808.

The ground-based ChM show a similar behaviour.
 Figure~\ref{elements_dy_gb} displays \cubi\, versus [C/Fe], 
[N/Fe], 
[O/Fe],
[Al/Fe], 
[Ni/Fe], 
and [Fe/H]. 
 [C/Fe] and [O/Fe] show strong anti-correlation with \cubi, with their respective Spearman coefficients being $r_S = -0.90$ and $r_S = -0.82$.
[N/Fe] and [Al/Fe] have prominent correlation with \cubi, with $r_S = +0.91$ and $r_S = +0.89$, respectively; demonstrating once more the strong dependency of the light elements with the location on the $y$ axis of the ChM.
Analogous to the result shown in Figure~\ref{elements_dy_hst}, the bottom panels of Figure \ref{elements_dy_gb} show the weak (or no) correlation of [Ni/Fe] and [Fe/H] with \cubi, 
corroborating that there are no significant variations of those abundances through the different populations of NGC\,2808.

The agreement between the relations of each analysed element with the ChM values shown in Figures~\ref{elements_dy_hst} and \ref{elements_dy_gb} for $HST$ and ground-based photometry, respectively,
demonstrates the effectiveness of the ChM constructed from ground based photometry in separating the stellar populations in GCs. 
The combined efforts from both $HST$ and ground based ChMs is an important tool to analyse the multiple population phenomenon in GCs over a wide field of view.

For a complete assessment of chemical variations along the ChMs, we assemble literature high resolution spectroscopic data and combine them with the recent information from our new ground based ChM.
Lithium is a crucial element to understand multiple populations. As it is easily destroyed by proton-capture in stellar environments, the observed abundances for this element depend on the interplay between several mechanisms that can decrease the internal and surface Li content of low-mass stars at different phases of their evolution.
The Li abundances along the ChM of NGC\,2808 has been discussed in \cite{marino/19} by using the ChM from {\it HST} photometry and Li abundances from \cite{dorazi/15}. They found lithium abundances only slightly changing from one population to another, with one population-E star being depleted in lithium by only $\sim$0.2~dex with respect to 
 1P stars.

\begin{figure}
    \centering
            \includegraphics[width=0.5\textwidth]{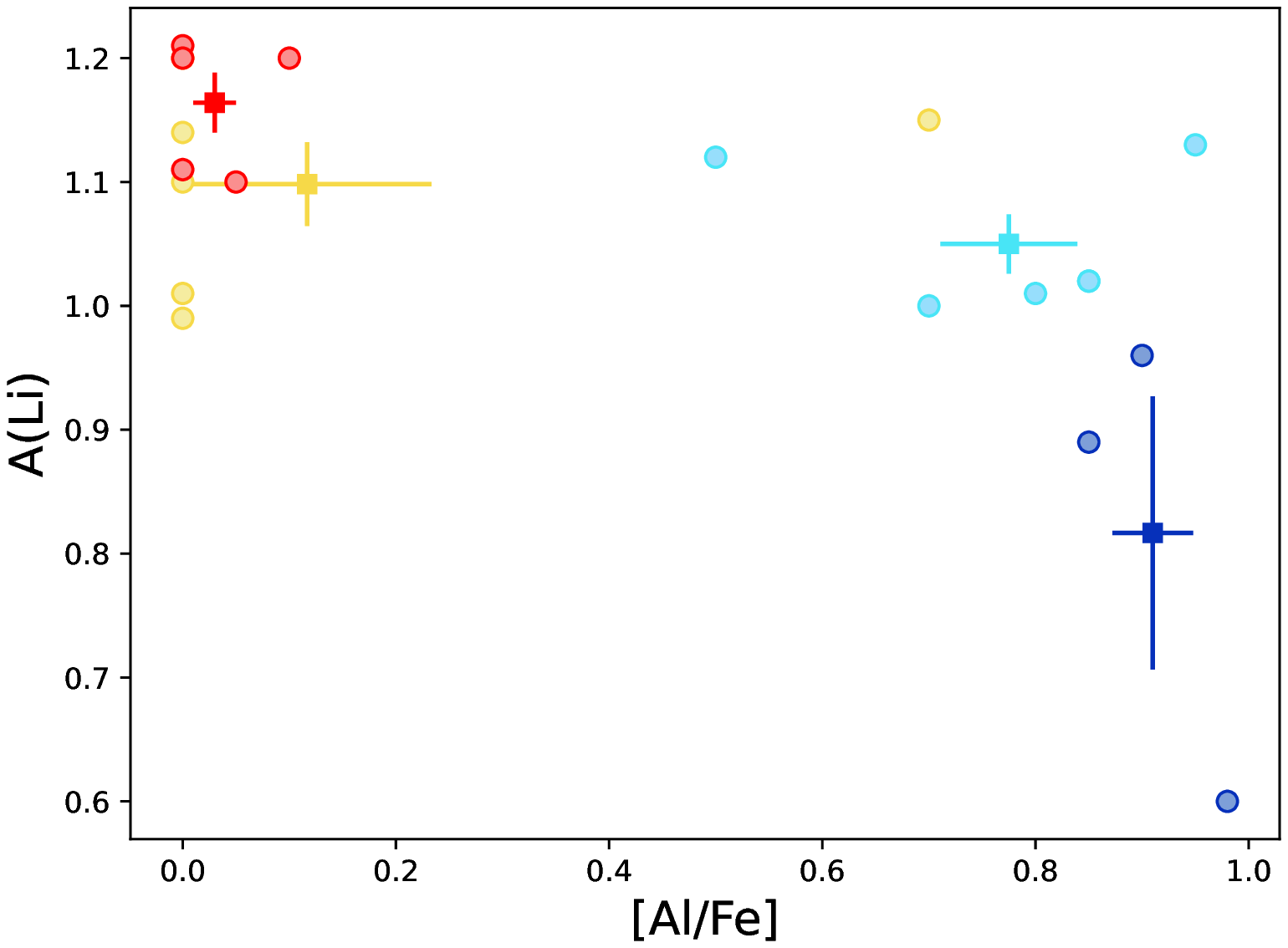}  
    \caption{[Al/Fe] versus NLTE A(Li) from \protect\cite{dorazi/15} with the corresponding population tagging according to our $HST$ and ground based ChMs (circles) compared to their averages abundances (squares). 
    }
    \label{chm_li}
\end{figure}

We take advantage of our newly-introduced ground-based ChM to investigate the Li abundances along this diagram. 
The \cite{dorazi/15} sample includes 12 stars with available ChM position from the large field of view of the ground-based photometry, plus 8 stars from the two $HST$ fields of view.
Since these stars are in a similar evolutionary stage, we are able to compare the [Al/Fe] and A(Li) according to their distinct populations. These stars are below the RGB bump, hence they do not experience yet extra mixing that induce another event of severe depletion in the Li content \cite[e.g.,][]{charbonnel_zahn/07}, the first strong episode of Li dilution occurs during the first dredge-up. 
Figure~\ref{chm_li} shows the A(Li) versus [Al/Fe] plot, using abundances from \cite{dorazi/15}, with the average values for each population highlighted. We observe the majority of 
1P stars showing lower content of Al and slightly higher Li than 2P stars. Most stars of population C have low Al, and Li similar to population D stars.
The three
population E stars\footnote{We note here that the Li abundances of the three E stars are not upper limits.} display high Al abundances and the lowest Li content. This result is in agreement with \cite{marino/19} for stars in the $HST$ ChM.


\begin{figure*}
\centering
\includegraphics[width=1\textwidth]{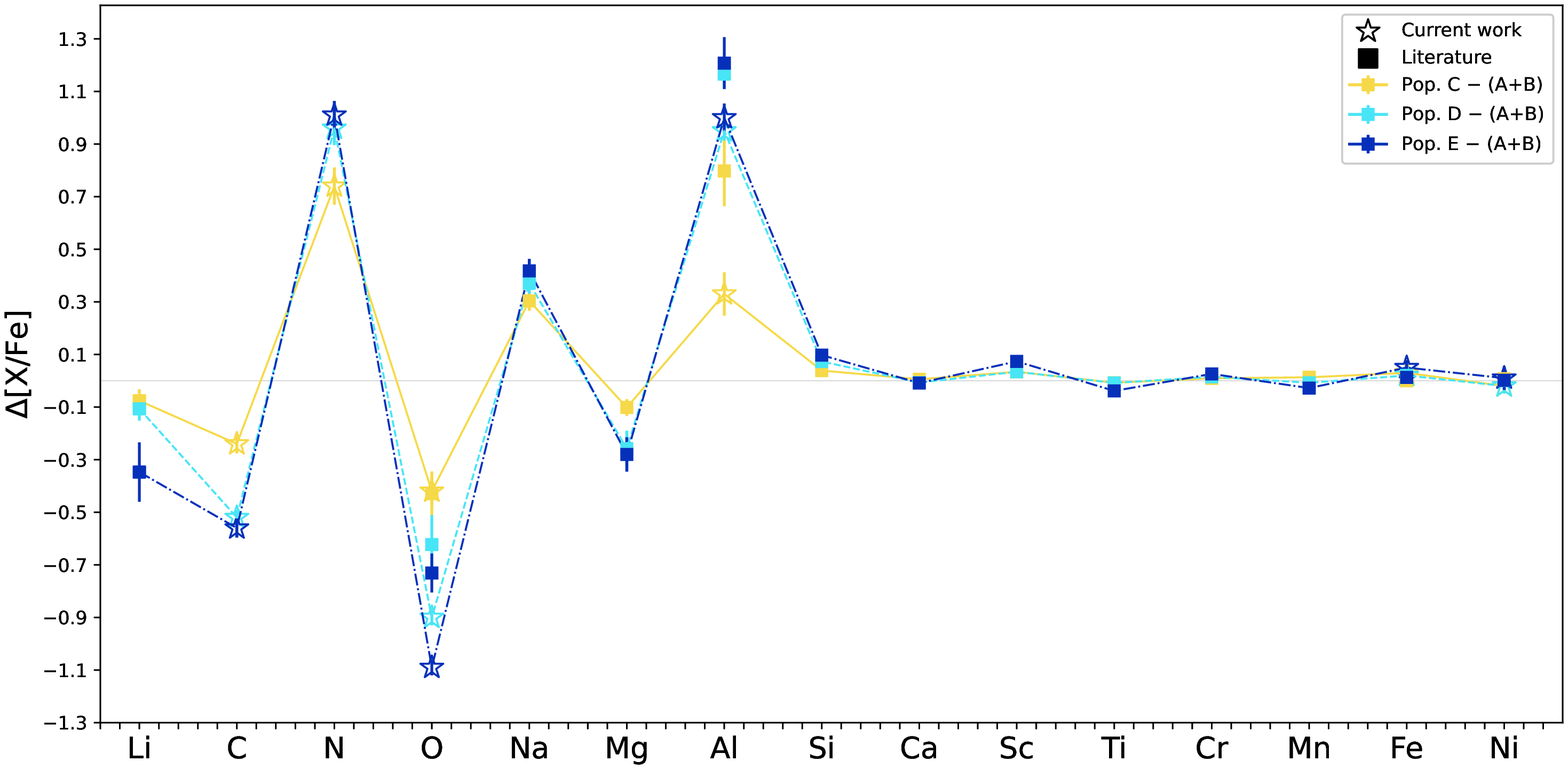}
\caption{Average chemical abundances from stars in populations C (yellow), D (cyan) and E (dark blue) with respect to the 1P (A+B), with our results (open stars) and data from \protect\cite{carretta/15}, \protect\cite{carretta/18} and \protect\cite{dorazi/15} (filled squares). All elements have $\Delta$[X/Fe] reported, with the exception of lithium ($\Delta$A(Li)) and iron ($\Delta$[Fe/H]).}
\label{xfe_literature}
\end{figure*}

 We also compile the abundances from \cite{carretta/15} and \cite{carretta/18} that have ChM information. Table \ref{mean_ab_pops} and Figure \ref{xfe_literature} summarises the average chemical abundances from \cite{carretta/15,carretta/18,dorazi/15}, in addition to our results, separated by the populations marked in the ChM.  

Following the outcome from our spectroscopic analysis, the literature chemical abundances from light elements, O, Na, Mg and Al , are the most sensitive to distinct ChM populations, while Si and Sc have little changes, and Ti and the iron peak elements, Cr, Mn, Fe, and Ni, present negligible variation within the different populations.


Finally, we compare our C, N and O abundances for which we have $HST$ ChM information 
to the [C/Fe], [N/Fe] and [O/Fe] values by \cite{milone/15} inferred from a photometric analysis based on the comparison between the observed diagrams with stellar models and synthetic spectra. The average abundances of each element in a given population are listed in Table~\ref{comps_ab_chm} with respect to population B.  
 In our work, 
  we did not separate populations A and B, 
although one star with $HST$ ChM information being consistent with population A, namely N2808\_1\_8\_0. For the sake of information, we report for this star abundances of: [Fe/H]$=-0.95$ (marginally higher than the average 1P value), [C/Fe]$=-0.25$, [N/Fe]$=-0.05$ and [O/Fe]$=+0.29$. 

The comparison between our average chemical abundances and the ones from \cite{milone/15} shows that:
\begin{itemize}
    \item we have more moderate values of $\Delta\rm{[C/Fe]}$, while \cite{milone/15} presented more extreme values for these quantities in populations D and E, though being compatible for population C;
    \item our results for $\Delta\rm{[N/Fe]}$ are compatible with the work of \cite{milone/15} for populations C, D and E;
    \item differently, our $\Delta\rm{[O/Fe]}$ are substantially larger 
    than the ones from \cite{milone/15}.
\end{itemize}



Results from \cite{milone/15} are entirely based on multi-band photometry. 
The  comparison between the N abundance from spectroscopic- and photometric-based analysis demonstrates that photometry is an excellent tool to assess the content of this element, which, together with helium, is the element that most affects the location of a star on the ChM. Our comparison shows that photometry allows to qualitatively distinguish stellar populations with different abundances of C and O.

\begin{table}
\centering
\caption{[C/Fe], [N/Fe] and [O/Fe] abundances for each population with respect to 1P stars. }
\label{comps_ab_chm}
\begin{tabular}{cccc}
\hline
 & \multicolumn{3}{c}{Our work} \\
Population                            & $\Delta\rm{[C/Fe]}$  & $\Delta\rm{[N/Fe]}$ & $\Delta\rm{[O/Fe]}$ \\
\hline
\hline                            
A+B                           & 0.0       & 0.0       & 0.0 \\
C                           & $-0.21\pm0.03$   & $0.66\pm0.10$   & $-0.45\pm0.06$  \\
D                           & $-0.50\pm0.03$   & $0.96\pm0.07$   & $-0.81\pm0.07$ \\
E                           & $-0.56\pm 0.06$   & $0.98\pm0.08$   & $-1.04\pm0.15$  \\
\hline
 & \multicolumn{3}{c}{\cite{milone/15}} \\
Population                            & $\Delta\rm{[C/Fe]}$  & $\Delta\rm{[N/Fe]}$ & $\Delta\rm{[O/Fe]}$ \\
\hline
\hline                            
A                           & $0.2\pm0.2$       & $0.0\pm0.1$       & $0.2\pm0.1$      \\    
B                           & 0.0       & 0.0      & 0.0        \\
C                           & $-0.1\pm0.1$       & $0.6\pm0.2$       & $0.0\pm0.1$        \\
D                           & $-0.8\pm0.3$       & $1.1\pm0.1$       & $-0.5\pm0.1$       \\
E                           & $-0.9\pm0.3$       & $1.2\pm0.2$       & $-0.7\pm0.1$      \\
\hline
\end{tabular}
\end{table}

\section{summary and conclusions}

This work investigates multiple stellar populations along the RGB and the AGB of NGC\,2808 by combining information from different techniques.
 We use the multi-band {\it HST} photometry and proper motions for stars in the $\sim 2.7 \times 2.7$ arcmin central field  \citep[e.g.][]{milone/15} and derived similar  high-precision photometry and proper motions of stars in an external field, $\sim$5.5 arcmin south-west from the cluster centre. 
 Moreover, we used $U,B,I$ photometry from ground-based facilities \citep[][]{stetson/19} and stellar proper motions from Gaia eDR3 \citep[][]{gaia_col_edr3/21} and GIRAFFE@VLT spectra. While stellar proper motions were used to identify probable cluster members, photometry was instrumental for two main results:
 
\begin{itemize}
    \item {\it HST} photometry allowed us to derive the classical $\Delta_{C \rm F275W,F336W,F438W}$ vs.\,$\Delta_{\rm F275W,F814W}$ ChM of RGB stars and detect the five main stellar populations of NGC\,2808 (A--E) for two different fields of view.
    

\item We also present a ChM constructed from ground based photometry alone. This new ChM, which is derived from  the $I$ vs.\,$B-I$ CMD and $I$ vs.\,$\Delta C_{\rm U,B,I}$ pseudo CMD of RGB stars, allows to identify multiple stellar populations along the RGB.
 The fact that accurate $U,B,I$ photometry is now available from wide-field telescopes opens the possibility to investigate multiple stellar populations over a wider field of view.
 \end{itemize}

To constrain the chemical composition of the stellar populations of NGC\,2808 we investigated 77 giant stars, including 70 RGB and 7 AGB stars, by using high resolution spectra from FLAMES/GIRAFFE at the Very Large Telescope. 
 We determined their stellar parameters and abundances for six elements: C, N, O, Al, Ni and Fe. Results on chemical abundances coupled with the {\it population assignment} based on the ChM can be summarised as follows:
 \begin{itemize}
     \item We detected large star-to-star variations in C, N, O, and Al. The maximum internal elemental variation range from $\sim$0.7~dex in carbon, to $\sim$1.1~dex in nitrogen, and aluminium, and up more than 1.3~dex in oxygen. 
     The abundances of these elements show the well-known (anti-)correlations, expected for these elements being involved in the high-temperature H-burning.
     These abundances are strongly dependent on the location of the stars on the ChM. On the other hand, iron and nickel remain essentially constant through the different stellar populations. 
     

     \item Hence, we found a remarkable correspondence between the abundances of RGB stars and their position on the ChMs. The [C/Fe] and [O/Fe] abundances increase with $\Delta_{C \rm F275W,F336W,F438W}$ and $\Delta_{C \rm U,B,I}$ while [N/Fe] and  [Al/Fe] anticorrelate with $\Delta_{C \rm F275W,F336W,F438W}$ and $\Delta_{C \rm U,B,I}$.

     \item We derived the chemical composition of stars in the main populations of NGC\,2808. The 1P stars (populations A$+$B) have lower content of aluminium and nitrogen in contrast with higher abundances of oxygen and carbon. While aluminium and nitrogen become gradually higher for populations C and D reaching their highest values in population E, oxygen and carbon have systematically lower abundances compared to population C to E. The two most He-enriched populations, namely D and E, do not show significant differences in light elements, except for O and Li, which is more depleted in the E stars. 

     \item The agreement between the relative variations in the chemical composition among the different populations, as obtained from the {\it classical} and ground-based ChM, strongly suggests that the latter will be a powerful tool for characterizing GC stellar populations in larger fields of view. 
     
 \end{itemize}
 
In addition, we assemble high resolution spectroscopy data from \cite{carretta/15}, \cite{carretta/18} and \cite{dorazi/15}  with the new information from our ground based ChM and observe the following:
 
 \begin{itemize}
     \item The light elements Li, O, Na, Mg and Al show the larger variations from populations C, D and E in respect to population A+B, in comparison with other elements analysed;
     \item Si and Sc show little changes  from populations C, D and E compared to 1P (populations A+B);
     \item while Ti and the iron peak elements Cr, Mn, Fe, Ni show negligible or no variation between the distinct populations.
 \end{itemize} 

    




The analysis of the chemical composition of seven AGB stars shows that 
they also span a range of [O/Fe] and [Al/Fe] that is comparable with that of the RGB stars. 
 AGB stars exhibit correlations between N and Al and C and O as well as the C-N and O-Al anticorrelations, in close analogy with what is observed for the RGB.
 The main difference is that the 1P stars that populate the AGB exhibit higher nitrogen abundances and lower carbon content than the RGB first population.

Intriguingly, we detect one AGB star (N2808\_2\_9\_wf) that is strongly depleted in oxygen and highly enhanced in aluminium. Based on the abundances of these elements, this AGB star would be associated to the helium rich stellar populations of NGC\,2808 (Y$\gtrsim$0.32). 
 Our discovery, together with previous spectroscopic evidence of an AGB star associated to the population D \citep[][]{marino/17} demonstrates that stars with high helium abundances can evolve into AGB. A similar conclusion comes from the ChMs of AGB stars, which reveals distinct populations of AGB stars that are associated to population D and possibly population E \citep[][]{marino/17, lagioia/21}.
These findings seem in contrast with the predictions of
evolutionary models of the helium-rich stars, which should skip the AGB phase \citep[e.g.][]{chantereau/16}.







\section*{acknowledgements}

This work has received funding from the European Research Council (ERC) under the European Union's Horizon 2020 research innovation programme (Grant Agreement ERC-StG 2016, No 716082 'GALFOR', PI: Milone, http://progetti.dfa.unipd.it/GALFOR).
 APM acknowledges support from MIUR through the FARE project R164RM93XW SEMPLICE (PI: Milone). MC, APM, and ED  have been supported by MIUR under PRIN program 2017Z2HSMF (PI: Bedin). AMA gratefully acknowledges support from the Swedish Research Council (VR 2020-03940).

%

\vspace{5mm}



\section*{Data availability}

The data underlying this article are available in the article and in its online supplementary material.



\bibliographystyle{mnras}
\bibliography{references} 

\begin{thebibliography}{}
\makeatletter
\relax
\def\mn@urlcharsother{\let\do\@makeother \do\$\do\&\do\#\do\^\do\_\do\%\do\~}
\def\mn@doi{\begingroup\mn@urlcharsother \@ifnextchar [ {\mn@doi@}
  {\mn@doi@[]}}
\def\mn@doi@[#1]#2{\def\@tempa{#1}\ifx\@tempa\@empty \href
  {http://dx.doi.org/#2} {doi:#2}\else \href {http://dx.doi.org/#2} {#1}\fi
  \endgroup}
\def\mn@eprint#1#2{\mn@eprint@#1:#2::\@nil}
\def\mn@eprint@arXiv#1{\href {http://arxiv.org/abs/#1} {{\tt arXiv:#1}}}
\def\mn@eprint@dblp#1{\href {http://dblp.uni-trier.de/rec/bibtex/#1.xml}
  {dblp:#1}}
\def\mn@eprint@#1:#2:#3:#4\@nil{\def\@tempa {#1}\def\@tempb {#2}\def\@tempc
  {#3}\ifx \@tempc \@empty \let \@tempc \@tempb \let \@tempb \@tempa \fi \ifx
  \@tempb \@empty \def\@tempb {arXiv}\fi \@ifundefined
  {mn@eprint@\@tempb}{\@tempb:\@tempc}{\expandafter \expandafter \csname
  mn@eprint@\@tempb\endcsname \expandafter{\@tempc}}}

\bibitem[\protect\citeauthoryear{{Alonso}, {Arribas}  \&
  {Mart{\'\i}nez-Roger}}{{Alonso} et~al.}{1999}]{alonso/99}
{Alonso} A.,  {Arribas} S.,   {Mart{\'\i}nez-Roger} C.,  1999, \mn@doi [\aaps]
  {10.1051/aas:1999521}, \href
  {https://ui.adsabs.harvard.edu/abs/1999A&AS..140..261A} {140, 261}

\bibitem[\protect\citeauthoryear{{Amarsi}, {Barklem}, {Asplund}, {Collet}  \&
  {Zatsarinny}}{{Amarsi} et~al.}{2018}]{amarsi/18}
{Amarsi} A.~M.,  {Barklem} P.~S.,  {Asplund} M.,  {Collet} R.,   {Zatsarinny}
  O.,  2018, \mn@doi [\aap] {10.1051/0004-6361/201832770}, \href
  {https://ui.adsabs.harvard.edu/abs/2018A&A...616A..89A} {616, A89}

\bibitem[\protect\citeauthoryear{{Amarsi}, {Nissen}  \&
  {Sk{\'u}lad{\'o}ttir}}{{Amarsi} et~al.}{2019}]{amarsi/19}
{Amarsi} A.~M.,  {Nissen} P.~E.,   {Sk{\'u}lad{\'o}ttir} {\'A}.,  2019, \mn@doi
  [\aap] {10.1051/0004-6361/201936265}, \href
  {https://ui.adsabs.harvard.edu/abs/2019A&A...630A.104A} {630, A104}

\bibitem[\protect\citeauthoryear{{Anderson}, {Bedin}, {Piotto}, {Yadav}  \&
  {Bellini}}{{Anderson} et~al.}{2006}]{anderson/06}
{Anderson} J.,  {Bedin} L.~R.,  {Piotto} G.,  {Yadav} R.~S.,   {Bellini} A.,
  2006, \mn@doi [\aap] {10.1051/0004-6361:20065004}, \href
  {https://ui.adsabs.harvard.edu/abs/2006A&A...454.1029A} {454, 1029}

\bibitem[\protect\citeauthoryear{{Arnould}, {Goriely}  \& {Jorissen}}{{Arnould}
  et~al.}{1999}]{arnould/99}
{Arnould} M.,  {Goriely} S.,   {Jorissen} A.,  1999, \aap, \href
  {https://ui.adsabs.harvard.edu/abs/1999A&A...347..572A} {347, 572}

\bibitem[\protect\citeauthoryear{{Asplund}, {Grevesse}, {Sauval}  \&
  {Scott}}{{Asplund} et~al.}{2009}]{asplund/09}
{Asplund} M.,  {Grevesse} N.,  {Sauval} A.~J.,   {Scott} P.,  2009, \mn@doi
  [\araa] {10.1146/annurev.astro.46.060407.145222}, \href
  {https://ui.adsabs.harvard.edu/abs/2009ARA&A..47..481A} {47, 481}

\bibitem[\protect\citeauthoryear{{Ballester}, {Modigliani}, {Boitquin},
  {Cristiani}, {Hanuschik}, {Kaufer}  \& {Wolf}}{{Ballester}
  et~al.}{2000}]{ballester/00}
{Ballester} P.,  {Modigliani} A.,  {Boitquin} O.,  {Cristiani} S.,  {Hanuschik}
  R.,  {Kaufer} A.,   {Wolf} S.,  2000, The Messenger, \href
  {https://ui.adsabs.harvard.edu/abs/2000Msngr.101...31B} {101, 31}

\bibitem[\protect\citeauthoryear{{Bastian} \& {Lardo}}{{Bastian} \&
  {Lardo}}{2018}]{bastian/2018}
{Bastian} N.,  {Lardo} C.,  2018, \mn@doi [\araa]
  {10.1146/annurev-astro-081817-051839}, \href
  {https://ui.adsabs.harvard.edu/abs/2018ARA&A..56...83B} {56, 83}

\bibitem[\protect\citeauthoryear{{Baumgardt} \& {Hilker}}{{Baumgardt} \&
  {Hilker}}{2018}]{baumgardt/18}
{Baumgardt} H.,  {Hilker} M.,  2018, \mn@doi [\mnras] {10.1093/mnras/sty1057},
  \href {https://ui.adsabs.harvard.edu/abs/2018MNRAS.478.1520B} {478, 1520}

\bibitem[\protect\citeauthoryear{{Bessell}}{{Bessell}}{1979}]{bessell/79}
{Bessell} M.~S.,  1979, \mn@doi [\pasp] {10.1086/130542}, \href
  {https://ui.adsabs.harvard.edu/abs/1979PASP...91..589B} {91, 589}

\bibitem[\protect\citeauthoryear{{Campbell} et~al.,}{{Campbell}
  et~al.}{2013}]{campbell/13}
{Campbell} S.~W.,  et~al., 2013, \mn@doi [\nat] {10.1038/nature12191}, \href
  {https://ui.adsabs.harvard.edu/abs/2013Natur.498..198C} {498, 198}

\bibitem[\protect\citeauthoryear{{Carretta}}{{Carretta}}{2015}]{carretta/15}
{Carretta} E.,  2015, \mn@doi [\apj] {10.1088/0004-637X/810/2/148}, \href
  {https://ui.adsabs.harvard.edu/abs/2015ApJ...810..148C} {810, 148}

\bibitem[\protect\citeauthoryear{{Carretta}, {Bragaglia}, {Gratton}, {Leone},
  {Recio-Blanco}  \& {Lucatello}}{{Carretta} et~al.}{2006}]{carretta/06}
{Carretta} E.,  {Bragaglia} A.,  {Gratton} R.~G.,  {Leone} F.,  {Recio-Blanco}
  A.,   {Lucatello} S.,  2006, \mn@doi [\aap] {10.1051/0004-6361:20054369},
  \href {https://ui.adsabs.harvard.edu/abs/2006A&A...450..523C} {450, 523}

\bibitem[\protect\citeauthoryear{{Carretta}, {D'Orazi}, {Gratton}  \&
  {Lucatello}}{{Carretta} et~al.}{2012}]{carretta/12}
{Carretta} E.,  {D'Orazi} V.,  {Gratton} R.~G.,   {Lucatello} S.,  2012,
  \mn@doi [\aap] {10.1051/0004-6361/201219277}, \href
  {https://ui.adsabs.harvard.edu/abs/2012A&A...543A.117C} {543, A117}

\bibitem[\protect\citeauthoryear{{Carretta}, {Bragaglia}, {Lucatello},
  {Gratton}, {D'Orazi}  \& {Sollima}}{{Carretta} et~al.}{2018}]{carretta/18}
{Carretta} E.,  {Bragaglia} A.,  {Lucatello} S.,  {Gratton} R.~G.,  {D'Orazi}
  V.,   {Sollima} A.,  2018, \mn@doi [\aap] {10.1051/0004-6361/201732324},
  \href {https://ui.adsabs.harvard.edu/abs/2018A&A...615A..17C} {615, A17}

\bibitem[\protect\citeauthoryear{{Castelli} \& {Kurucz}}{{Castelli} \&
  {Kurucz}}{2004}]{castelli/04}
{Castelli} F.,  {Kurucz} R.~L.,  2004, ArXiv Astrophysics e-prints, \href
  {http://adsabs.harvard.edu/abs/2004astro.ph..5087C} {}

\bibitem[\protect\citeauthoryear{{Chantereau}, {Charbonnel}  \&
  {Meynet}}{{Chantereau} et~al.}{2016}]{chantereau/16}
{Chantereau} W.,  {Charbonnel} C.,   {Meynet} G.,  2016, \mn@doi [\aap]
  {10.1051/0004-6361/201628418}, \href
  {https://ui.adsabs.harvard.edu/abs/2016A&A...592A.111C} {592, A111}

\bibitem[\protect\citeauthoryear{{Charbonnel} \& {Zahn}}{{Charbonnel} \&
  {Zahn}}{2007}]{charbonnel_zahn/07}
{Charbonnel} C.,  {Zahn} J.~P.,  2007, \mn@doi [\aap]
  {10.1051/0004-6361:20077274}, \href
  {https://ui.adsabs.harvard.edu/abs/2007A&A...467L..15C} {467, L15}

\bibitem[\protect\citeauthoryear{{Code}}{{Code}}{1969}]{code/69}
{Code} A.~D.,  1969, \mn@doi [\pasp] {10.1086/128809}, \href
  {https://ui.adsabs.harvard.edu/abs/1969PASP...81..475C} {81, 475}

\bibitem[\protect\citeauthoryear{{Cordoni}, {Milone}, {Marino}, {Di
  Criscienzo}, {D'Antona}, {Dotter}, {Lagioia}  \& {Tailo}}{{Cordoni}
  et~al.}{2018}]{cordoni/18}
{Cordoni} G.,  {Milone} A.~P.,  {Marino} A.~F.,  {Di Criscienzo} M.,
  {D'Antona} F.,  {Dotter} A.,  {Lagioia} E.~P.,   {Tailo} M.,  2018, \mn@doi
  [\apj] {10.3847/1538-4357/aaedc1}, \href
  {https://ui.adsabs.harvard.edu/abs/2018ApJ...869..139C} {869, 139}

\bibitem[\protect\citeauthoryear{{D'Antona}, {Bellazzini}, {Caloi}, {Pecci},
  {Galleti}  \& {Rood}}{{D'Antona} et~al.}{2005}]{dantona/05}
{D'Antona} F.,  {Bellazzini} M.,  {Caloi} V.,  {Pecci} F.~F.,  {Galleti} S.,
  {Rood} R.~T.,  2005, \mn@doi [\apj] {10.1086/431968}, \href
  {https://ui.adsabs.harvard.edu/abs/2005ApJ...631..868D} {631, 868}

\bibitem[\protect\citeauthoryear{{D'Orazi} et~al.,}{{D'Orazi}
  et~al.}{2015}]{dorazi/15}
{D'Orazi} V.,  et~al., 2015, \mn@doi [\mnras] {10.1093/mnras/stv612}, \href
  {https://ui.adsabs.harvard.edu/abs/2015MNRAS.449.4038D} {449, 4038}

\bibitem[\protect\citeauthoryear{{Dotter}, {Chaboyer}, {Jevremovi{\'c}},
  {Kostov}, {Baron}  \& {Ferguson}}{{Dotter} et~al.}{2008}]{dotter/08}
{Dotter} A.,  {Chaboyer} B.,  {Jevremovi{\'c}} D.,  {Kostov} V.,  {Baron} E.,
  {Ferguson} J.~W.,  2008, \mn@doi [\apjs] {10.1086/589654}, \href
  {https://ui.adsabs.harvard.edu/abs/2008ApJS..178...89D} {178, 89}

\bibitem[\protect\citeauthoryear{{Gaia Collaboration} et~al.,}{{Gaia
  Collaboration} et~al.}{2021}]{gaia_col_edr3/21}
{Gaia Collaboration} et~al., 2021, \mn@doi [\aap]
  {10.1051/0004-6361/202039657}, \href
  {https://ui.adsabs.harvard.edu/abs/2021A&A...649A...1G} {649, A1}

\bibitem[\protect\citeauthoryear{{Gratton}, {Carretta}, {Claudi}, {Lucatello}
  \& {Barbieri}}{{Gratton} et~al.}{2003}]{gratton/03a}
{Gratton} R.~G.,  {Carretta} E.,  {Claudi} R.,  {Lucatello} S.,   {Barbieri}
  M.,  2003, \mn@doi [\aap] {10.1051/0004-6361:20030439}, \href
  {https://ui.adsabs.harvard.edu/abs/2003A&A...404..187G} {404, 187}

\bibitem[\protect\citeauthoryear{{Gratton}, {Sneden}  \& {Carretta}}{{Gratton}
  et~al.}{2004}]{gratton/04}
{Gratton} R.,  {Sneden} C.,   {Carretta} E.,  2004, \mn@doi [\araa]
  {10.1146/annurev.astro.42.053102.133945}, \href
  {https://ui.adsabs.harvard.edu/abs/2004ARA&A..42..385G} {42, 385}

\bibitem[\protect\citeauthoryear{{Gratton}, {D'Orazi}, {Bragaglia}, {Carretta}
  \& {Lucatello}}{{Gratton} et~al.}{2010}]{gratton/10}
{Gratton} R.~G.,  {D'Orazi} V.,  {Bragaglia} A.,  {Carretta} E.,   {Lucatello}
  S.,  2010, \mn@doi [\aap] {10.1051/0004-6361/201015405}, \href
  {https://ui.adsabs.harvard.edu/abs/2010A&A...522A..77G} {522, A77}

\bibitem[\protect\citeauthoryear{{Gratton}, {Lucatello}, {Carretta},
  {Bragaglia}, {D'Orazi}  \& {Momany}}{{Gratton} et~al.}{2011}]{gratton/11}
{Gratton} R.~G.,  {Lucatello} S.,  {Carretta} E.,  {Bragaglia} A.,  {D'Orazi}
  V.,   {Momany} Y.~A.,  2011, \mn@doi [\aap] {10.1051/0004-6361/201117690},
  \href {https://ui.adsabs.harvard.edu/abs/2011A&A...534A.123G} {534, A123}

\bibitem[\protect\citeauthoryear{{Gratton}, {Bragaglia}, {Carretta}, {D'Orazi},
  {Lucatello}  \& {Sollima}}{{Gratton} et~al.}{2019}]{gratton/2019}
{Gratton} R.,  {Bragaglia} A.,  {Carretta} E.,  {D'Orazi} V.,  {Lucatello} S.,
   {Sollima} A.,  2019, \mn@doi [\aapr] {10.1007/s00159-019-0119-3}, \href
  {https://ui.adsabs.harvard.edu/abs/2019A&ARv..27....8G} {27, 8}

\bibitem[\protect\citeauthoryear{{Greggio} \& {Renzini}}{{Greggio} \&
  {Renzini}}{1990}]{greggio/90}
{Greggio} L.,  {Renzini} A.,  1990, \mn@doi [\apj] {10.1086/169384}, \href
  {https://ui.adsabs.harvard.edu/abs/1990ApJ...364...35G} {364, 35}

\bibitem[\protect\citeauthoryear{{Harris}}{{Harris}}{1996}]{harris/96}
{Harris} W.~E.,  1996, \mn@doi [\aj] {10.1086/118116}, \href
  {https://ui.adsabs.harvard.edu/abs/1996AJ....112.1487H} {112, 1487}

\bibitem[\protect\citeauthoryear{{Heiter} et~al.,}{{Heiter}
  et~al.}{2021}]{heiter/21}
{Heiter} U.,  et~al., 2021, \mn@doi [\aap] {10.1051/0004-6361/201936291}, \href
  {https://ui.adsabs.harvard.edu/abs/2021A&A...645A.106H} {645, A106}

\bibitem[\protect\citeauthoryear{{Hong}, {Lim}, {Chung}, {Kim}, {Han}  \&
  {Lee}}{{Hong} et~al.}{2021}]{hong/21}
{Hong} S.,  {Lim} D.,  {Chung} C.,  {Kim} J.,  {Han} S.-I.,   {Lee} Y.-W.,
  2021, \mn@doi [\aj] {10.3847/1538-3881/ac0ce6}, \href
  {https://ui.adsabs.harvard.edu/abs/2021AJ....162..130H} {162, 130}

\bibitem[\protect\citeauthoryear{{Jang} et~al.,}{{Jang} et~al.}{2022}]{jang/22}
{Jang} S.,  et~al., 2022, \mn@doi [\mnras] {10.1093/mnras/stac3086}, \href
  {https://ui.adsabs.harvard.edu/abs/2022MNRAS.517.5687J} {517, 5687}

\bibitem[\protect\citeauthoryear{{Karakas} \& {Lattanzio}}{{Karakas} \&
  {Lattanzio}}{2003}]{karakas_latanzio/03}
{Karakas} A.~I.,  {Lattanzio} J.~C.,  2003, \mn@doi [\pasa] {10.1071/AS03010},
  \href {https://ui.adsabs.harvard.edu/abs/2003PASA...20..279K} {20, 279}

\bibitem[\protect\citeauthoryear{{Lagioia}, {Milone}, {Marino}, {Cordoni}  \&
  {Tailo}}{{Lagioia} et~al.}{2019}]{lagioia/2019}
{Lagioia} E.~P.,  {Milone} A.~P.,  {Marino} A.~F.,  {Cordoni} G.,   {Tailo} M.,
   2019, \mn@doi [\aj] {10.3847/1538-3881/ab45f2}, \href
  {https://ui.adsabs.harvard.edu/abs/2019AJ....158..202L} {158, 202}

\bibitem[\protect\citeauthoryear{{Lagioia} et~al.,}{{Lagioia}
  et~al.}{2021}]{lagioia/21}
{Lagioia} E.~P.,  et~al., 2021, \mn@doi [\apj] {10.3847/1538-4357/abdfcf},
  \href {https://ui.adsabs.harvard.edu/abs/2021ApJ...910....6L} {910, 6}

\bibitem[\protect\citeauthoryear{{Latour} et~al.,}{{Latour}
  et~al.}{2019}]{latour/19}
{Latour} M.,  et~al., 2019, \mn@doi [\aap] {10.1051/0004-6361/201936242}, \href
  {https://ui.adsabs.harvard.edu/abs/2019A&A...631A..14L} {631, A14}

\bibitem[\protect\citeauthoryear{{Legnardi} et~al.,}{{Legnardi}
  et~al.}{2022}]{legnardi/22}
{Legnardi} M.~V.,  et~al., 2022, \mn@doi [\mnras] {10.1093/mnras/stac734},
  \href {https://ui.adsabs.harvard.edu/abs/2022MNRAS.513..735L} {513, 735}

\bibitem[\protect\citeauthoryear{{Marino}, {Villanova}, {Piotto}, {Milone},
  {Momany}, {Bedin}  \& {Medling}}{{Marino} et~al.}{2008}]{marino/08}
{Marino} A.~F.,  {Villanova} S.,  {Piotto} G.,  {Milone} A.~P.,  {Momany} Y.,
  {Bedin} L.~R.,   {Medling} A.~M.,  2008, \mn@doi [\aap]
  {10.1051/0004-6361:200810389}, \href
  {https://ui.adsabs.harvard.edu/abs/2008A&A...490..625M} {490, 625}

\bibitem[\protect\citeauthoryear{{Marino} et~al.,}{{Marino}
  et~al.}{2014}]{marino/14}
{Marino} A.~F.,  et~al., 2014, \mn@doi [\mnras] {10.1093/mnras/stt1993}, \href
  {https://ui.adsabs.harvard.edu/abs/2014MNRAS.437.1609M} {437, 1609}

\bibitem[\protect\citeauthoryear{{Marino} et~al.,}{{Marino}
  et~al.}{2017}]{marino/17}
{Marino} A.~F.,  et~al., 2017, \mn@doi [\apj] {10.3847/1538-4357/aa7852}, \href
  {https://ui.adsabs.harvard.edu/abs/2017ApJ...843...66M} {843, 66}

\bibitem[\protect\citeauthoryear{{Marino} et~al.,}{{Marino}
  et~al.}{2019a}]{marino/19}
{Marino} A.~F.,  et~al., 2019a, \mn@doi [\mnras] {10.1093/mnras/stz1415}, \href
  {https://ui.adsabs.harvard.edu/abs/2019MNRAS.487.3815M} {487, 3815}

\bibitem[\protect\citeauthoryear{{Marino} et~al.,}{{Marino}
  et~al.}{2019b}]{marino/2019b}
{Marino} A.~F.,  et~al., 2019b, \mn@doi [\apj] {10.3847/1538-4357/ab53d9},
  \href {https://ui.adsabs.harvard.edu/abs/2019ApJ...887...91M} {887, 91}

\bibitem[\protect\citeauthoryear{{M{\'e}sz{\'a}ros} et~al.,}{{M{\'e}sz{\'a}ros}
  et~al.}{2020}]{meszaros/20}
{M{\'e}sz{\'a}ros} S.,  et~al., 2020, \mn@doi [\mnras] {10.1093/mnras/stz3496},
  \href {https://ui.adsabs.harvard.edu/abs/2020MNRAS.492.1641M} {492, 1641}

\bibitem[\protect\citeauthoryear{{Milone} \& {Marino}}{{Milone} \&
  {Marino}}{2022}]{milone/2022}
{Milone} A.~P.,  {Marino} A.~F.,  2022, \mn@doi [Universe]
  {10.3390/universe8070359}, \href
  {https://ui.adsabs.harvard.edu/abs/2022Univ....8..359M} {8, 359}

\bibitem[\protect\citeauthoryear{{Milone} et~al.,}{{Milone}
  et~al.}{2012}]{milone/2012}
{Milone} A.~P.,  et~al., 2012, \mn@doi [\aap] {10.1051/0004-6361/201016384},
  \href {https://ui.adsabs.harvard.edu/abs/2012A&A...540A..16M} {540, A16}

\bibitem[\protect\citeauthoryear{{Milone} et~al.,}{{Milone}
  et~al.}{2015}]{milone/15}
{Milone} A.~P.,  et~al., 2015, \mn@doi [\apj] {10.1088/0004-637X/808/1/51},
  \href {https://ui.adsabs.harvard.edu/abs/2015ApJ...808...51M} {808, 51}

\bibitem[\protect\citeauthoryear{{Milone} et~al.,}{{Milone}
  et~al.}{2017}]{milone/17}
{Milone} A.~P.,  et~al., 2017, \mn@doi [\mnras] {10.1093/mnras/stw2531}, \href
  {https://ui.adsabs.harvard.edu/abs/2017MNRAS.464.3636M} {464, 3636}

\bibitem[\protect\citeauthoryear{{Milone} et~al.,}{{Milone}
  et~al.}{2018}]{milone/18}
{Milone} A.~P.,  et~al., 2018, \mn@doi [\mnras] {10.1093/mnras/sty2573}, \href
  {https://ui.adsabs.harvard.edu/abs/2018MNRAS.481.5098M} {481, 5098}

\bibitem[\protect\citeauthoryear{{Nordlander} \& {Lind}}{{Nordlander} \&
  {Lind}}{2017}]{nordlander/17}
{Nordlander} T.,  {Lind} K.,  2017, \mn@doi [\aap]
  {10.1051/0004-6361/201730427}, \href
  {https://ui.adsabs.harvard.edu/abs/2017A&A...607A..75N} {607, A75}

\bibitem[\protect\citeauthoryear{{Pancino} et~al.,}{{Pancino}
  et~al.}{2017}]{pancino/2017}
{Pancino} E.,  et~al., 2017, \mn@doi [\aap] {10.1051/0004-6361/201730474},
  \href {https://ui.adsabs.harvard.edu/abs/2017A&A...601A.112P} {601, A112}

\bibitem[\protect\citeauthoryear{{Pasquini} et~al.,}{{Pasquini}
  et~al.}{2002}]{pasquini/02}
{Pasquini} L.,  et~al., 2002, The Messenger, \href
  {https://ui.adsabs.harvard.edu/abs/2002Msngr.110....1P} {110, 1}

\bibitem[\protect\citeauthoryear{{Piotto} et~al.,}{{Piotto}
  et~al.}{2007}]{piotto/07}
{Piotto} G.,  et~al., 2007, \mn@doi [\apjl] {10.1086/518503}, \href
  {https://ui.adsabs.harvard.edu/abs/2007ApJ...661L..53P} {661, L53}

\bibitem[\protect\citeauthoryear{{Placco}, {Sneden}, {Roederer}, {Lawler}, {Den
  Hartog}, {Hejazi}, {Maas}  \& {Bernath}}{{Placco} et~al.}{2021}]{placco/21}
{Placco} V.~M.,  {Sneden} C.,  {Roederer} I.~U.,  {Lawler} J.~E.,  {Den Hartog}
  E.~A.,  {Hejazi} N.,  {Maas} Z.,   {Bernath} P.,  2021, \mn@doi [Research
  Notes of the American Astronomical Society] {10.3847/2515-5172/abf651}, \href
  {https://ui.adsabs.harvard.edu/abs/2021RNAAS...5...92P} {5, 92}

\bibitem[\protect\citeauthoryear{{Renzini}}{{Renzini}}{1990}]{renzini/90}
{Renzini} A.,  1990, in {Fabbiano} G.,  {Gallagher} J.~S.,   {Renzini} A.,
  eds,  Astrophysics and Space Science Library Vol. 160, Windows on Galaxies.
  p.~255, \mn@doi{10.1007/978-94-009-0543-6_32}

\bibitem[\protect\citeauthoryear{{Smith} \& {Norris}}{{Smith} \&
  {Norris}}{1993}]{smith_norris/93}
{Smith} G.~H.,  {Norris} J.~E.,  1993, \mn@doi [\aj] {10.1086/116418}, \href
  {https://ui.adsabs.harvard.edu/abs/1993AJ....105..173S} {105, 173}

\bibitem[\protect\citeauthoryear{{Sneden}}{{Sneden}}{1973}]{sneden/73}
{Sneden} C.~A.,  1973, PhD thesis, THE UNIVERSITY OF TEXAS AT AUSTIN.

\bibitem[\protect\citeauthoryear{{Stetson}, {Pancino}, {Zocchi}, {Sanna}  \&
  {Monelli}}{{Stetson} et~al.}{2019}]{stetson/19}
{Stetson} P.~B.,  {Pancino} E.,  {Zocchi} A.,  {Sanna} N.,   {Monelli} M.,
  2019, \mn@doi [\mnras] {10.1093/mnras/stz585}, \href
  {https://ui.adsabs.harvard.edu/abs/2019MNRAS.485.3042S} {485, 3042}

\bibitem[\protect\citeauthoryear{{Wiese}, {Fuhr}  \& {Deters}}{{Wiese}
  et~al.}{1996}]{wiese/96}
{Wiese} W.~L.,  {Fuhr} J.~R.,   {Deters} T.~M.,  1996, {Atomic transition
  probabilities of carbon, nitrogen, and oxygen : a critical data compilation}

\makeatother
\end{thebibliography}








\bsp	
\label{lastpage}
\end{document}